\documentclass{ieeeaccess}
\usepackage{cite}
\usepackage{amsmath,amssymb,amsfonts}
\usepackage{algorithmic}
\usepackage{graphicx}
 \usepackage{xspace}
\usepackage{url}
\usepackage{textcomp}
\def\BibTeX{{\rm B\kern-.05em{\sc i\kern-.025em b}\kern-.08em
    T\kern-.1667em\lower.7ex\hbox{E}\kern-.125emX}}
\begin{document}
\history{Date of publication xxxx 00, 0000, date of current version xxxx 00, 0000.}
\doi{10.1109/ACCESS.2017.DOI}

\title{Brain Network Dynamics and Multiscale Modelling of Neurodegenerative Disorders: A Review}
\author{\uppercase{Hina Shaheen}\authorrefmark{1}
\uppercase{Roderick Melnik}\authorrefmark{2}}
\address[1]{Faculty of Science, University of Manitoba, Winnipeg, MB R3T 2N2, Canada}
\address[2]{MS2Discovery Interdisciplinary Research Institute, Wilfrid Laurier University, Waterloo, ON N2L 3C5, Canada}

\tfootnote{The authors are grateful to the NSERC and the CRC Program for their support. This research was enabled in part by support provided by SHARCNET \url{(www. sharcnet.ca)} and Digital Research Alliance of Canada \url{(www.alliancecan.ca)}.}

\markboth
{Author \headeretal: Preparation of Papers for IEEE TRANSACTIONS and JOURNALS}
{Author \headeretal: Preparation of Papers for IEEE TRANSACTIONS and JOURNALS}

\corresp{Corresponding author: Hina Shaheen (e-mail: hina.shaheen@umanitoba.ca).}

\begin{abstract}
It is essential to understand the complex structure of the human brain to develop new treatment approaches for neurodegenerative disorders (NDDs). This review paper comprehensively discusses the challenges associated with modelling the complex brain networks and dynamic processes involved in NDDs, particularly Alzheimer's disease (AD), Parkinson's disease (PD), and cortical spreading depression (CSD). We investigate how the brain's biological processes and associated multiphysics interact and how this influences the structure and functionality of the brain. We review the literature on brain network models and dynamic processes, highlighting the need for sophisticated mathematical and statistical modelling techniques. Specifically, we 
go through large-scale brain network models relevant to AD and PD, highlighting the pathological mechanisms and potential therapeutic strategies investigated in the literature.
Additionally, we investigate the propagation of CSD in the brain and its implications for neurological disorders. Furthermore, we discuss how data-driven approaches and artificial neural networks refine and validate models related to NDDs. Overall, this review underscores the significance of coupled multiscale models in deciphering disease mechanisms, offering potential avenues for therapeutic development and advancing our understanding of pathological brain dynamics.
\end{abstract}

\begin{keywords}
Neurodegenerative diseases, Brain networks, Dynamic processes, Multiphysics modelling, Biochemical activities, Pathological mechanisms, Therapeutic strategies, Mathematical modelling, Statistical analysis and Bayesian inference techniques, Large-scale brain network models, Data-driven approaches, Validation, Inverse problems, AI tools, Systems neurosciences
\end{keywords}

\titlepgskip=-15pt

\maketitle

\section{Introduction}
\label{sec:introduction}
\PARstart{T}{he} human brain represents the peak of complexity in the natural world. The brain, defined by complicated connections and geographically distributed biological components, mostly neurons and glial cells, organizes numerous tasks required for human survival. This neuronal complexity highlights the brain's ability to integrate, coordinate, and interpret sensory information, control visceral and musculoskeletal actions, and perform higher cognitive tasks such as memory, creativity, and problem-solving \cite{carter2019human}. The human brain, weighing over $1400$ grams on average, combines with the spinal cord to create the central nervous system (CNS), containing an estimated $100$ billion neurons. It accounts for around $2\%$ of the total body weight in adults \cite{parvizi2021functional,ziegler2017school}.

The human brain is divided into five primary sections that trace its embryonic origins: myelencephalon, metencephalon, mesencephalon, diencephalon, and telencephalon \cite{shaheen2024multiscale}. These divisions, crucial for understanding its structural organization, serve as the foundation for functional assessments and diagnostic modalities, prominently magnetic resonance imaging (MRI) \cite{herregods2023blurring,daroff2014encyclopedia}. Emphasizing the significance of cerebral anatomy, the cortex, a thin layer of neural tissue enshrouding the brain's surface, epitomizes the epicentre of advanced cognitive processes. Divided into two hemispheres and four lobes—frontal, parietal, temporal, and occipital—the cortex orchestrates a symphony of mental functions such as perception, language, memory, and executive control \cite{nazlee2023age,gonzalez2023neurocognitive,bruner2023parietal,berron2020medial,rehman2023neuroanatomy}. Beyond its structural complexity, the human cortex exemplifies a hub of neuronal intricacy, housing millions of interconnected neurons in networks pivotal for information processing and cognitive functioning \cite{miller2022natural,fletcher2022new,kok2022cognitive}. Supported by a rich milieu of glial cells, the brain's energy demands are met through a delicate balance of oxygen and glucose, facilitated by the intricate interplay of cerebrospinal fluid, blood vessels, and the blood-brain barrier \cite{shaheen2023astrocytic,shekhar2023potential}. Moreover, the brain's functional repertoire extends beyond localized networks to encompass large-scale brain networks, crucial for higher-order cognitive processes, including attentional control, social cognition, and self-referential thinking \cite{oishi2019developmental,nikrahan2023theory,bressler2010large,khodagholy2022large}.

Notably, the human brain, an intricate tapestry of interconnected neurons, represents one of biology's most complex and dynamic systems \cite{dutta2024unsolved,lopez2024digging,seguin2023brain,papo2023does}. At its core, this complexity is governed by networks that operate across various scales - from molecular interactions to large-scale neural circuits. Studying these networks, especially in the context of neurodevelopmental and neurodegenerative disorders (NDDs), is pivotal in understanding the etiology and progression of these conditions. This review paper aims to provide a comprehensive overview of the recent advancements in the multiscale modelling of brain networks and their application in analyzing dynamic processes in NDDs. Multiscale modelling is an interdisciplinary approach that combines data and techniques from neuroscience, physics, mathematics, and computer science \cite{ramaswamy2024data}. This approach has been instrumental in bridging the gap between microscopic neuronal functions and macroscopic brain activities. In NDDs, such as AD, PD, and CSD, disruptions at various scales of brain organization can lead to profound effects on cognition and behaviour \cite{acharya2024brain}. Thus, understanding these multiscale interactions is crucial for developing effective diagnostic and therapeutic strategies. Importantly, models focused on either the micro or macro scales, often overlooking the critical mesoscale processes that mediate these interactions. For instance, Lorenzi et al. \cite{lorenzi2023multi} developed a multi-layer mean-field model of the cerebellum that incorporates microstructural elements and population-specific dynamics, offering a comprehensive framework for studying cerebellar function at multiple scales. The importance of such detailed connectivity is further highlighted by Palesi et al. \cite{palesi2020importance}, who demonstrated how cerebellar connectivity profoundly influences simulated brain dynamics. Additionally, Goldman et al. \cite{goldman2019bridging} bridged the gap between single-neuron dynamics and global brain states, emphasizing the critical role of mesoscale processes in linking micro and macro scales. Similarly, Tesler et al. \cite{tesler2024multiscale} focused on the hippocampus CA1, employing multiscale modeling to elucidate the complex neuronal dynamics within this region. Finally, Overwiening et al. \cite{overwiening2023multi} conducted a multi-scale study of the thalamus, revealing how state-dependent responsiveness at the neuronal level impacts broader brain functions. Integrating these models into whole-brain dynamic simulators like The Virtual Brain (TVB) marks a major advancement in multiscale modelling, allowing for more detailed and reliable simulations that effectively bridge the micro and macro scales \cite{sanz2015mathematical}.

In recent years, machine learning (ML) has emerged as a powerful tool for analyzing complex biological data and refining models of brain connectivity \cite{vatansever2021artificial,Alber2019Integrating}. Through techniques such as artificial neural networks (ANNs) and convolutional neural networks (CNNs), researchers are now able to process large datasets and make more accurate predictions about disease progression. These approaches are particularly beneficial for NDDs, where early detection and predictive modelling are critical to improving patient outcomes. In this review, we aim to integrate recent advances in ML with the broader field of brain network modelling to provide a comprehensive overview of current challenges and opportunities. Understanding brain connectivity is essential for modelling the progression of NDDs. Brain connectivity can be divided into two key types: structural connectivity and practical (functional) connectivity. Structural connectivity refers to the physical connections between different regions of the brain, often mapped using techniques such as diffusion tensor imaging (DTI) \cite{shaheen2023astrocytic}. This provides a ‘wiring diagram’ of the brain, revealing how different areas are linked by neural pathways \cite{keresztes2022introducing}.

In contrast, practical connectivity refers to the functional relationships between brain regions, which are dynamic and can change over time \cite{shibasaki2008human}. These connections are typically identified using functional MRI (fMRI) or other neuroimaging techniques that capture the brain’s activity during rest or tasks. Practical connectivity reflects how regions of the brain communicate and interact in real-time, and it often evolves as a result of neurodegeneration. Distinguishing between these two types of connectivity is crucial for understanding how diseases like AD and PD progress. For example, while structural damage in the brain is often irreversible, practical connectivity can sometimes be restored or compensated for through therapeutic interventions. To fully understand the progression of NDDs, it is necessary to consider the brain at multiple biological scales. At the smallest scale, molecular processes such as protein misfolding and aggregation play a critical role in the onset of diseases like AD \cite{shaheen2023astrocytic}. At the cellular level, these molecular abnormalities disrupt the function of neurons and glial cells, leading to the breakdown of neuronal circuits \cite{shaheen2021neuron}.

As the disease progresses, these disruptions propagate to larger-scale brain networks, affecting entire regions of the brain and leading to cognitive and motor deficits. Multiscale modelling aims to integrate these various levels of organization into a single framework, allowing researchers to simulate how changes at the molecular level can lead to widespread brain dysfunction \cite{Alber2019Integrating}. In particular, we highlight recent advances in coupled multiscale models that link molecular, cellular, and network-level processes \cite{perdikaris2016multiscale}. These models provide insights into how neurodegenerative disorders spread through the brain and offer potential pathways for therapeutic intervention \cite{xia2022bayesian}. By considering the brain as a system of interacting networks at multiple scales, we can develop more accurate and comprehensive models of disease progression. Recent advances in machine learning have opened up new possibilities for improving the accuracy and precision of brain network models. By utilizing large datasets from sources such as the Alzheimer's Disease Neuroimaging Initiative (ADNI) and the Human Connectome Project (HCP), machine learning algorithms can identify patterns and trends that are not immediately apparent through traditional modelling techniques \cite{shaheen2023astrocytic,shaheen2023data}. In particular, CNNs and RNNs have proven effective in predicting disease progression and identifying biomarkers associated with neurodegenerative disorders \cite{ahmed2020artificial,bishara2023state}. These data-driven approaches complement traditional mathematical and statistical models, allowing for a more comprehensive understanding of how diseases like AD and PD develop and spread throughout the brain. By integrating ML with existing models of brain connectivity, researchers can refine their predictions and develop more effective treatment strategies \cite{shaheen2024neural}.

Section II begins by exploring the fundamental concepts of brain networks and their relevance in neuroscientific research. This includes an overview of the different scales at which brain networks operate, ranging from single neurons and synapses to whole-brain networks. Following this, we utilize the methodologies employed in multiscale modelling. These methods encompass computational models that simulate neural processes, neuroimaging techniques for mapping brain networks, and statistical tools for analyzing complex data, as described in Sections III and IV. A significant portion of this review is dedicated to discussing the application of multiscale models in the study of NDDs in Section V. We highlight key findings that illustrate how alterations in small-scale neuronal dynamics can lead to large-scale network dysfunctions in these disorders. Moreover, we examine how these models have enhanced our understanding of the progression and variability of NDD symptoms, as seen in Sections VI and VII.

Finally, in Section VIII, we address the challenges and future directions in the field. This includes integrating multiscale data, developing more sophisticated models that can accurately mimic brain functions, and translating these findings into clinical practice. Our goal is to provide a thorough and insightful exploration of the current landscape of multiscale modelling in brain research, particularly in the context of NDDs, and to stimulate further research in this promising and rapidly evolving field.

In this study, the term "network" is used to describe interconnected structures at various scales within the brain, ranging from large-scale brain networks, which encompass multiple brain regions and their interactions, to mesoscale circuits and microcircuits at the neuronal level. These networks are essential for understanding the complex dynamics of brain function and the progression of neurodegenerative disorders. Our decision to organize the discussion by first addressing large-scale brain networks before moving to smaller-scale networks and neuronal circuits was intentional. This structure allows for a comprehensive overview, starting from the broader, more integrative aspects of brain connectivity and gradually narrowing down to the specific details of individual neuronal interactions. By following this top-down approach, we aim to provide a clear and logical progression that reflects the hierarchical organization of the brain itself.
\section{Large-scale brain networks}\label{lbrainnetwork}
In recent years, the recognition of the brain's composition into various large-scale networks has garnered significant attention \cite{oishi2019developmental,nikrahan2023theory,bressler2010large,khodagholy2022large,greene2016development,vogel2010development}. Central to network neuroscience is the fundamental concept of brain networks, envisioned as systems comprising interconnected components represented as nodes and edges within a graph structure \cite{bassett2017network,uddin2022controversies,papo2023does}. Over the past decades, various methodologies have been employed to delineate and characterize these large-scale brain networks in neuroimaging data \cite{park2013structural,schoonheim2022network,churchland1992computational,braun2018maps}.

The scale at which these brain networks are investigated is crucial, with most studies focusing on macroscale linkages between brain regions due to limitations in data resolution \cite{sporns2022structure,bassett2018nature}. Understanding the cognitive processes underlying human brain function necessitates an appreciation of large-scale brain structure, moving beyond simplistic regional mappings toward recognizing the collaborative role of interconnected brain areas within networks \cite{bressler2010large,downing2001cortical,kanwisher1997fusiform,lynn2019physics}. Anatomical investigations have historically provided insights into the brain's structural organization, while advancements in noninvasive neuroimaging technologies have revolutionized our ability to probe structure and function \cite{friston2009modalities,bandettini2012functional}.

Large-scale brain modelling has significantly advanced our understanding of spatiotemporal neural dynamics, primarily focusing on the cerebral cortex \cite{aqil2021graph,kunze2016transcranial,oishi2019developmental,molnar2019new,shaheen2022multiscale}. These models, often integrating biophysical measurement processes, aim to simulate neuroimaging modalities commonly utilized in clinical neuroscience, such as electroencephalography (EEG), magnetoencephalography (MEG), and functional magnetic resonance imaging (fMRI) \cite{oishi2019developmental,sanz2015mathematical}. By capturing the dynamics of brain activity across multiple spatial and temporal scales, these models offer insights into the complex organization and interactions of brain networks \cite{sporns2013network,sporns2011human,bardsley2018computational}.

Here, we discuss a multiscale computational brain network model to clarify the temporal and spatio-temporal dynamics of the human brain at the network level. According to this methodology, the brain comprises linked networks that function on different temporal and geographical scales. Depending on the needs of cognition, these networks dynamically interact on small and large scales \cite{presigny2022colloquium,d2022quest}. By exploring these interactions, we aim to deepen our understanding of the intricate organization and functioning of the human brain.

\subsection{Brain network concept}
Brain networks can be classified according to their functional interaction or anatomical connection. Anatomical connections between neurons—local synapses that connect neurons—are the foundation of structural networks. The process by which neurons receive impulses at their dendrites and send them unidirectionally as action potentials through the axon is explained by Santiago Ramón y Cajal's dynamic polarization theory \cite{sabbatini2003neurons,jones2015understanding}. Neuronal populations with distinct structural organization or connectivity patterns can be modelled as network nodes, with long axon routes connecting them serving as network edges.

Co-dependent activity in several brain areas under varying functional or behavioural situations is referred to as functional interconnection. The human connectome is a large-scale network made up of linked anatomical regions and pathways that can be mapped at near-millimetre resolution using noninvasive imaging techniques \cite{sporns2012discovering,sporns2005human}. The connectome, which defines large-scale neural dynamics and records them as practical and functional connectivity patterns, offers a comprehensive map of the structural connections in the brain \cite{friston2009modalities,smith2012future}. While functional connectivity studies statistical patterns of dynamic interactions across regions, effective connection searches for networks of causal influence.

\subsection{Brain graph}
The human brain is one of the world's most complex networks, and studies on it have increased dramatically in recent years \cite{sporns2022structure,sporns2022graph,bernhardt2015network}. This growth is fueled by graph theory and network neuroscience advancements, offering insights into brain structure and function \cite{mears2016network}. A mathematical foundation for modelling the interactions between the constituents of a brain network is provided by graph theory \cite{farahani2019application,papo2023does}, typically applied to study functional or structural connectivity, with functional connectivity being the focus in human neuroscience \cite{farahani2019application}.

A brain graph comprises nodes (vertices) and edges, representing pairwise relationships between brain regions. Graphs may be directed or undirected, with weighted edges indicating the strength of connections. Brain networks, or connectomes, visualize the cerebral associations between anatomical brain areas, typically categorized into gray matter and white matter \cite{palombo2020sandi,keresztes2022introducing}. High-angular resolution diffusion imaging datasets are commonly used to construct brain graphs \cite{szalkai2019high}, with numerous datasets publicly available \cite{tadic2019functional}.

Topological features of brain networks, such as small-worldness and modularity, provide insights into complex brain dynamics and cognitive processes \cite{deco2015rethinking,bassett2017network,sporns2013network}. Understanding changes in these features under various conditions may elucidate fundamental processes underlying cognition and neurological disorders \cite{braun2018maps,fornito2015connectomics}. Neuroimaging techniques, including fMRI and DTI, are vital in mapping the human connectome, offering insights into brain physiology and the function \cite{farahani2019application,lacy2022cortical}. The brain graphs can be extracted from diffusion tensor MRI data of healthy subjects, facilitating the analysis of complex brain dynamics \cite{szalkai2017parameterizable,mcnab2013human,kerepesi2016direct}.

\subsection{Large-scale brain networks and NDDs}
NDDs are a diverse group of diseases causing progressive damage to both central and peripheral nervous systems, including AD, various dementias, PD, and Huntington's Disease (HD) \cite{Seeley2010Neurodegenerative, Erkkinen2018Clinical}. These diseases seem to target large-scale neural networks in the brain, a hypothesis emerging from fMRI advances, but not yet thoroughly tested in living humans\cite{Seeley2010Neurodegenerative}. NDDs result in cognitive, behavioural, and motor impairments characterized by misfolded proteins in vulnerable neurons, spreading damage to connected regions. This spread may relate to neural network dysfunction \cite{Seeley2010Neurodegenerative, Buckner2005Molecular, Palop2006Network, Meyer1992Organizational, McGahan2020Mathematical}. The study by Seeley et al. aimed to investigate this in living humans, focusing on early-onset dementia syndromes and using clinical criteria for diagnosis \cite{Seeley2009Neurodegenerative}.

The spread of neurodegeneration may be due to neuron vulnerability in synaptic zones, axonal transport deficits, or propagation of misfolded proteins along neural pathways \cite{Seeley2010Neurodegenerative, Palop2006Network}. This is supported by the observation that connected brain regions have defined axonal associations \cite{sporns2005human, Supekar2008Network}. Large-scale brain networks, crucial in understanding NDDs, are yet to influence neurosurgical perspectives fully. Advances in imaging and analytical tools like graph theory are essential in this exploration \cite{Dadario2022Should, sporns2022structure}. Currently, studies employ diffusion imaging and tractography to map the structural networks of disease proteins and neuronal susceptibilities on a network-specific basis. This approach aims to connect structural networks with brain dynamics in health and disease, enhancing understanding of brain functionality and mental disorders.

\section{NDD biomarkers and their inclusion into modelling processes} \label{biomarker}

A biomarker is a specific characteristic measured to indicate healthy biological processes, harmful biological processes, or responses to treatment \cite{frank2003clinical}. In NDDs, biomarkers are crucial for diagnosis, disease progression monitoring, and treatment effectiveness evaluation. Common NDD biomarkers include imaging biomarkers (MRI, PET), biochemical biomarkers (cerebrospinal fluid, blood-based), and clinical assessments (cognitive, motor)\cite{Giampietri2022Fluid, Young2020Imaging,jama1}. These biomarkers can be used to identify people who may be at risk for NDDs, as well as to predict how the disease will advance and how well a therapy will work \cite{Ranson2023Harnessing, Ewen2021Biomarkers}. Intense learning integrates biomarkers into models, revealing complex relationships between biomarkers and disease outcomes\cite{Peng2021Multiscale}.

NDDs, notably AD and PD, are significant causes of morbidity and mortality worldwide\cite{Babich2022Phytotherapeutic}. AD biomarkers include $A\beta$ and $\tau$ proteins, while $\alpha$-synuclein is a PD biomarker associated with Lewy bodies in the brain\cite{shaheen2021deep}. Mathematical Modelling, including data-driven models, has become vital in understanding NDD pathophysiology and analyzing clinical outcomes\cite{Mengi2021Artificial, vendel2019need}. These models turn physical and physiological concepts into solvable equations, simulating various scenarios for research and hypothesis testing. Models explaining drug pharmacology in the central nervous system are crucial for understanding brain and ocular NDD pathogenesis \cite{vendel2019need, Guidoboni2020Neurodegenerative, Sacco2019Comprehensive, Zhang2020Data}. Mathematical and computational approaches have also advanced the understanding of brain functional connectivity in NDDs\cite{vosoughi2020mathematical, Marzetti2019Brain, Battaglia2020Functional}. PD models, for instance, are categorized into Asyn aggregation, pathogenesis, and pathology propagation models\cite{Bakshi2019Mathematical}. These models vary in complexity and approach, ranging from ordinary differential equations (ODEs) to network models integrated with ODEs\cite{Tandon2022Pathway}. Reliable biomarkers are essential for early PD diagnosis, as clinical symptoms alone make early detection challenging\cite{Surguchov2022Biomarkers}. Understanding biomarkers related to brain mapping and projections is critical in AD and PD research.

\section{Advances in Neuroconnectivity: From Mapping to Functional Geometry in NDDs}\label{brain_connectomes}
This section discusses the organization of brain networks at different scales, from large to small. Large-scale networks encompass functional connections across broad brain regions, whereas small-scale networks involve local neuronal groups or brain areas \cite{Rubinov2010Complex, bassett2017network, Suarez2020Linking, Van2013Network,pal2022nonlocal}. The brain connectome, mapping all neural connections, includes structural and functional connectivity, with new imaging techniques like diffusion MRI enhancing our understanding \cite{Van2013Network, Suarez2020Linking}.

Mapping and projections within the brain connectome are essential for understanding the network and its cognitive processes \cite{Schilling2022Prevalence, Del2017Mapping}. Breaking down the vast large-scale network into smaller, more comprehensible parts helps analyze brain functions and disorders \cite{Keller2022Hierarchical, shaheen2022multiscale, Litwinczuk2022Combination}. In NDDs like AD and PD, biomarkers such as $A\beta$ and $\tau$ proteins in AD, and $\alpha$-synuclein in PD, are linked to these brain mappings \cite{Andica2020Mr, Cummings2019Role, Oliveira2021Alpha, shaheen2022multiscale}. The complexities of large networks in NDDs require a reduction to more minor scales for detailed analysis. Age-related changes in structural networks are often mirrored in functional ones \cite{fjell2016brain, Seidler2010Motor}. Intrinsic neuronal activity, crucial in resting states, enables information exchange and assists in task execution \cite{Zimmermann2016Structural}.

This study examines PD as a multifactorial neurodegenerative disease where aging, environmental, and genetic factors play roles in dopamine deficiency and neurodegeneration \cite{McGahan2020Mathematical}. Age-related biological changes in the brain, from synaptic structures to macrostructural variations, are significant in aging and pathology differentiation \cite{Schultz2017Phases, Fjell2017Relationship}. Machine learning (ML) applications in aging and neurological disorders analysis represent recent computational advancements \cite{Fjell2017Relationship,centofante2023specific}. Deep Brain Stimulation (DBS), effective in various neurological conditions like PD \cite{lozano2019deep}, benefits from optimization in patient selection and stimulation settings. Models like the Rubin-Terman model simulate DBS and can be extended to consider healthy and pathological states, neural activity responses to stimulation, and therapeutic outcome biomarkers \cite{Mestre2020Brain, Lu2019Application}. Reaction-diffusion equations are essential in modelling physical processes. An inverse partial differential equation is produced by adding a spatial component to these time-dependent nonlinear ordinary differential equations \cite{Kaltenbacher2020Inverse}. While the "inverse problem" focuses on identifying a model whose solutions match empirical or real-world data, the "direct problem" entails solving a particular differential equation or system \cite{Isakov2006Inverse}. The usefulness and difficulties of inverse issues in neuroscience, in particular with regard to NDDs, will be covered in the next Section (\ref{inverse}).

\subsection{Inverse problems} \label{inverse}
Inverse problems, prevalent in various scientific and engineering fields, focus on deducing system parameters from indirect observations \cite{Kaltenbacher2020Inverse}. Mathematically, they involve reconstructing a signal 
$f_{true}\in X$  from data $ g \in Y$ given as follows \cite{adler2017solving}:
\begin{equation}\label{Ie}
    g=\mathcal{T}(f_{true})+\delta g
\end{equation}
Topological vector spaces $X$ and $Y$ are involved in the equation (\ref{Ie}) above, where the forward operator $\mathcal{T}:X \rightarrow Y$ defines how a signal creates data without noise. The data's noise component is represented by a single sample in the $Y$-valued random variable $\delta_g \in Y$ \cite{adler2017solving}.

In neuroscience, inverse problems entail inferring neural activity from EEG or fMRI data. Large-scale brain network analysis is challenging due to coarse signals but advances in hemodynamic studies and high-density electroencephalography have made progress \cite{Uhlirova2016Roadmap, Liu2017Detecting}. Conversely, smaller-scale networks offer more precision with technologies like two-photon imaging \cite{Ronzitti2017Recent}. Machine learning techniques are increasingly applied in brain network modelling to address the challenge of linking large-scale brain activity (macro-scale) with more localized, detailed processes (mesoscale). Specifically, these techniques are valuable in solving the inverse problem, where researchers aim to infer the underlying brain dynamics from observed data. These techniques allow for the analysis of complex datasets and extracting meaningful patterns that help bridge the gap between different levels of brain organization. Deep learning is instrumental in solving inverse problems in neurodegenerative disorders. It surpasses traditional methods in accuracy and can incorporate constraints, but it requires extensive data and can be complex to interpret \cite{Raissi2019Physics, Zuo2022Deep, Bergen2019Machine}. Combining multiscale modelling and ML provides a robust approach for analyzing complex systems, like brain networks in neurodegenerative diseases. This synergy is transforming numerous scientific and engineering domains \cite{Peng2021Multiscale, Zampieri2019Machine}. ML's impact is notable in biological sciences and is considered one of the major medical breakthroughs due to its data-driven network architecture \cite{Alber2019Integrating, Peng2020Multiscale}. ML and multiscale modelling complement each other; ML can integrate physics-based principles for managing ill-posed problems, while multiscale modelling can employ ML for surrogate Modelling and system dynamics analysis \cite{Peng2021Multiscale}. In biological sciences, ordinary differential equations are commonly used for observations at various biological levels, as opposed to partial differential equations that describe more complex spatial fluctuations \cite{shaheen2021deep, shaheen2022multiscale, Alber2019Integrating}. Modelling in multi-dimensional parametric spaces, therefore, involves significant challenges in uncertainty quantification.

\section{Multiscale brain modelling and Neurodegenerative disorders}
NDDs present a complex challenge in biomedical research, affecting millions worldwide \cite{agnati2018brain}. These diseases, including AD, PD, and others, involve progressive degeneration of neurons and brain function. Multiscale brain modelling offers a promising approach to understanding the underlying mechanisms of NDDs \cite{andjelkovic2020topology}. By integrating information across different spatial and temporal scales, multiscale models provide a comprehensive view of how alterations at the cellular and network levels contribute to disease progression \cite{arendt2001alzheimer}. These models enable researchers to simulate and analyze the dynamics of large-scale brain networks affected by NDDs, shedding light on the complex interactions between various brain regions and their implications for disease onset and progression. By incorporating data from molecular, cellular, and systems-level studies, multiscale brain models offer insights into the intricate interplay of genetic, environmental, and physiological factors underlying NDDs \cite{andjelkovic2020topology}.
Moreover, multiscale brain modelling facilitates the identification of potential biomarkers for early disease detection and the development of novel therapeutic interventions. By elucidating the dynamic changes in brain structure and function associated with NDDs, these models contribute to advancing personalized medicine approaches tailored to individual patients' needs. Large-scale networks and the structure of the brain are moulded by interaction and change appropriately \cite{leistritz2015time}. Functional segregation and integration are fundamental principles where discrete brain areas handle specific tasks, while all higher-order abilities depend on precise communication across areas \cite{coronel2023whole, dai2023eight, guo2023functional}. Brain capabilities are integrated through locally and globally coupled neural networks \cite{sporns2013network}. Brain network organization supports segregation and specialization, with nodes facilitating effective communication \cite{cohen2016segregation}. Multiscale brain modelling integrates data, mathematical models, and computational techniques to understand brain function \cite{fjell2016brain}.

Temporal distribution refers to how brain functions spread over time, from rapid sensory processing to slower decision-making \cite{shinn2023functional}. Spatiotemporal distribution involves different brain regions engaging in various tasks at other times \cite{jin2022photoacoustic}. Mathematical modelling aids in understanding brain systems and testing hypotheses, utilizing techniques such as stochastic PDEs and ODEs \cite{shaheen2021mathematical}. ODE models help understand brain network disruptions and diseases like Parkinson's \cite{bardsley2018computational}. Multiscale models investigate brain processes across temporal and spatial scales, offering insights into healthy and pathological states \cite{horstemeyer2021multiscale}. Inverse problems in neuroimaging involve reconstructing neural activity from measured data, crucial for understanding brain dynamics \cite{hansen2017spatio}. Bayesian inference enhances the accuracy of neural data analysis by quantifying uncertainty and incorporating prior knowledge \cite{kass2023identification}. Combining low and high-resolution models in a multi-fidelity approach bridges macroscopic and microscopic insights into brain processes \cite{murray2017working}. Molecular dynamics simulations explore finer scales, aiding in understanding diseases like Alzheimer's \cite{jimenez2023macromolecular}. Brain imaging techniques like MRI provide insights into neurodegenerative diseases by detecting structural and functional changes \cite{vecchio2017small}. Resting-state fMRI is commonly used for functional connectivity analysis \cite{yu2018fused}.

Understanding brain mechanics across scales is crucial, though it's challenging \cite{perdikaris2016visualizing}. Advanced mathematical techniques like profound learning offer promise in solving high-dimensional PDEs, enhancing medical image analysis and neuroscience research \cite{han2018solving,weinan2017deep}. AI tools, including deep learning, drive NDDs research \cite{zitnik2019machine}. Human brain modelling aids in understanding behaviours and improving human-machine systems \cite{lu2018mathematical}. Neuronal communication and activity are modelled using integrate-and-fire models and small-world networks \cite{dumont2017stochastic, kim2018effect,thieu2022coupled}. Mathematical models assist in understanding complex phenomena like drug distribution within the brain \cite{vendel2019need}. These models, such as compartmental models, simulate the pharmacokinetics of drugs, predicting how they are absorbed, distributed, metabolized, and excreted within the brain \cite{linninger2009mathematical, alavijeh2005drug}. This is particularly important for optimizing drug delivery across the blood-brain barrier and ensuring therapeutic agents reach their target regions, especially in treating NDDs. Additionally, mathematical models help explore brain function by simulating neural network dynamics, showing how small cellular changes can impact overall brain activity. For example, stochastic models reveal how random neural fluctuations can lead to metastable states, offering insights into the mechanisms of cognitive functions and dysfunctions \cite{pietras2022mesoscopic,shaheen2023data}. Mathematical modelling is crucial in unravelling the complexities of brain function and dysfunction \cite{jimenez2019metastable}. In NDDs like AD, microglial chemotactic signalling and function are essential for brain development and maintaining neuronal networks \cite{colonna2017microglia}. Recent evidence suggests microglia's involvement in modulating blood flow, neurovascular coupling, and other functions \cite{csaszar2022microglia}. Viewing the brain as a self-organized dynamic system presents challenges but offers insights into its complexity \cite{kozma2017cinematic,tadic2021self,tadic2024fundamental}. Understanding brain functions from the structural connectome is a significant challenge \cite{bargmann2013connectome,benkarim2022riemannian,winding2023connectome,lariviere2023brainstat}. Important advancements in diagnosis and therapy have resulted from the importance of large-scale brain networks in the pathophysiology of NDDs \cite{colonna2017microglia}. Further investigation into the intricate connections between brain networks and NDDs holds promise for bettering the prognosis of those who are impacted. NDDs, particularly AD, PD, and CSD, show promise for biomarker integration into modelling procedures, enhancing their analysis \cite{csaszar2022microglia}.

\subsection{Modelling Alzheimer's Disease: Computational Approaches and Brain Networks}
AD is a prominent CNS disorder characterized by memory loss, cognitive impairment, and various neurological and psychiatric symptoms \cite{zhang2020mathematical}. Protein aggregates, observed in postmortem brain tissues of NDDs, are central to AD pathology \cite{fornari2019prion,fornari2020spatially}. AD's etiology is linked to impaired neuronal-astrocytic functions in memory-related brain regions, leading to the formation of extracellular amyloid plaques (APS) and intracellular neurofibrillary tangles (NFTs) composed of $\beta$-amyloid peptide ($A\beta$) and tau protein ($\tau$), respectively. Astrocytes, critical in synapse development and immune response, exhibit altered reactivity in AD, contributing to neuroinflammation and protein aggregation \cite{shaheen2021neuron,thuraisingham2022kinetic,bennett2021activated,shaheen2024multiscale,shaheen2024neural}. The prion-like hypothesis suggests the systematic spread of misfolded proteins via axonal pathways, similar to prion diseases, contributing to AD progression \cite{smethurst2022role,fornari2019prion,fornari2020spatially, prusiner1998prions, jucker2013self}. Dysregulation of intracellular calcium ($Ca^{2+}$) in astrocytes further exacerbates synaptic dysfunction in AD \cite{barros2018multi,wang2018brain}. The amyloid hypothesis implicates $A\beta$ accumulation as a primary driver of synaptic failure and neuronal death in AD \cite{selkoe2016amyloid}. The complex interplay between $A\beta$, astrocytes, $Ca^{2+}$, and $\tau$ underscores the multifaceted nature of AD pathology, necessitating ongoing research and improved care strategies for this global health concern \cite{buskila2019generating,barros2018multi,wang2018brain,selkoe2016amyloid}. 

Several notable mathematical and statistical models have been developed to understand various aspects of AD \cite{shaheen2024multiscale}. One such model is the Smoluchowski model proposed by Murphy and Pallitto, which focuses on the kinetics of $A\beta$ fibril formation, providing insights into the aggregation process \cite{pallitto2001mathematical,shaheen2023astrocytic}. Additionally, Bertsch et al. introduced a model for the accumulation and distribution of $A\beta$ in brain slice geometry, aiding in the understanding of spatial dynamics in AD pathology \cite{bertsch2017alzheimer,pal2022influence}. Hao et al. developed a mathematical model to investigate the kinetics of $A\beta$ aggregation, shedding light on the underlying mechanisms of plaque formation \cite{hao2016mathematical,shaheen2021neuron}. Grasso et al. proposed a molecular dynamics-based model to study the interactions between $A\beta$ peptides and potential therapeutic agents, offering insights into drug development \cite{grasso2020molecular}. Kuznetsov et al. developed a model to elucidate the formation of $A\beta$ oligomers and their role in AD pathology, contributing to our understanding of early disease mechanisms \cite{kuznetsov2018formation}. Dear et al. proposed a catalytic model to investigate the impact of $A\beta$ on neuronal function and survival, providing insights into potential therapeutic targets \cite{dear2020catalytic}. These mathematical and statistical models are crucial in advancing our understanding of AD pathogenesis and may guide the development of novel therapeutic interventions.

AD research has also utilized deep learning algorithms to analyze various data types and aid in the diagnosis, prognosis, and understanding of disease mechanisms. Convolutional neural networks (CNNs) have been used to evaluate neuroimaging data, such as MRI and PET scans, to identify the structural and functional alterations linked to AD \cite{vc2023comprehensive}. Recurrent neural networks (RNNs) have been applied to analyze longitudinal data, such as cognitive assessments over time, to predict disease progression and identify early biomarkers of AD \cite{wang2018predictive}. Artificial brain images have been produced using generative adversarial networks (GANs), which have helped to supplement small datasets and increase the resilience of deep learning models \cite{bowles2018gan}. Furthermore, attention mechanisms have been incorporated into deep learning architectures to prioritize relevant information in large-scale omics data, such as genomics and proteomics, facilitating the discovery of novel biomarkers and therapeutic targets for AD \cite{dragomir2018network}. Additionally, transfer learning approaches have been used to increase performance on AD-related tasks, such as text-based diagnosis or prediction from electronic health records, by using pre-trained deep learning models on huge datasets from related tasks, such as natural language processing \cite{de2020artificial}. In addition, recent advancements in using the “Virtual Brain” (TVB) simulator have been significant in modelling approaches to AD. TVB has shown great promise in simulating network-specific parameters relevant to neurodegenerative dementias. For example, TVB simulations can reveal critical insights into network-specific changes associated with AD, particularly concerning metabolic alterations in brain networks \cite{monteverdi2023virtual}. Profound learning advancements promise to enhance our understanding of AD and develop more accurate diagnostic tools and effective treatments.

\subsubsection{Computational Models in Alzheimer's Disease}
The advent of computational modelling has revolutionized the study of AD, enabling researchers to unravel the intricacies of its pathogenesis and progression. These models encompass various scales and aspects, from molecular interactions to neural network dynamics, offering a multifaceted view of the disease. As shown in Fig. \ref{fig1a}, some computational modelling descriptions are as follows. 
\Figure[t!]()[width=0.95\textwidth]{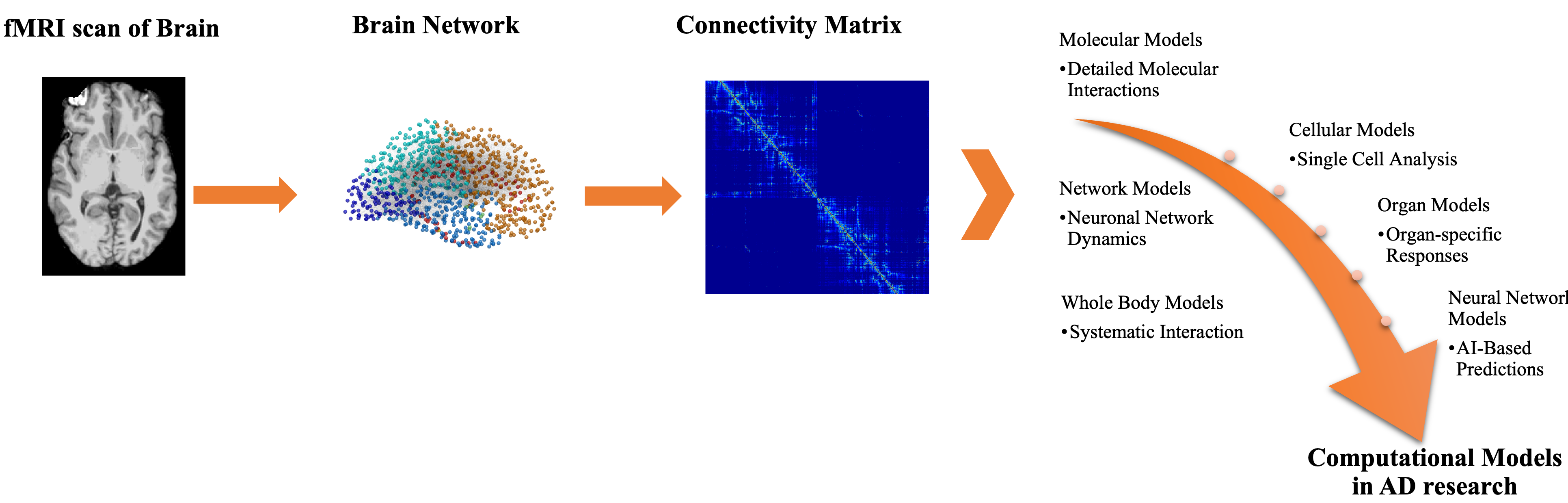}
   {(Color online) Computational models generated from the functional and effective connectivity patterns of the fMRI scan of the brain.\label{fig1a}}

1. Molecular and Cellular Models: At the molecular level, computational models simulate the interactions between different proteins implicated in AD, such as amyloid-beta and tau. These models help understand the aggregation of these proteins and their subsequent neurotoxic effects \cite{vosoughi2020mathematical}. Cellular models further extend this understanding by simulating neuronal cell responses to these aggregations, providing insights into the mechanisms of neuronal degeneration and death characteristic of AD \cite{slanzi2020vitro}. 

2. Neural Network Models: Beyond individual cells, computational models of neural networks in AD illustrate how pathological changes disrupt neural communication \cite{maestu2021neuronal}. These models focus on how synaptic dysfunction and neuronal loss lead to the breakdown of neural networks, correlating with the cognitive deficits observed in AD patients \cite{kashyap2019synapse}. By simulating neural circuits, these models offer insights into the functional changes occurring in the brain, thereby elucidating the relationship between neural degeneration and cognitive decline.

3. Integration of Multiscale Data: A significant strength of computational models lies in their ability to integrate data across different biological scales \cite{butcher2004systems,shaheen2022multiscale}. This integration is critical for understanding AD, as the disease involves complex molecular, cellular, and neural interactions. Computational models, therefore, serve as a nexus for multiscale data, offering a more holistic view of the disease's progression and impacts \cite{yang2021bicoss}.

4. Advancing Research and Therapeutic Strategies: The use of computational models in AD research extends beyond academic understanding \cite{argyle2023out,shaheen2021analysis,shaheen2021mathematical,shaheen2022multiscale,shaheen2023bayesian}. These models are increasingly instrumental in identifying potential therapeutic targets and in simulating the effects of new drugs \cite{shaheen2023data,shaheen2023bayesian}. By providing a virtual testing ground for novel treatments, computational models can potentially accelerate the development of effective therapies for AD.

\subsubsection{Spatial and Temporal Aspects in Alzheimer's Modelling}
AD modelling takes on added dimensions when incorporating spatial and temporal aspects, offering a dynamic view of the disease's progression and its impact on brain structure and function. Fig. \ref{fig:2} illustrates AD's spatial and temporal progression patterns. This figure shows the progression over time (from early to end-stage) and how different brain areas are affected.

1. Spatial Modelling: Understanding Structural Changes: Spatial models in AD research focus on the structural alterations in the brain caused by the disease \cite{argyle2023out,shaheen2022multiscale,pal2022influence,shaheen2023bayesian}. These models analyze changes in brain volume, cortical thickness, and the integrity of white matter tracts. Advanced imaging techniques like MRI are often used to gather spatial data, which computational models then utilize to map the disease's progression \cite{ajabani2023predicting,shaheen2023data,shaheen2023astrocytic}. This spatial modelling is crucial for understanding how AD affects different brain regions, often highlighting early hippocampus and entorhinal cortex changes critical for memory formation.

2. Temporal Modelling: Capturing Disease Progression: Temporal models add the dimension of time to AD research, allowing scientists to track the progression of the disease \cite{shaheen2021deep,shaheen2021neuron}. These models are vital for understanding the chronological sequence of events in AD pathology, from the initial accumulation of amyloid-beta plaques to the eventual cognitive decline. Temporal models help predict the disease's trajectory, essential for early diagnosis and intervention.

3. Bridging Spatial and Temporal Data: Integrating spatial and temporal data in computational models provides a comprehensive understanding of AD. This integration allows researchers to observe the disease's current state and predict future changes in brain structure and function \cite{agha2021vitro,acharya2024brain,agnati2018brain}. This holistic approach is crucial for developing targeted therapies that address specific stages of the disease \cite{ahmed2020artificial,albahri2023systematic}. Understanding AD's spatial and temporal aspects is vital for research and clinical practice. These models can guide clinicians in tailoring interventions and monitoring patient disease progression, offering a more personalized approach to treatment.

\Figure[t!]()[width=0.95\textwidth]{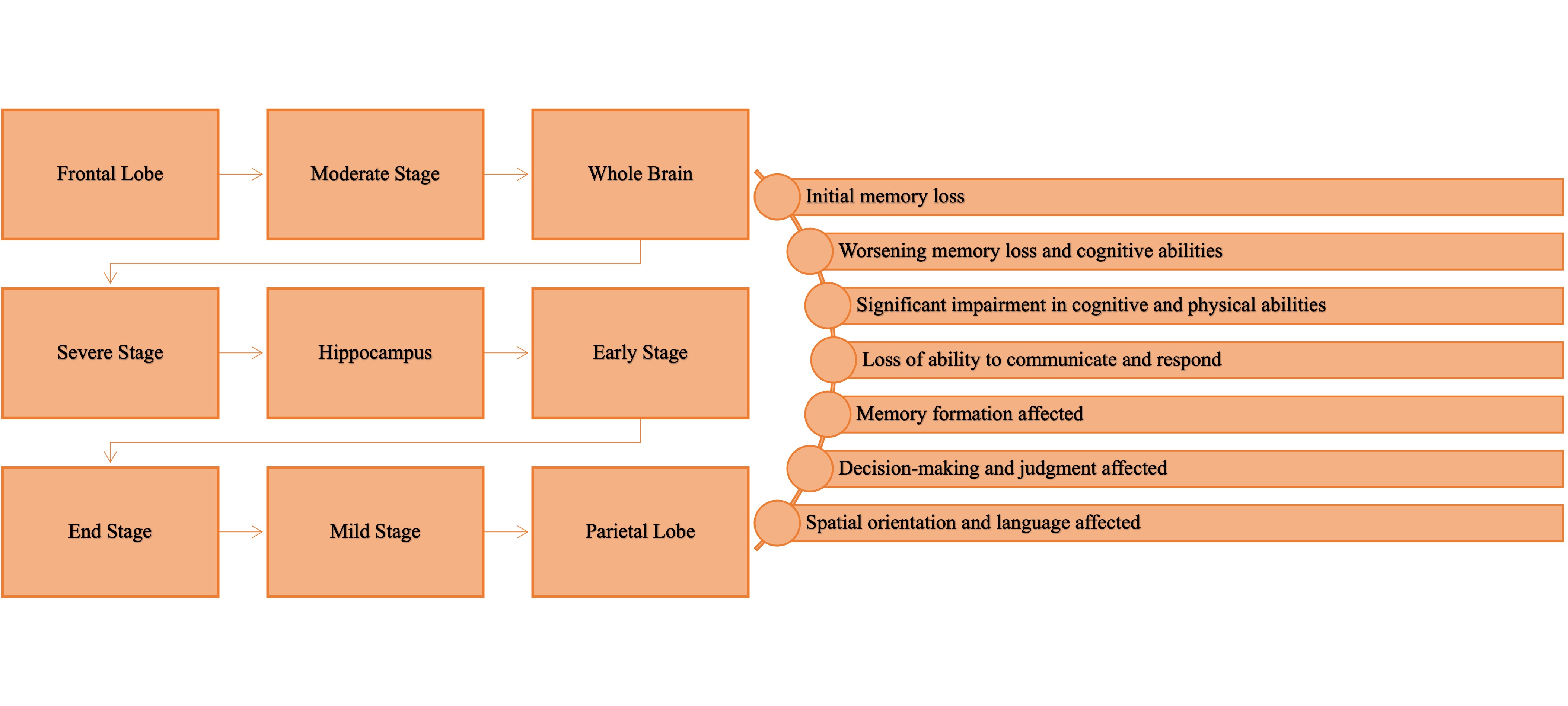} 
   {(Color online) The spatial and temporal progression patterns of AD in the brain.\label{fig:2}}
\subsubsection{Brain Networks and Alzheimer's Disease}
The study of brain networks in AD provides crucial insights into how the disease impacts cognitive functions. Analyzing brain networks entails examining the relationships and exchanges between distinct brain areas crucial to various mental functions. In AD, these networks are disrupted, leading to the characteristic symptoms of memory loss and cognitive decline.

1. Impact of Alzheimer's on Brain Network Connectivity and Functionality: AD significantly alters the brain's network connectivity and functionality. The disturbance of the default mode network (DMN), which is engaged in memory and self-referential thought while the brain is at rest, is one of the first indications of AD \cite{valera2021multimodal,wei2021self}. As AD progresses, other networks, such as the salience and executive networks, also exhibit diminished connectivity. This disruption correlates with the severity of cognitive symptoms, illustrating the impact of network degradation on mental capabilities.

2. Incorporating Brain Network Changes in Computational Models: Computational models in AD research have evolved to incorporate changes in brain networks. These models replicate how various brain areas interact with one another and how AD disease affects them \cite{pal2022influence,shaheen2022multiscale,shaheen2023astrocytic}. By integrating neuroimaging data, such as fMRI and PET scans, computational models can visualize the alterations in network connectivity and predict how these changes progress over time. This approach is invaluable for understanding the systemic nature of AD and its impact on brain functionality.

\subsubsection{Integrating Spatial, Temporal, and Network Models}
Integrating spatial, temporal, and network approaches in computational models represents a significant advancement in Alzheimer's research \cite{golriz2020challenges}. By combining these perspectives, researchers can develop comprehensive models that provide a more accurate and holistic understanding of the disease. Spatial models highlight the structural changes, temporal models capture the progression over time, and network models elucidate the functional impairments in the brain \cite{veitch2019understanding}.

From the perspective of computational neuroscience, the intricate cognitive functions affected by AD and similar disorders are understood to arise from the amalgamation of ongoing dynamic functional processes at micro and mesoscales within a relatively constant spatial structure \cite{park2013structural}. This implies that higher mental functions result from computations conducted through the dynamic integration of local processing units at these scales. These integrated components might be thought of as network topologies inside of hierarchical adaptive architectures or large-scale ensembles of coordinated neural activity \cite{bullmore2009complex}. These network models can be analyzed as bigraphs representing momentary brain states, which evolve to form recognizable static functional network structures. Within this framework, specific network configurations are associated with distinct classes of cognitive abilities \cite{sporns2021dynamic,shine2019human}. Thus, spatially mapping these network configurations may offer insights into the brain networks crucial for optimizing perception, cognition, and behaviour. Moreover, these networks and cognitive abilities are implicated in neurodegenerative brain diseases.

Traditional regional approaches to understanding the brain-behaviour relationship in clinical settings are being supplanted by functional network approaches \cite{fox2018mapping}. Functional network topologies may be characterized using a low-dimensional manifold, despite the seemingly impossible job of indexing several brain state configurations due to their high dimensionality \cite{korhonen2021principles}. This suggests that the combinations of brain states may be represented in a lower-dimensional space so that a vector in this space serves as the primary characterization for each condition. This manifold's neurotransmitter-modifiable activity may be related to different cognitive capacities. Although movement neuroscience frequently uses low-dimensional concepts, computational functions relevant to clinical neurodegenerative disorders, such as perception, cognition, and behaviour, function at a different level \cite{shine2021computational}.

Rather than associating impairment in specific cognitive functions with particular brain regions, a model using trajectories in a continuous or discrete manifold can be utilized to represent network topologies linked to high-level cognitive abilities \cite{shaheen2022multiscale,shaheen2023bayesian,shaheen2023astrocytic}. A particular dementia condition may only be specifically linked to disruption in a subset of this manifold. These models can aid in the explanation of the similarities between illness states' altered functional connectivity patterns and brain atrophy. Uncertainty surrounds the underlying neurobiology that permits AD and kindred illnesses to specifically impair some cognitive functions, brain networks, or areas while sparing others. The paradox of syndromic diversity is a problem that arises from this variety in selectivity, which causes a range of cognitive symptoms among various forms of AD dementia \cite{jones2022computational}. Examining this variation and the underlying causes of it should shed light on the neurobiology behind syndromic variations.  According to recent studies, variations in the macroscale functional pathophysiology of AD, as opposed to only the molecular level, may contribute to the variety of mental capacities impacted by AD \cite{sintini2021tau, jones2022computational,shaheen2023bayesian}. This is consistent with broad ideas of network failure in neuropsychiatric disorders and emphasizes the role of large-scale network dynamics in disease causation. Further development of AD models incorporating large-scale network physiology requires a complete model that connects mental capacities, brain networks, and neuroanatomy. Nonetheless, proof of the specific degeneration of areas or modes within the manifold outlined must be provided.

1. Significance of Integrated Models: These integrated models are crucial for understanding the complex nature of Alzheimer's and its progression. They allow researchers to observe the current state of the disease and forecast its future development. This comprehensive view is essential for identifying critical intervention stages and developing strategies to delay or prevent the onset of severe symptoms.

2. Future Directions and Multiscale Modelling: The potential of multiscale Modelling in AD research is immense. As computational power and data collection techniques advance, these models will become even more sophisticated, offering finer resolutions and more accurate predictions. The future of AD research lies in harnessing the power of these integrated models to uncover new therapeutic targets and personalize treatment strategies based on individual disease progression patterns.

\subsection{Parkinson's disease (PD)}
Intraneuronal Lewy bodies (LBs) and Lewy neurites accumulate synuclein, which causes dopaminergic neurons in the substantia nigra pars compacta to die \cite{hijaz2020initiation,vaz2022extracellular,goloborshcheva2020reduced}. This is the hallmark of PD. Approximately $1\%$ of people over 65 have PD, which is the second most prevalent neurological condition worldwide. Symptoms include bradykinesia, resting tremors, and muscle stiffness. Filamentous $\alpha$-synuclein ($\alpha$-syn), which accumulates as a result of improper phosphorylation in Parkinson's disease patients, is the main component of LBs \cite{spillantini1998alpha}. The development of PD is primarily driven by misfolding and aggregation of $\alpha$-syn, but other processes that are also involved include neuroinflammation \cite{hirsch2009neuroinflammation}, mitochondrial dysfunction \cite{bose2016mitochondrial} and aberrant protein clearance systems \cite{goedert2013100,mckinnon2014ubiquitin,testa2011germline}. Interestingly, the intracellular $Ca^{+2}$ concentration controls the release of $\alpha$-syn into the extracellular environment or packs it into exosomes via the endosome route. This modulation of various physiological events has been linked to acute neurodegenerative disorders, including PD \cite{yu2020potential,shaheen2021mathematical,jain2019neuroprotection}. \cite{yu2020potential,shaheen2021mathematical,jain2019neuroprotection}. DBS has shown efficacy in treating PD \cite{lozano2019deep}. Globally, PD has seen a significant increase in disability-adjusted life years (DALYs) and deaths since $2000$, with nearly $90,000$ new cases diagnosed annually in the USA alone, according to the World Health Organization (WHO).

PD research relies on various mathematical and statistical models, each providing valuable insights into different aspects of the disease. Compartmental models, such as those proposed by Kunze et al., divide the brain into distinct compartments representing different neuronal populations or regions affected by PD pathology \cite{kunze2016transcranial}. These models simulate the dynamics of neurotransmitter release, protein aggregation, and neuronal activity, aiding in understanding disease progression and evaluating therapeutic interventions. Stochastic models incorporate random fluctuations to capture the inherent variability in PD progression and treatment response \cite{liu2020stochastic}. By accounting for uncertainty, these models provide insights into individual disease trajectories and the efficacy of personalized treatment strategies. Bayesian models integrate data from diverse sources, such as clinical assessments, neuroimaging, and genetic analyses, using probabilistic frameworks \cite{vu2012progression,khanna2018using,shaheen2023bayesian,shaheen2023data}. These models enable the estimation of disease progression rates, identification of predictive biomarkers, and optimization of treatment decisions based on individual patient characteristics. Moreover, network models characterize the complex interactions between brain regions and neuronal circuits implicated in PD pathophysiology \cite{alexander2004biology,amartumur2024neuropathogenesis}. These models use graph theory and network analysis to identify critical nodes and pathways contributing to disease progression, guiding the development of targeted interventions.

ML models, including support vector machines (SVMs), deep neural networks (DNNs), and random forests, utilize large-scale datasets to uncover hidden patterns and predictors of PD progression and treatment response. Some notable studies demonstrate the potential of ML in personalized medicine and precision healthcare for PD patients \cite{ahmed2019machine,johnson2021precision,gupta2023perspective}. Markov models describe transitions between disease states or treatment outcomes based on probabilistic rules \cite{dams2011modelling,siebert2012state,li2024cost}. These models assess the cost-effectiveness of interventions, optimize treatment strategies, and simulate long-term disease trajectories in PD populations. Population-based models, as seen in the previous studies, integrate epidemiological data with mathematical frameworks to estimate disease prevalence, incidence, and mortality rates \cite{kopec2010validation,chiavenna2019estimating,chowell2016mathematical,chen2014modeling}. By projecting future trends in PD burden, these models inform public health policies and resource allocation for PD management and prevention.

1. Alpha-Synuclein Aggregation: A fundamental molecular change in PD is the abnormal accumulation of alpha-synuclein, a protein primarily found in the brain. These accumulations, known as Lewy bodies, are considered a hallmark of PD \cite{shaheen2022multiscale,shaheen2021deep}. They disrupt normal cellular processes, leading to neuronal damage and death. Studies have shown that the misfolding of alpha-synuclein can cause a toxic gain of function, contributing to neurodegeneration \cite{dauer2003parkinson,deffains2019parkinsonism}. Research into how these proteins aggregate and cause cell death is ongoing, focusing on their structure, the conditions promoting aggregation, and how this process can be targeted therapeutically.

2. Dopamine Neuron Loss: The degeneration of dopaminergic neurons in the substantia nigra pars compacta is a crucial component of the pathogenesis of PD. This loss is responsible for the characteristic motor symptoms of PD, such as bradykinesia, rigidity, and tremors \cite{mamelak2018parkinson,park2013structural}. The degeneration of these neurons is thought to be linked not only to alpha-synuclein pathology but also to mitochondrial dysfunction, oxidative stress, and neuroinflammation \cite{picca2020mitochondrial}. Understanding the interconnected pathways leading to the death of dopaminergic neurons is essential for developing effective treatments.

3. Modelling Molecular Dynamics: In recent years, computational models have become invaluable in simulating the molecular dynamics of PD \cite{shaheen2022multiscale,mamelak2018parkinson,deffains2019parkinsonism}. These models allow for the exploration of alpha-synuclein interactions at a molecular level, providing insights into how misfolding occurs and how it might be inhibited. Molecular dynamics simulations, for example, offer a detailed view of the structural changes in alpha-synuclein and their impact on neuronal cells.

4. Protein Misfolding in PD: Computational modelling has also been crucial in understanding the process of protein misfolding and aggregation in PD \cite{tan2009protein}. These models help identify potential binding sites for therapeutic molecules, predict the aggregation propensity of alpha-synuclein, and understand the influence of genetic mutations on protein structure and function \cite{vidovic2022alpha}. Advanced computational techniques, like ML algorithms, are being integrated to predict disease progression and responses to potential treatments based on protein folding patterns.

5. Neuronal Dysfunction and Death:
 Computational models play a pivotal role in elucidating the mechanisms of neuronal dysfunction and death in PD. These models integrate data from various sources, including genetic, molecular, and cellular studies, to simulate the neuronal pathways involved in PD. They provide insights into the cascade of events triggered by alpha-synuclein aggregation, mitochondrial dysfunction, oxidative stress, and neuroinflammation \cite{muddapu2019computational}. These models are also instrumental in simulating the specific pathways leading to the death of dopaminergic neurons. By integrating various cellular stressors observed in PD, such as ER stress, mitochondrial impairments, and disruptions in calcium homeostasis, researchers can predict how these factors converge to cause neuronal death. To develop strategies for protecting neurons or restoring their function in PD, it's crucial to understand how disruptions in neural circuits contribute to the disease. Neural circuit models can interpret these disruptions, guiding targeted interventions aimed at preserving neuronal health and functionality, as described next.

\subsubsection{Neural Circuit Models}
1. Altered Neural Circuits in PD:
PD significantly affects brain circuits, particularly those within the basal ganglia, a group of nuclei associated with various functions, including motor control. Fig. \ref{fig:p} shows the list of computational models generated from the multiscale brain network of the cortex-basal ganglia thalamus system. 

\Figure[t!]()[width=0.95\textwidth]{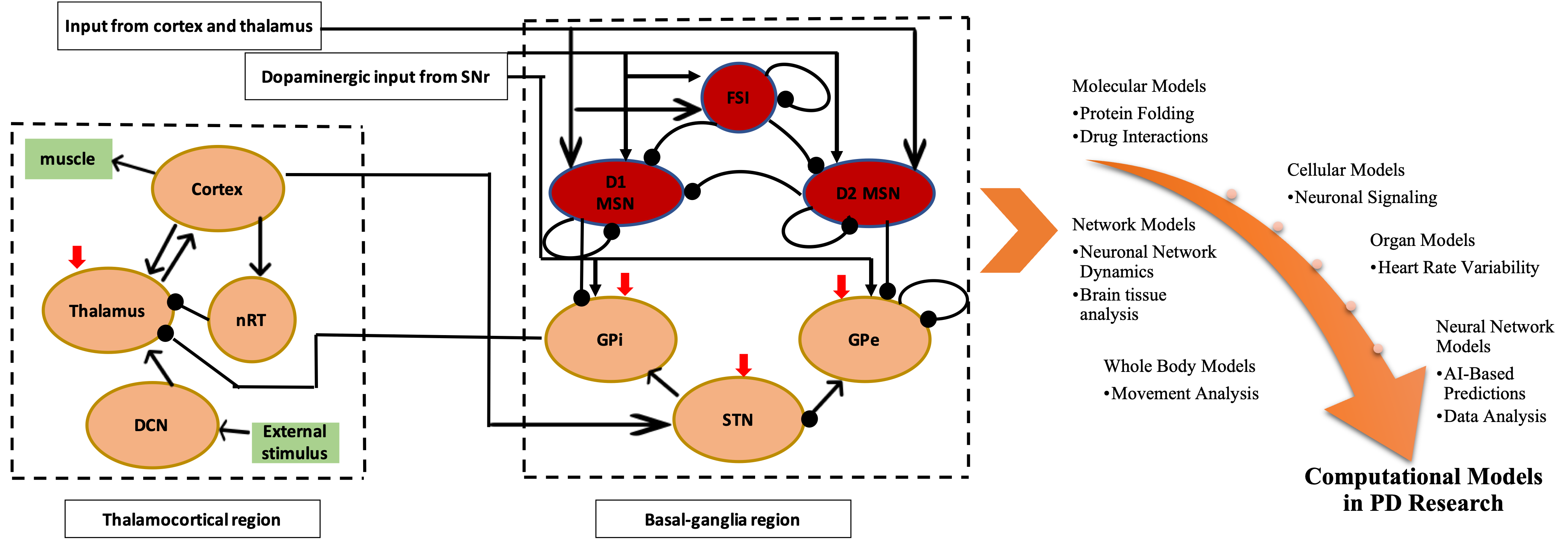} 
   {(Color online) Computational models describing possible mechanisms for cortex-basal ganglia thalamus system in Parkinson's Disease \cite{shaheen2022dbs}.\label{fig:p}}

\subsubsection{Molecular and Cellular Level Models}
In PD, the degeneration of dopaminergic neurons in the substantia nigra leads to a disruption of the normal dopamine signalling in the basal ganglia. This disruption causes an imbalance between the direct (stimulatory) and indirect (inhibitory) pathways of movement control \cite{shaheen2022multiscale,shaheen2021deep}. The net effect is an increased inhibitory output from the basal ganglia to the thalamus and a decreased motor cortex activation, leading to motor symptoms of PD such as bradykinesia, rigidity, and tremor.

2. Computational Models of Circuit Dysfunction:
Computational models have been developed to simulate the altered neural activity observed in PD. These models aim to replicate the changes in firing patterns, synaptic plasticity, and neurotransmitter levels within the basal ganglia circuits. By altering parameters such as dopamine levels, synaptic weights, and neuronal excitability, these models can mimic the motor symptoms of PD \cite{shaheen2022multiscale}. They also help in understanding the underlying mechanisms of DBS, a therapeutic intervention used in PD, by simulating the effects of electrical stimulation on neural circuits.

\subsubsection{Brain Network and Systems Level Models}

1. Large-Scale Brain Network Changes:
In PD, changes are not limited to local circuits but also affect large-scale brain networks. Neuroimaging studies, including functional MRI (fMRI) and PET scans, have revealed alterations in various brain networks associated with both motor and non-motor symptoms of PD \cite{barber2017neuroimaging,bargmann2013connectome}. These include changes in the cortico-basal ganglia-thalamo-cortical loop, the default mode network, and the executive control network. The disruption in these networks correlates with the severity of PD symptoms and can even precede the onset of the classic motor symptoms.

2. Modelling Network Dynamics:
Computational models at the network and system levels simulate the interactions between different brain regions in PD. These models incorporate data from neuroimaging studies to construct network-based models replicating brain activity dynamics in PD. They help understand how molecular and cellular changes translate into alterations in large-scale brain networks. Such models are crucial in studying PD's progression and evaluating the impact of interventions like pharmacological treatments and DBS on brain network functionality.

\subsubsection{Integrative and Multiscale Models}
1. Bridging Scales:
To comprehensively understand PD, models that integrate multiple scales are necessary. These multiscale models combine data from the molecular, cellular, circuit, and system levels to form a cohesive picture of the disease. By bridging these scales, these models can trace the pathophysiological progression from molecular alterations, such as alpha-synuclein aggregation, to changes in large-scale brain networks and the emergence of clinical symptoms.

2. Systems Biology Approaches:
Systems biology models in PD adopt a holistic view, incorporating diverse data types and analyses to understand the disease's progression and response to treatments. These models often use computational techniques to analyze large datasets, including genomics, proteomics, and neuroimaging data. By integrating these various data types, systems biology models can identify critical molecular and cellular pathways implicated in PD, predict disease progression, and propose novel therapeutic targets. They represent a promising approach in pursuing personalized medicine for PD, where treatments could be tailored to the specific molecular and physiological profile of individual patients.

\subsection{Cortical spreading depression (CSD)}
Initially observed by Leão in 1944, CSD manifests as a slow wave propagating across the cerebral cortex, disrupting brain homeostasis and temporarily impairing synaptic transmission \cite{leao1944spreading}. Linked to various clinical conditions, including migraine, ischemic stroke, and epilepsy, CSD plays a significant role in the pathogenesis of neurological disorders \cite{gerardo2017computational,cozzolino2018understanding}. Neuronal excitability, characterized by action potentials and refractory periods, underlies CSD initiation and propagation \cite{chamanzar2018algorithm}. Numerous biophysical and neurochemical factors, including ion channels, neurotransmitter concentrations, and glial activity, contribute to CSD dynamics \cite{tuckwell2013stochastic,tuckwell1981ion,chang2013mathematical,duman2012synaptic,ellingsrud2022validating}. Mathematical models have been instrumental in elucidating CSD mechanisms and exploring factors influencing its onset, propagation, and cortical depolarization \cite{shibasaki2008human,huang2011simplified,kager2002conditions, shapiro2001osmotic,chang2013mathematical}. 

One recent mathematical model based on partial differential equations has been utilized to describe the spatial and temporal dynamics of CSD initiation, propagation, and termination within the cerebral cortex \cite{shaheen2021analysis}. This model incorporates key biophysical parameters such as ion fluxes, membrane potentials, and neurotransmitter concentrations to simulate CSD dynamics accurately \cite{tuckwell2013stochastic}. Another recent statistical model developed by Shukla et al. employs Bayesian inference techniques to analyze experimental data and infer underlying mechanisms of CSD initiation and propagation \cite{shukla2022molecular}. By integrating data from multiple sources, including electrophysiological recordings and imaging studies, this model provides insights into the complex interactions between neural and glial elements during CSD events \cite{shukla2022molecular}. Furthermore, ML approaches have been increasingly applied to CSD research for predictive modelling and pattern recognition tasks. For instance, Gerardo et al. utilized deep learning algorithms to analyze high-dimensional neural data and identify characteristic features associated with CSD onset and propagation. Their approach demonstrated the potential of ML techniques in capturing complex spatiotemporal patterns underlying CSD dynamics \cite{kroos2017patient}. thereby enhancing our understanding of the biological basis of CSD. This integration of ML can provide deeper insights into the mechanisms driving CSD and inform strategies for therapeutic interventions, as described below.

\Figure[t!]()[width=0.95\textwidth]{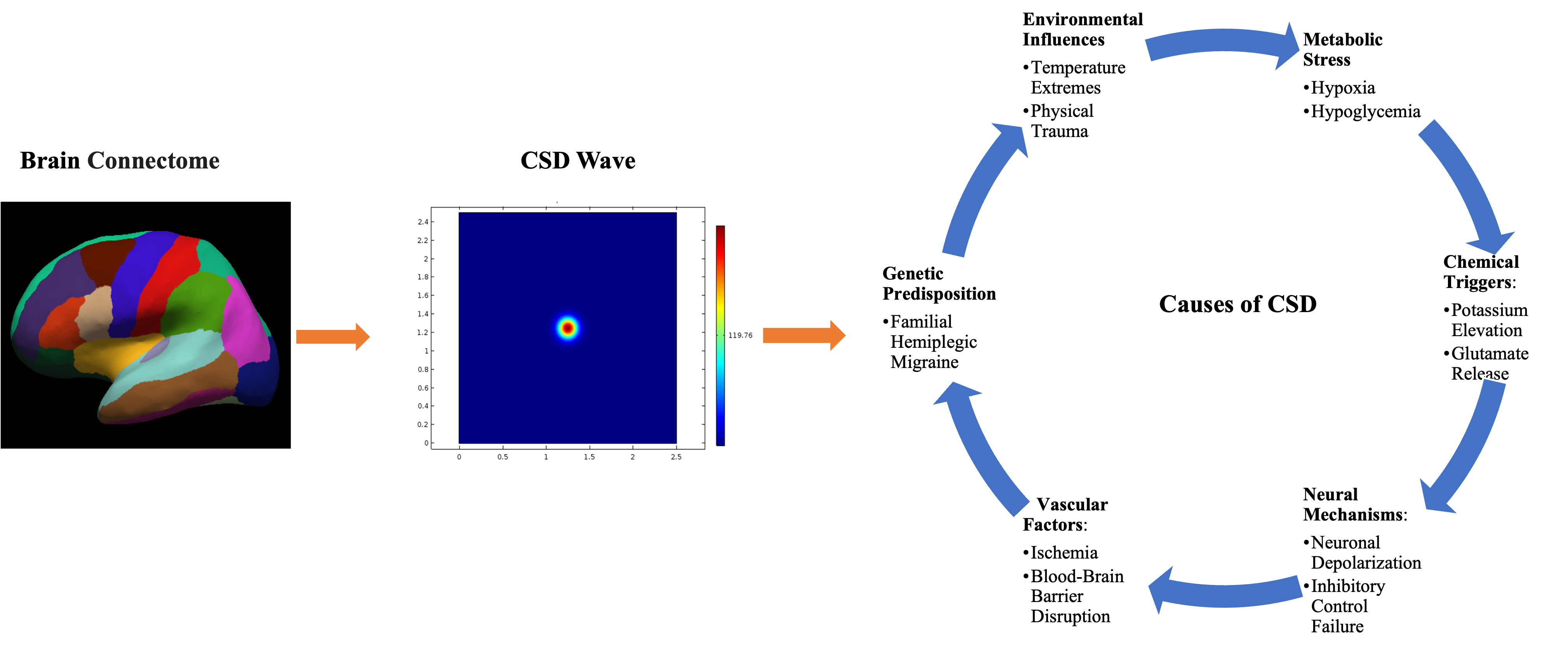}
   {(Color online) Causes of Cortical Spreading Depression (CSD) in human brain.\label{figcsd}}
\subsubsection{Biological Basis of Cortical Spreading Depression (CSD)}
 
1. Cellular Mechanisms:
CSD is characterized by a wave of depolarization followed by a period of suppressed neural activity. The critical cellular event in CSD is a massive ion flux, especially potassium ($K^+$) and calcium ($Ca^{2+}$) \cite{shaheen2021analysis}. During CSD, there is an efflux of $K^+$ from neurons and an influx of $Ca^{2+}$ into cells. This ion imbalance triggers a cascade of events, including the release of various neurotransmitters and a change in neuronal membrane potentials. These changes further contribute to the spreading nature of CSD, as the ion imbalance in one neuron can affect its neighbours. Fig. \ref{figcsd} shows the factors responsible for CSD in the brain.

2. Glial Contribution:
Glial cells, particularly astrocytes, play a crucial role in the initiation and propagation of CSD. Astrocytes are involved in ion homeostasis in the brain and can influence neuronal activity. During CSD, astrocytes can contribute to the propagation of the wave through their network, facilitating the spread of ion fluxes and neurotransmitter imbalances \cite{rovegno2018role}. This glial-neuronal interaction is a crucial aspect of CSD dynamics and is a focus of ongoing research.

\subsubsection{Computational Modelling of CSD}
1. Single-Cell Models:
Single-cell models in CSD focus on the intracellular processes of neurons and glial cells \cite{saetra2021electrodiffusive,xu2020mathematical}. These models aim to replicate the cellular responses during CSD, such as ion fluxes, neurotransmitter release, and changes in membrane potentials. By simulating these processes, single-cell models help understand the initiation and early propagation mechanisms of CSD at the cellular level.

2. Network Models:
Network models of CSD consider how the phenomenon propagates through neural networks \cite{reshodko1975computer,pettersen2012extracellular,desroches2019modeling,de2020neurons}. These models simulate the spread of CSD across interconnected neurons, considering factors like synaptic connectivity and network topology. Such models are essential for understanding how localized cellular events can lead to widespread neural depression characteristic of CSD. Network models of CSD focus on the propagation of this wave-like event through intricate networks of neurons, capturing the essence of how CSD evolves across the cerebral cortex \cite{plenz2021self}. These models are designed to simulate the movement of CSD by considering key factors such as synaptic connectivity—the links between neurons—and the overall topology of the neural network, which includes the structure and arrangement of these neurons \cite{sun2021future,chaisangmongkon2017computing}. This approach is critical because it allows researchers to comprehend how CSD, starting as localized cellular events, can escalate into a widespread and coordinated neural depression. The detailed simulation of synaptic interactions and neural network pathways in these models sheds light on the precise mechanisms underlying CSD, offering valuable insights into its pathophysiology. Fig. \ref{fig:csd2} shows the computational models for CSD in the brain and its applications.

Recent advancements in network modelling of CSD have integrated more sophisticated computational techniques and neurological insights \cite{einevoll2013modelling,shaheen2021analysis,oyetunde2018leveraging}. These recent models often employ complex algorithms to more accurately replicate the dynamics of neural interactions and the spread of CSD. For example, some of the newer models incorporate nonlinear dynamics to represent more realistically how neurons behave under stress conditions like those during CSD. These models also consider the plasticity of synaptic connections—how the strength and efficacy of synapses can change—which is a crucial factor in understanding CSD's progression and effects. Additionally, there has been a focus on integrating multimodal data, including electrophysiological and imaging data, into these models, enhancing their accuracy and relevance \cite{michelson2018multi}.
\Figure[t!]()[width=0.5\textwidth]{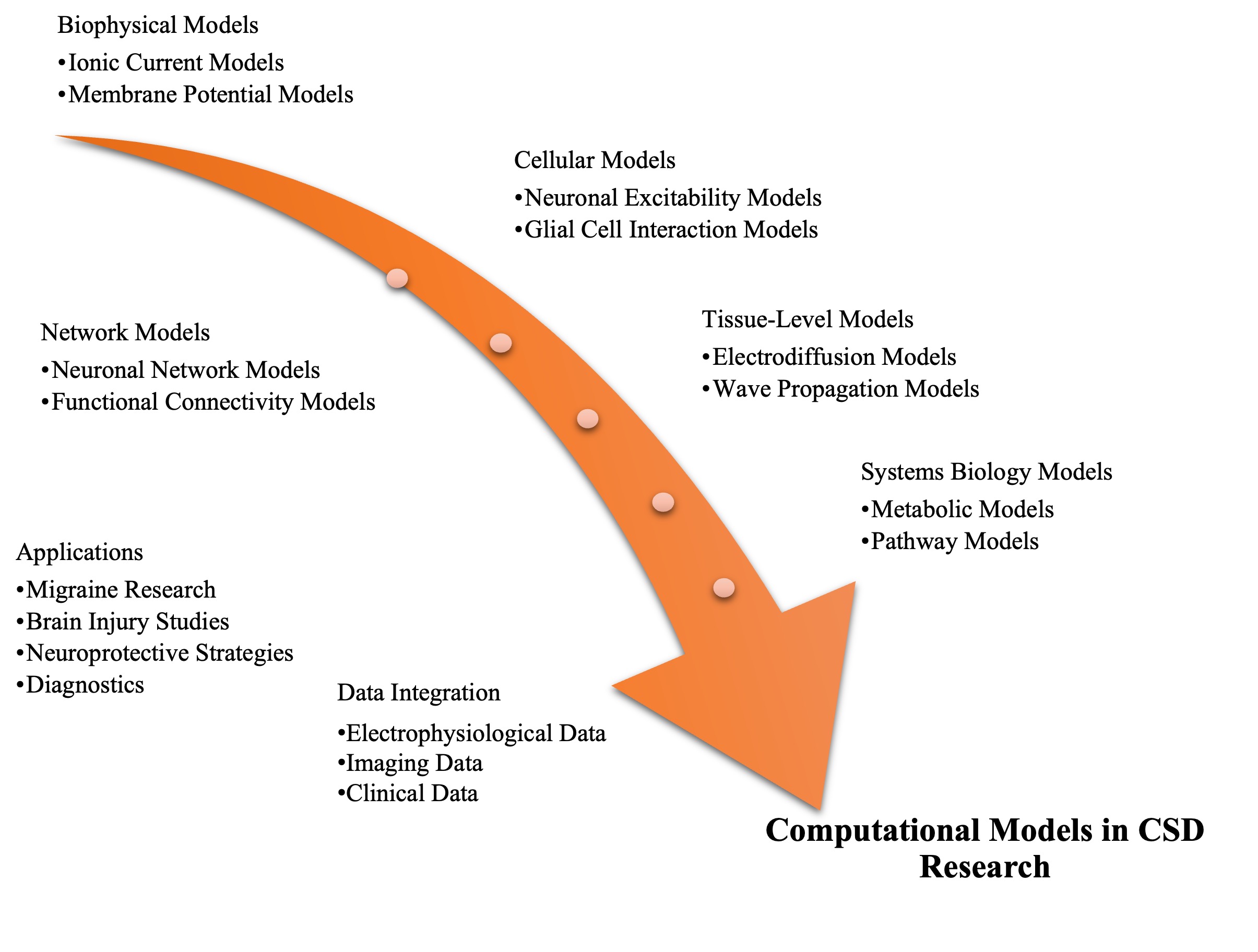} 
   {(Color online) Computational models of CSD along with data integration and applications in NDDs research.} \cite{shaheen2022dbs}.\label{fig:csd2}
   
Moreover, the latest models have started exploring network topology's implications in greater depth \cite{michelson2018multi}. Research indicates that the specific arrangement and connectivity patterns of neurons greatly influence the propagation of CSD \cite{nwagbo2024review}. Advanced network models now include detailed mapping of these connections and study how variations in network structures can impact the spread of CSD \cite{dhollander2021fixel}. This aspect is essential in understanding individual differences in CSD susceptibility and manifestation, as variations in neural network topology can lead to significant differences in how CSD events are experienced \cite{al2020giving}. Overall, the continuous evolution of network models of CSD is proving instrumental in unravelling the complex dynamics of this neurological event, paving the way for targeted therapeutic strategies and enhanced understanding of related neurological conditions.

\subsubsection{Multiscale Models of CSD}
1. Integrating Scales:
Multiscale models of CSD aim to integrate information across various biological scales – from molecular dynamics (like ion channel behaviour) to cellular responses, up to the network-level phenomena \cite{hampel2018revolution}. These models strive to provide a comprehensive understanding of CSD, linking molecular and cellular changes to the large-scale neural network behaviour seen in the phenomenon \cite{sanacora2022stressed}. At the molecular level, these models focus on understanding the intricacies of ion channel behaviours, which are crucial for neuronal activity. Ion channels, the gatekeepers of neuronal signalling, undergo various changes during CSD, including alterations in their conduction properties and distributions across the cell membrane \cite{shaheen2021analysis,torres2020mir}. By simulating these molecular dynamics, scientists can begin to decode how minute changes at this level can instigate more significant cellular responses. This molecular perspective is essential, as it lays the foundation for understanding the initial stages of CSD and how it can escalate to affect entire networks of neurons.

At the cellular scale, multiscale models of CSD search into how individual neurons and glial cells respond to and contribute to the spreading wave of depression \cite{devor2012neuronal}. This involves understanding the cellular mechanisms like synaptic transmission, neuronal excitability, and glial modulation, all altered during CSD. These models integrate data on how cells interact with each other and their immediate environment, taking into account factors like neurotransmitter release, uptake, and the role of astrocytes in modulating synaptic activity. By mapping these cellular-level responses, researchers can understand how CSD propagates through neuronal circuits. This level of modelling is crucial in identifying potential therapeutic targets, as it bridges the gap between molecular changes and the more significant network dynamics. On a broader scale, multiscale models extend to encompass the network-level phenomena observed in CSD \cite{poort2016texture}. This involves examining how networks of neurons and synapses interact to propagate the wave of depression across cortical areas. At this level, the focus shifts to understanding the emergent properties of neural circuits, such as synchronization, network connectivity, and the influence of different brain regions on the progression of CSD. These large-scale models are crucial for interpreting how localized molecular and cellular events can lead to widespread network disruptions, as seen in CSD. 

2. Systems Biology Approach:
Models that use a systems biology framework in CSD research incorporate various types of data – genetic, molecular, and electrophysiological – to gain a comprehensive understanding of the phenomenon \cite{poort2016texture,tuckwell2013stochastic,torres2020mir,devor2012neuronal}. These models can help identify potential targets for intervention and understand the complex interplay of different biological processes involved in CSD. They are precious in elucidating the broader implications of CSD in neurological diseases and identifying potential therapeutic strategies.

\section{Combining Stochastic and Deterministic Approaches in NDD Modelling}
Stochastic and deterministic characteristics are used in hybrid multiscale models of NDDs such as AD, PD, and CSD. This strategy aims to decipher the intricacies of NDDs of different sizes. Its main objectives are understanding disease processes, identifying risk factors, and predicting disease development \cite{rajula2020comparison}. This study focuses on constructing multiscale models that encapsulate random elements and predictable trends, illuminating disease mechanisms. In these multiscale NDD models, stochastic methods model random molecular or genetic interactions, while deterministic models handle neuronal network behaviours and pathological spread. These models work together to provide insights into the disease's impact on brain regions and circuits. The strength of hybrid models lies in their ability to encapsulate the inherent randomness and observable patterns in NDDs. These models are pivotal in studying the development of pathological processes, finding therapeutic targets, and examining treatment effects. For instance, if a particular treatment is hypothesized to stabilize ion channels and reduce the likelihood of CSD initiation, the stochastic model can test this hypothesis under different conditions. The deterministic model can then be used to simulate how this treatment might alter the spread of depolarization, providing insights into its potential therapeutic impact. For example, migraine aura triggers, such as spontaneous neurotransmitter release, can be modelled using a Poisson process, which describes random, independent events over time \cite{alstadhaug2007periodicity}. Once a CSD event is triggered, deterministic models predict its propagation through the cortical surface, particularly in visual or sensory processing areas. This helps develop interventions that may interrupt or modulate the spread of the aura. Genetic variations that increase the likelihood of CSD initiation can be modelled using a branching process, which captures the proliferation of random genetic mutations or expressions through neuronal networks \cite{mubayi2019studying}. The deterministic model then predicts how these genetic variations affect the spatial and temporal propagation of CSD, informing personalized approaches to prevent or mitigate CSD in individuals with a higher genetic risk. The varying effects of anesthetic agents, which may influence neurotransmitter release and ion channel function, can be simulated using a semi-Markov process, capturing the random time intervals between transitions in different states, such as awake, sedated, and deeply anesthetized \cite{ball2000stochastic}. The deterministic component simulates how these state changes impact CSD propagation, providing insights into optimizing anesthetics to minimize the risk of triggering CSD during procedures \cite{de2014joint}. Inflammation, which can randomly influence CSD initiation, can be modelled using time-reversible Markov chains to describe fluctuations in cytokine levels that affect neuronal stability \cite{cipresso2023affects}. The deterministic model predicts how inflammation-induced CSD events spread through the brain, guiding the development of anti-inflammatory treatments to reduce CSD in conditions like traumatic brain injury. Additionally, therapeutic cooling, which affects neural activity at the molecular level, can be modelled using Brownian motion to understand the microscopic effects of cooling on the likelihood of CSD initiation. The field is dynamic, with methodologies evolving based on specific NDDs and research requirements.

Neural network dynamics in NDDs are often illustrated through spiking neuron interactions, where the stochastic nature at the neuronal level influences entire network properties \cite{dumont2017stochastic}. AD, CSD, and PD are characterized by progressive cognitive decline. The high failure rate of drug trials in these areas underscores the urgency for effective treatments. Mathematical, computational, and statistical methods are vital for enhancing clinical trial designs and understanding failures, often attributed to limited knowledge of AD pathogenesis and inadequate trial designs \cite{Sevigny2016Antibody, Merchant2019Proposed,corti2024structure}. When combined with deterministic models that predict the average disease progression based on established biological mechanisms, these approaches can help optimize trial designs by identifying more precise inclusion criteria, stratifying patient populations, and refining endpoints \cite{shaheen2023bayesian,dumont2017stochastic}. For example, a combination of stochastic and deterministic models might be used to simulate the effects of different dosages of a therapeutic agent on both the molecular level (e.g., amyloid-beta aggregation) and the clinical level (e.g., cognitive decline), leading to better-informed decisions about dosage and treatment schedules in trials. These models can also be employed to anticipate potential adverse effects or lack of efficacy in specific patient subgroups, thus reducing the likelihood of trial failures and improving the overall success rate of AD clinical trials \cite{rajula2020comparison,shaheen2023data,fornari2020spatially}. Modelling AD progression and clinical trial simulations are crucial to addressing these challenges. While AD's complex nature limits robust model creation, efforts have been made to develop stochastic models at the molecular level \cite{albahri2023systematic, hadjichrysanthou2018development, frisoni2022probabilistic}.

\section{Systems-Based Approaches and Uncertainty Quantification in NDD Modelling}
Systems-based approaches play a pivotal role in neurodegenerative disease (NDD) research. They involve creating comprehensive models integrating various biological systems to better understand diseases like Alzheimer's and Parkinson's. Such models often combine genetic, molecular, and environmental data, providing a holistic view of disease mechanisms. A key component in these approaches is uncertainty quantification, which addresses the inherent unpredictability of biological systems and experimental data. By quantifying this uncertainty, researchers can improve the reliability of their models, leading to more accurate predictions and effective strategies for managing and treating NDDs.

\subsection{Neural Network Models in Neuroscience and Applications of Artificial Neural Networks (ANNs)}
The present study explores the application of Artificial Neural Networks (ANNs) in understanding AD, utilizing patient data from the ADNI database. ANNs have increasingly become a prominent modelling system in neuroscience, offering new approaches to challenges in brain research\cite{Yamins2016Using, Kriegeskorte2015Deep, Sussillo2009Generating, Barak2013Fixed} There are three key reasons why ANNs are particularly valuable for neuroscientists. Firstly, they offer new modelling strategies for tackling complex real-life brain functions, often not adequately captured by traditional computational models. Secondly, ANNs can better represent the diversity of activity patterns in neural populations, addressing the limitations of simpler models, especially in complex areas like the prefrontal cortex. Thirdly, ANNs not only model biological systems but also probe more profound neuroscience questions, such as understanding the underlying reasons for certain brain functions\cite{Yang2020Artificial}

The aim here is to showcase the potential of sophisticated biological nervous system models. Nervous system network models strive to replicate the complexity and functionality of biological systems, providing insights into neuron behaviour and cognitive processes\cite{Lima2022Comprehensive, Sarishma2022Review, Battaglia2022Functional}. In contrast, ANNs, central to AI and ML, mimic the structure of biological networks but focus more on computation and generalization rather than biological accuracy\cite{Wang2022Hierarchical, Peng2021Multiscale, Bihl2023Artificial, Thukroo2022Review,ghebrehiwet2024revolutionizing,storm2024integrative}. ANNs learn through training, adjusting weights in a network of interconnected artificial neurons.

However, despite these potentialities, ANNs come with notable limitations, particularly in their application to neuroscience. One significant limitation is their alignment with biological realism. While ANNs are inspired by the brain's structure, they are fundamentally computational models that do not fully capture the intricate biological processes underlying neural activity. This can lead to a gap between the computational efficiency of ANNs and the biological accuracy needed to truly understand brain function. Furthermore, ANNs require large datasets and substantial computational resources, which may not always be feasible in neuroscience research, mainly when dealing with rare or hard-to-acquire data. The interpretability of ANNs is also a concern, as the models often function as "black boxes," making it challenging to understand the exact features or patterns driving their predictions \cite{liang2021explaining}. This lack of transparency can hinder the ability to draw meaningful biological conclusions from ANN-based analyses. Generally, ANNs are more robust and often outperform other computational tools in solving various problems across seven key categories: classification, regression, clustering, dimensionality reduction, anomaly detection, reinforcement learning, and generative modelling \cite{basheer2000artificial}.

While nervous system network models and ANNs seek to understand neural network behaviour, they have distinct focuses. The former goes through the biological mechanisms and the relationship between brain activity and behaviour, with detailed physiological modelling. ANNs, however, prioritize computational efficiency and problem-solving capabilities, often sacrificing biological detail for broader applicability. ANNs have incredibly advanced AI and ML, while nervous system network models have deepened our understanding of the brain's functions.

\subsection{Fundamentals and Techniques in ANNs}
In ANNs, artificial neurons function as computational nodes, receiving inputs, performing weighted computations, and generating outputs. After being weighted, these inputs pass through an activation function to produce the neuron's output. ANNs are structured into an input layer, one or more hidden layers for computations, and an output layer for final results. The network's structure, comprising neuron arrangements and connections, dictates the information flow and computation complexity. Various architectures like feedforward, recurrent, and convolutional neural networks are tailored for specific data types and tasks.

ANNs undergo training to learn from data, adjusting connection weights to minimize a loss function that measures the deviation between the network's output and the expected result. Backpropagation is a standard method for adjusting weights, where the error gradient is propagated backward through the network.

The topology of biological neural networks, which can provide insight into brain activity, is replicated in custom ANNs to correspond with neurobiological characteristics. Studying memory, attention, and decision-making can be facilitated by including characteristics such as recurrent connections, which can mimic brain information-processing processes. Custom learning rules, such as Hebbian learning, can mirror synaptic plasticity, enriching the neurobiological relevance of ANNs. These networks can also include biological constraints like anatomical patterns or neurophysiological characteristics, providing insights into neurological phenomena and validating brain mechanism theories.

ANNs can be trained using experimental data to model and analyze brain activities and cognitive processes, utilizing neuroimaging, electrophysiological recordings, or behavioural tests. The present study employs ANNs to address inverse Ordinary Differential Equations (ODEs) challenges, like estimating unknown parameters in systems represented by observed data, such as ADNI data for AD patients\cite{irimata2020fundamental, raket2020statistical, davenport2023neurodegenerative}. The process involves data collection, constructing a forward ODE model, generating training data, designing the ANN architecture, training the network, and using the trained ANN to approximate inverse solutions. ANNs offer a data-driven approach for complex inverse ODE problems, capable of estimating unknowns where analytical methods fall short.

Moreover, statistical and mathematical methods such as Approximate Bayesian Computation (ABC) and Monte Carlo (MC) methods are used to estimate uncertainties in ANN predictions, providing probabilistic outputs\cite{adhikari2022exploiting}. These techniques are essential for training, optimizing, regularizing, and quantifying uncertainties in ANNs.

\subsection{Distributed Brain Network Models}

The human brain, a complex network with interconnected regions functioning in a coordinated manner, continually processes and transmits data across its spatially distributed yet functionally linked parts. Understanding the connection between neural areas and cognitive functions is crucial, particularly how the brain's functional connections, structured efficiently in a small-world network, relate to cognitive abilities like intelligence \cite{farahani2019application, Van2009Efficiency}.

Emerging evidence shows that cognitive and affective processes involve distributed neural systems, including cerebellar subregions, whose integration in a range of brain functions is yet to be fully understood \cite{Farzan2016Enhancing}. Neuroscience aims to connect neural regions with cognitive computations, going beyond mapping cognitive processes to neural entities. Network neuroscience provides detailed representations of brain network architectures, allowing cognitive neuroscientists to merge these structural insights with functional maps. This mapping enables a deeper mechanistic understanding by establishing causal relationships between neuronal components and functions, as demonstrated in studies utilizing network coding models \cite{Ito2020Discovering}. 

Intelligence, a widely studied cognitive ability and a strong predictor of long-term life success, varies among individuals. Studies suggest that humans possess genetically determined intelligence constraints, yet learning and education can augment cognitive abilities \cite{Dubois2018Distributed}. Moreover, understanding neural substrates is critical in areas like depression diagnosis and treatment, as explored in brain network models \cite{Li2018Brain}. Similarly, categorization, which involves processing sensory information into discrete behavioural classes, is another cognitive function that these network models aim to unravel \cite{Feng2021Distributed}. 

Physics-informed ML (PIML), incorporating physics-based principles into ML, has recently been developed to learn from small and noisy datasets \cite{Xu2022Physics, Nabati2023Real, lozano2019deep}. PIML is particularly useful in AD patient data analysis, integrating patient-specific data with physiological processes represented by ODEs \cite{Whittington2018Spatiotemporal}. 

Systems-based approaches and uncertainty quantification are integral in NDD modelling \cite{shaheen2024multiscale}. Traditional reductionist methods often fail to capture the systemic complexity of NDDs, whereas systems-based approaches consider interactions among system components \cite{Chen2022Challenges, Bender2021Artificial, Ding2020Towards}. These approaches aid in understanding disease progression, identifying key factors, and suggesting therapeutic targets in integrating different data types\cite{Niarakis2022Addressing, Schafer2021Bayesian}. 

Uncertainty quantification addresses the variability and limitations in biological systems due to factors like system complexity and individual differences \cite{bilgel2019predicting, frisoni2022probabilistic, ghazi2021robust}. It involves statistical methods to assess the reliability of models and manage uncertainties. This process supports the development of reliable and clinically relevant models.

Integrating these approaches in NDD modelling enhances our understanding of disease mechanisms, accounts for individual variability, and provides realistic assessments of model confidence levels. Such integrations facilitate personalized medicine and support clinical and drug development decisions in NDDs. This study introduces network coding models optimized for task performance and biological realism, employing an ABC optimization scheme for model inversion without assuming local linearity. Bayesian inference measures instability by evaluating the likelihood of model parameters given experimental data, offering a suitable approach for complex models like ours \cite{Beaumont2019Approximate, Christopher2018Parameter}. This "likelihood-free" method simplifies solving dynamic systems by avoiding the numerical integration of equations.
\section{Conclusion}
Brain-inspired computing, an evolving field at the intersection of neuroscience and computer science, draws from the principles and mechanisms of the human brain to enhance computational models and algorithms. This approach is particularly relevant in the realm of complex or hybrid systems where traditional computational methods may fall short. Complex systems, by nature, are multifaceted and comprise multiple interacting components. These systems often exhibit behaviours that are not easily predictable from the properties of individual elements. Brain-inspired computing brings a new perspective to understanding and modelling these systems. By mimicking neural processes, such as parallel processing and adaptive learning, brain-inspired algorithms can effectively navigate the intricate dynamics of complex systems.

Brain-inspired techniques like deep learning and neural networks have revolutionized machine learning (ML). These techniques utilize layered structures of algorithms, loosely modelling the brain's neural networks, to process data in sophisticated ways. The application of these models has proven particularly effective in pattern recognition, predictive analytics, and decision-making processes, which are essential in managing complex systems. One of the critical advantages of brain-inspired computing in ML is its ability to handle vast and varied data sets, learn from them, and make predictions or decisions based on incomplete or uncertain information. This capability is critical in complex systems with intricate variables and interdependencies.

Furthermore, by combining physical and digital elements, hybrid systems can particularly benefit from brain-inspired ML approaches. In these systems, the adaptability and flexibility of brain-like algorithms can enhance interaction between digital and physical components, leading to more efficient system operation and better integration of diverse data sources. Moreover, brain-inspired ML approaches also contribute to advancing Artificial Intelligence, pushing the boundaries of autonomous systems, robotics, and cognitive computing. By incorporating aspects such as learning from experience, adapting to new situations, and even exhibiting forms of 'intuition,' these systems get closer to replicating human-like decision-making processes. In conclusion, brain-inspired computing and ML in complex and hybrid systems present a promising avenue for tackling the challenges posed by these intricate environments. Drawing inspiration from the most advanced known system for processing information—the human brain—these approaches offer new tools and paradigms for understanding, modelling, and managing complex systems. The ongoing advancement in this field is expected to have profound implications across various domains, including healthcare, finance, environmental science, and beyond.

\section*{Acknowledgment}

The authors thank the NSERC and the CRC Program for their support. This research was enabled in part by support provided by SHARCNET \url{(www. sharcnet.ca)} and Digital Research Alliance of Canada \url{(www.alliancecan.ca)}.

\bibliography{IEEEabrv.bib, IEEEexample.bib}{}

\begin{thebibliography}{100}
\providecommand{\url}[1]{#1}
\csname url@samestyle\endcsname
\providecommand{\newblock}{\relax}
\providecommand{\bibinfo}[2]{#2}
\providecommand{\BIBentrySTDinterwordspacing}{\spaceskip=0pt\relax}
\providecommand{\BIBentryALTinterwordstretchfactor}{4}
\providecommand{\BIBentryALTinterwordspacing}{\spaceskip=\fontdimen2\font plus
\BIBentryALTinterwordstretchfactor\fontdimen3\font minus
  \fontdimen4\font\relax}
\providecommand{\BIBforeignlanguage}[2]{{%
\expandafter\ifx\csname l@#1\endcsname\relax
\typeout{** WARNING: IEEEtran.bst: No hyphenation pattern has been}%
\typeout{** loaded for the language `#1'. Using the pattern for}%
\typeout{** the default language instead.}%
\else
\language=\csname l@#1\endcsname
\fi
#2}}
\providecommand{\BIBdecl}{\relax}
\BIBdecl

\bibitem{carter2019human}
R.~Carter, \emph{The human brain book: An illustrated guide to its structure,
  function, and disorders}.\hskip 1em plus 0.5em minus 0.4em\relax Penguin,
  2019.

\bibitem{parvizi2021functional}
A.~Parvizi-Fard, M.~Amiri, D.~Kumar, M.~M. Iskarous, and N.~V. Thakor, ``A
  functional spiking neuronal network for tactile sensing pathway to process
  edge orientation,'' \emph{Scientific Reports}, vol.~11, no.~1, pp. 1--16,
  2021.

\bibitem{ziegler2017school}
D.~Ziegler, ``In-school neurological reparative therapy for traumatized
  students,'' \emph{Optimizing Learning Outcomes: Proven Brain-centric,
  Trauma-Sensitive Practices}, pp. 150--165, 2017.

\bibitem{shaheen2024multiscale}
H.~Shaheen, ``Multiscale modelling of brain networks and the analysis of
  dynamic processes in neurodegenerative disorders,'' \emph{Theses and
  Dissertations (Comprehensive). 2604. https://scholars.wlu.ca/etd/2604}, 2024.

\bibitem{herregods2023blurring}
N.~Herregods, R.~G. Lambert, E.~Schiettecatte, J.~Dehoorne, T.~Renson,
  F.~Laloo, T.~Van Den~Berghe, L.~B. Jans, and J.~L. Jaremko, ``Blurring and
  irregularity of the subchondral cortex in pediatric sacroiliac joints on t1
  images: Incidence of normal findings that can mimic erosions,''
  \emph{Arthritis Care \& Research}, vol.~75, no.~1, pp. 190--197, 2023.

\bibitem{daroff2014encyclopedia}
R.~B. Daroff and M.~J. Aminoff, \emph{Encyclopedia of the Neurological
  Sciences}.\hskip 1em plus 0.5em minus 0.4em\relax Academic Press, 2014.

\bibitem{nazlee2023age}
N.~Nazlee, G.~D. Waiter, and A.-L. Sandu, ``Age-associated sex and asymmetry
  differentiation in hemispheric and lobar cortical ribbon complexity across
  adulthood: A uk biobank imaging study,'' \emph{Human Brain Mapping}, vol.~44,
  no.~1, pp. 49--65, 2023.

\bibitem{gonzalez2023neurocognitive}
H.~A. Gonz{\'a}lez-Usigli, G.~G. Ortiz, C.~Charles-Ni{\~n}o, M.~A.
  Mireles-Ram{\'\i}rez, F.~P. Pacheco-Mois{\'e}s, B.~M. d.~G. Torres-Mendoza,
  J.~d.~J. Hern{\'a}ndez-Cruz, D.~L. d.~C. Delgado-Lara, and L.~J.
  Ram{\'\i}rez-Jirano, ``Neurocognitive psychiatric and neuropsychological
  alterations in \mbox{Parkinson’s} disease: A basic and clinical approach,''
  \emph{Brain Sciences}, vol.~13, no.~3, p. 508, 2023.

\bibitem{bruner2023parietal}
E.~Bruner, A.~Battaglia-Mayer, and R.~Caminiti, ``The parietal lobe evolution
  and the emergence of material culture in the human genus,'' \emph{Brain
  Structure and Function}, vol. 228, no.~1, pp. 145--167, 2023.

\bibitem{berron2020medial}
D.~Berron, D.~van Westen, R.~Ossenkoppele, O.~Strandberg, and O.~Hansson,
  ``Medial temporal lobe connectivity and its associations with cognition in
  early \mbox{Alzheimer's} disease,'' \emph{Brain}, vol. 143, no.~4, pp.
  1233--1248, 2020.

\bibitem{rehman2023neuroanatomy}
A.~Rehman and Y.~Al~Khalili, ``Neuroanatomy, occipital lobe,'' in
  \emph{StatPearls [Internet]}.\hskip 1em plus 0.5em minus 0.4em\relax
  StatPearls Publishing, 2023.

\bibitem{miller2022natural}
C.~T. Miller, D.~Gire, K.~Hoke, A.~C. Huk, D.~Kelley, D.~A. Leopold, M.~C.
  Smear, F.~Theunissen, M.~Yartsev, and C.~M. Niell, ``Natural behavior is the
  language of the brain,'' \emph{Current Biology}, vol.~32, no.~10, pp.
  R482--R493, 2022.

\bibitem{fletcher2022new}
A.~Fletcher and M.~Benveniste, ``A new method for training creativity:
  narrative as an alternative to divergent thinking,'' \emph{Annals of the New
  York Academy of Sciences}, vol. 1512, no.~1, pp. 29--45, 2022.

\bibitem{kok2022cognitive}
A.~Kok, ``Cognitive control, motivation and fatigue: A cognitive neuroscience
  perspective,'' \emph{Brain and Cognition}, vol. 160, p. 105880, 2022.

\bibitem{shaheen2023astrocytic}
H.~Shaheen, S.~Pal, and R.~Melnik, ``Astrocytic clearance and fragmentation of
  toxic proteins in \mbox{Alzheimer’s} disease on large-scale brain
  networks,'' \emph{Physica D: Nonlinear Phenomena}, p. 133839, 2023.

\bibitem{shekhar2023potential}
N.~Shekhar, S.~Tyagi, S.~Rani, and A.~K. Thakur, ``Potential of capric acid in
  neurological disorders: An overview,'' \emph{Neurochemical Research},
  vol.~48, no.~3, pp. 697--712, 2023.

\bibitem{oishi2019developmental}
K.~Oishi, J.~Chotiyanonta, D.~Wu, M.~I. Miller, S.~Mori, K.~Oishi
  \emph{et~al.}, ``Developmental trajectories of the human embryologic brain
  regions,'' \emph{Neuroscience Letters}, vol. 708, p. 134342, 2019.

\bibitem{nikrahan2023theory}
G.~R. Nikrahan, ``Theory of brain complexity and marital behaviors: The
  application of complexity science and neuroscience to explain the
  complexities of marital behaviors,'' \emph{Frontiers in Human Neuroscience},
  vol.~17, p.~83, 2023.

\bibitem{bressler2010large}
S.~L. Bressler and V.~Menon, ``Large-scale brain networks in cognition:
  emerging methods and principles,'' \emph{Trends in Cognitive Sciences},
  vol.~14, no.~6, pp. 277--290, 2010.

\bibitem{khodagholy2022large}
D.~Khodagholy, J.~J. Ferrero, J.~Park, Z.~Zhao, and J.~N. Gelinas,
  ``Large-scale, closed-loop interrogation of neural circuits underlying
  cognition,'' \emph{Trends in Neurosciences}, 2022.

\bibitem{dutta2024unsolved}
T.~Dutta and A.~Bandyopadhyay, ``Unsolved mysteries of the mind and the brain:
  Fractal brain hypothesis,'' in \emph{Emotion, Cognition and Silent
  Communication: Unsolved Mysteries}.\hskip 1em plus 0.5em minus 0.4em\relax
  Springer, 2024, pp. 1--40.

\bibitem{lopez2024digging}
S.~L{\'o}pez-Ortiz, G.~Caruso, E.~Emanuele, H.~Men{\'e}ndez,
  S.~Pe{\~n}{\'\i}n-Grandes, C.~S. Guerrera, F.~Caraci, R.~Nistic{\`o},
  A.~Lucia, A.~Santos-Lozano \emph{et~al.}, ``Digging into the intrinsic
  capacity concept: Can it be applied to alzheimer’s disease?''
  \emph{Progress in Neurobiology}, vol. 234, p. 102574, 2024.

\bibitem{seguin2023brain}
C.~Seguin, O.~Sporns, and A.~Zalesky, ``Brain network communication: concepts,
  models and applications,'' \emph{Nature reviews neuroscience}, vol.~24,
  no.~9, pp. 557--574, 2023.

\bibitem{papo2023does}
D.~Papo and J.~Buld{\'u}, ``Does the brain behave like a (complex) network? i.
  dynamics,'' \emph{Physics of life reviews}, vol.~48, pp. 47--98, 2023.

\bibitem{ramaswamy2024data}
S.~Ramaswamy, ``Data-driven multiscale computational models of cortical and
  subcortical regions,'' \emph{Current opinion in neurobiology}, vol.~85, p.
  102842, 2024.

\bibitem{acharya2024brain}
P.~Acharya, N.~Y. Choi, S.~Shrestha, S.~Jeong, and M.-Y. Lee, ``Brain
  organoids: A revolutionary tool for modeling neurological disorders and
  development of therapeutics,'' \emph{Biotechnology and Bioengineering}, vol.
  121, no.~2, pp. 489--506, 2024.

\bibitem{lorenzi2023multi}
R.~M. Lorenzi, A.~Geminiani, Y.~Zerlaut, M.~De~Grazia, A.~Destexhe, C.~A.
  Gandini Wheeler-Kingshott, F.~Palesi, C.~Casellato, and E.~D’Angelo, ``A
  multi-layer mean-field model of the cerebellum embedding microstructure and
  population-specific dynamics,'' \emph{PLOS Computational Biology}, vol.~19,
  no.~9, p. e1011434, 2023.

\bibitem{palesi2020importance}
F.~Palesi, R.~M. Lorenzi, C.~Casellato, P.~Ritter, V.~Jirsa, C.~A. Gandini
  Wheeler-Kingshott, and E.~D’angelo, ``The importance of cerebellar
  connectivity on simulated brain dynamics,'' \emph{Frontiers in Cellular
  Neuroscience}, vol.~14, p. 240, 2020.

\bibitem{goldman2019bridging}
J.~S. Goldman, N.~Tort-Colet, M.~Di~Volo, E.~Susin, J.~Bout{\'e}, M.~Dali,
  M.~Carlu, T.-A. Nghiem, T.~G{\'o}rski, and A.~Destexhe, ``Bridging single
  neuron dynamics to global brain states,'' \emph{Frontiers in systems
  neuroscience}, vol.~13, p.~75, 2019.

\bibitem{tesler2024multiscale}
F.~Tesler, R.~M. Lorenzi, A.~Ponzi, C.~Casellato, F.~Palesi, D.~Gandolfi, C.~A.
  Gandini Wheeler~Kingshott, J.~Mapelli, E.~D'Angelo, M.~Migliore
  \emph{et~al.}, ``Multiscale modeling of neuronal dynamics in hippocampus
  ca1,'' \emph{Frontiers in Computational Neuroscience}, vol.~18, p. 1432593,
  2024.

\bibitem{overwiening2023multi}
J.~Overwiening, F.~Tesler, D.~Guarino, and A.~Destexhe, ``A multi-scale study
  of thalamic state-dependent responsiveness,'' \emph{bioRxiv}, pp. 2023--12,
  2023.

\bibitem{sanz2015mathematical}
P.~Sanz-Leon, S.~A. Knock, A.~Spiegler, and V.~K. Jirsa, ``Mathematical
  framework for large-scale brain network modeling in the virtual brain,''
  \emph{Neuroimage}, vol. 111, pp. 385--430, 2015.

\bibitem{vatansever2021artificial}
S.~Vatansever, A.~Schlessinger, D.~Wacker, H.~{\"U}. Kaniskan, J.~Jin, M.-M.
  Zhou, and B.~Zhang, ``Artificial intelligence and machine learning-aided drug
  discovery in central nervous system diseases: State-of-the-arts and future
  directions,'' \emph{Medicinal research reviews}, vol.~41, no.~3, pp.
  1427--1473, 2021.

\bibitem{Alber2019Integrating}
M.~Alber, A.~B. Tepole, W.~R. Cannon, S.~De, S.~Dura-Bernal, K.~Garikipati,
  G.~Karniadakis, W.~W. Lytton, P.~Perdikaris, L.~Petzold \emph{et~al.},
  ``Integrating machine learning and multiscale modeling—perspectives,
  challenges, and opportunities in the biological, biomedical, and behavioral
  sciences,'' \emph{NPJ Digital Medicine}, vol.~2, no.~1, pp. 1--11, 2019.

\bibitem{keresztes2022introducing}
L.~Keresztes, E.~Sz{\"o}gi, B.~Varga, and V.~Grolmusz, ``Introducing and
  applying newtonian blurring: an augmented dataset of 126,000 human
  connectomes at braingraph. org,'' \emph{Scientific Reports}, vol.~12, no.~1,
  p. 3102, 2022.

\bibitem{shibasaki2008human}
H.~Shibasaki, ``Human brain mapping: hemodynamic response and
  electrophysiology,'' \emph{Clinical Neurophysiology}, vol. 119, no.~4, pp.
  731--743, 2008.

\bibitem{shaheen2021neuron}
H.~Shaheen, S.~Singh, and R.~Melnik, ``A neuron-glial model of exosomal release
  in the onset and progression of alzheimer's disease,'' \emph{Frontiers in
  Computational Neuroscience}, vol.~15, p. 653097, 2021.

\bibitem{perdikaris2016multiscale}
P.~Perdikaris, L.~Grinberg, and G.~E. Karniadakis, ``Multiscale modeling and
  simulation of brain blood flow,'' \emph{Physics of Fluids}, vol.~28, no.~2,
  p. 021304, 2016.

\bibitem{xia2022bayesian}
Y.~Xia and N.~Zabaras, ``Bayesian multiscale deep generative model for the
  solution of high-dimensional inverse problems,'' \emph{Journal of
  Computational Physics}, vol. 455, p. 111008, 2022.

\bibitem{shaheen2023data}
H.~Shaheen, R.~Melnik, and S.~Sundeep, ``Data-driven stochastic model for
  quantifying the interplay between amyloid-beta and calcium levels in
  alzheimer's disease,'' \emph{Statistical Analysis and Data Mining: The ASA
  Data Science Journal}, vol.~17, no.~2, p. e11679, 2024.

\bibitem{ahmed2020artificial}
Z.~Ahmed, K.~Mohamed, S.~Zeeshan, and X.~Dong, ``Artificial intelligence with
  multi-functional machine learning platform development for better healthcare
  and precision medicine,'' \emph{Database}, vol. 2020, p. baaa010, 2020.

\bibitem{bishara2023state}
D.~Bishara, Y.~Xie, W.~K. Liu, and S.~Li, ``A state-of-the-art review on
  machine learning-based multiscale modeling, simulation, homogenization and
  design of materials,'' \emph{Archives of Computational Methods in
  Engineering}, vol.~30, no.~1, pp. 191--222, 2023.

\bibitem{shaheen2024neural}
H.~Shaheen and R.~Melnik, ``Neural dynamics in parkinson’s disease:
  Integrating machine learning and stochastic modelling with connectomic data.
  in: Franco, l., de mulatier, c., paszynski, m., krzhizhanovskaya, v.v.,
  dongarra, j.j., sloot, p.m.a. (eds) computational science iccs 2024. lecture
  notes in computer science.'' in \emph{International Conference on
  Computational Science}, vol. 14835.\hskip 1em plus 0.5em minus 0.4em\relax
  Springer, Cham, 2024, pp. 46--60.

\bibitem{greene2016development}
D.~J. Greene, C.~N. Lessov-Schlaggar, and B.~L. Schlaggar, ``Development of the
  brain’s functional network architecture,'' in \emph{Neurobiology of
  Language}.\hskip 1em plus 0.5em minus 0.4em\relax Elsevier, 2016, pp.
  399--406.

\bibitem{vogel2010development}
A.~C. Vogel, J.~D. Power, S.~E. Petersen, and B.~L. Schlaggar, ``Development of
  the brain’s functional network architecture,'' \emph{Neuropsychology
  Review}, vol.~20, pp. 362--375, 2010.

\bibitem{bassett2017network}
D.~S. Bassett and O.~Sporns, ``Network neuroscience,'' \emph{Nature
  Neuroscience}, vol.~20, no.~3, pp. 353--364, 2017.

\bibitem{uddin2022controversies}
L.~Q. Uddin, R.~F. Betzel, J.~R. Cohen, J.~S. Damoiseaux, F.~De~Brigard, S.~B.
  Eickhoff, A.~Fornito, C.~Gratton, E.~M. Gordon, A.~R. Laird \emph{et~al.},
  ``Controversies and progress on standardization of large-scale brain network
  nomenclature,'' \emph{Network Neuroscience}, pp. 1--111, 2022.

\bibitem{park2013structural}
H.-J. Park and K.~Friston, ``Structural and functional brain networks: from
  connections to cognition,'' \emph{Science}, vol. 342, no. 6158, p. 1238411,
  2013.

\bibitem{schoonheim2022network}
M.~M. Schoonheim, T.~A. Broeders, and J.~J. Geurts, ``The network collapse in
  multiple sclerosis: An overview of novel concepts to address disease
  dynamics,'' \emph{NeuroImage: Clinical}, p. 103108, 2022.

\bibitem{churchland1992computational}
P.~S. Churchland and T.~J. Sejnowski, \emph{The computational brain}.\hskip 1em
  plus 0.5em minus 0.4em\relax MIT Press, 1992.

\bibitem{braun2018maps}
U.~Braun, A.~Schaefer, R.~F. Betzel, H.~Tost, A.~Meyer-Lindenberg, and D.~S.
  Bassett, ``From maps to multi-dimensional network mechanisms of mental
  disorders,'' \emph{Neuron}, vol.~97, no.~1, pp. 14--31, 2018.

\bibitem{sporns2022structure}
O.~Sporns, ``Structure and function of complex brain networks,''
  \emph{Dialogues in Clinical Neuroscience}, 2022.

\bibitem{bassett2018nature}
D.~S. Bassett, P.~Zurn, and J.~I. Gold, ``On the nature and use of models in
  network neuroscience,'' \emph{Nature Reviews Neuroscience}, vol.~19, no.~9,
  pp. 566--578, 2018.

\bibitem{downing2001cortical}
P.~E. Downing, Y.~Jiang, M.~Shuman, and N.~Kanwisher, ``A cortical area
  selective for visual processing of the human body,'' \emph{Science}, vol.
  293, no. 5539, pp. 2470--2473, 2001.

\bibitem{kanwisher1997fusiform}
N.~Kanwisher, J.~McDermott, and M.~M. Chun, ``The fusiform face area: a module
  in human extrastriate cortex specialized for face perception,'' \emph{Journal
  of Neuroscience}, vol.~17, no.~11, pp. 4302--4311, 1997.

\bibitem{lynn2019physics}
C.~W. Lynn and D.~S. Bassett, ``The physics of brain network structure,
  function and control,'' \emph{Nature Reviews Physics}, vol.~1, no.~5, pp.
  318--332, 2019.

\bibitem{friston2009modalities}
K.~J. Friston, ``Modalities, modes, and models in functional neuroimaging,''
  \emph{Science}, vol. 326, no. 5951, pp. 399--403, 2009.

\bibitem{bandettini2012functional}
P.~A. Bandettini, ``Functional mri: A confluence of fortunate circumstances,''
  \emph{Neuroimage}, vol.~61, no.~2, pp. A3--A11, 2012.

\bibitem{aqil2021graph}
M.~Aqil, S.~Atasoy, M.~L. Kringelbach, and R.~Hindriks, ``Graph neural fields:
  A framework for spatiotemporal dynamical models on the human connectome,''
  \emph{PLoS Computational Biology}, vol.~17, no.~1, p. e1008310, 2021.

\bibitem{kunze2016transcranial}
T.~Kunze, A.~Hunold, J.~Haueisen, V.~Jirsa, and A.~Spiegler, ``Transcranial
  direct current stimulation changes resting state functional connectivity: A
  large-scale brain network modeling study,'' \emph{Neuroimage}, vol. 140, pp.
  174--187, 2016.

\bibitem{molnar2019new}
Z.~Moln{\'a}r, G.~J. Clowry, N.~{\v{S}}estan, A.~Alzu'bi, T.~Bakken, R.~F.
  Hevner, P.~S. H{\"u}ppi, I.~Kostovi{\'c}, P.~Rakic, E.~Anton \emph{et~al.},
  ``New insights into the development of the human cerebral cortex,''
  \emph{Journal of Anatomy}, vol. 235, no.~3, pp. 432--451, 2019.

\bibitem{shaheen2022multiscale}
H.~Shaheen, S.~Pal, and R.~Melnik, ``Multiscale co-simulation of deep brain
  stimulation with brain networks in neurodegenerative disorders,'' \emph{Brain
  Multiphysics}, vol.~3, p. 100058, 2022.

\bibitem{sporns2013network}
O.~Sporns, ``Network attributes for segregation and integration in the human
  brain,'' \emph{Current Opinion in Neurobiology}, vol.~23, no.~2, pp.
  162--171, 2013.

\bibitem{sporns2011human}
------, ``The human connectome: a complex network,'' \emph{Annals of the new
  York Academy of Sciences}, vol. 1224, no.~1, pp. 109--125, 2011.

\bibitem{bardsley2018computational}
J.~M. Bardsley, \emph{Computational Uncertainty Quantification for Inverse
  Problems: An Introduction to Singular Integrals}.\hskip 1em plus 0.5em minus
  0.4em\relax SIAM, 2018.

\bibitem{presigny2022colloquium}
C.~Presigny and F.~D.~V. Fallani, ``Colloquium: Multiscale modeling of brain
  network organization,'' \emph{Reviews of Modern Physics}, vol.~94, no.~3, p.
  031002, 2022.

\bibitem{d2022quest}
E.~D’Angelo and V.~Jirsa, ``The quest for multiscale brain modeling,''
  \emph{Trends in Neurosciences}, 2022.

\bibitem{sabbatini2003neurons}
R.~M. Sabbatini, ``Neurons and synapses. the history of its discovery,''
  \emph{Brain and Mind.}, vol.~17, pp. 1--6, 2003.

\bibitem{jones2015understanding}
E.~Y. Jones, ``Understanding cell signalling systems: paving the way for new
  therapies,'' \emph{Philosophical Transactions of the Royal Society A:
  Mathematical, Physical and Engineering Sciences}, vol. 373, no. 2036, p.
  20130155, 2015.

\bibitem{sporns2012discovering}
O.~Sporns, \emph{Discovering the human connectome}.\hskip 1em plus 0.5em minus
  0.4em\relax MIT Press, 2012.

\bibitem{sporns2005human}
O.~Sporns, G.~Tononi, and R.~K{\"o}tter, ``The human connectome: a structural
  description of the human brain,'' \emph{PLoS Comput Biol}, vol.~1, no.~4, p.
  e42, 2005.

\bibitem{smith2012future}
S.~M. Smith, ``The future of fmri connectivity,'' \emph{Neuroimage}, vol.~62,
  no.~2, pp. 1257--1266, 2012.

\bibitem{sporns2022graph}
O.~Sporns, ``Graph theory methods: applications in brain networks,''
  \emph{Dialogues in Clinical Neuroscience}, 2022.

\bibitem{bernhardt2015network}
B.~C. Bernhardt, L.~Bonilha, and D.~W. Gross, ``Network analysis for a network
  disorder: the emerging role of graph theory in the study of epilepsy,''
  \emph{Epilepsy \& Behavior}, vol.~50, pp. 162--170, 2015.

\bibitem{mears2016network}
D.~Mears and H.~B. Pollard, ``Network science and the human brain: using graph
  theory to understand the brain and one of its hubs, the amygdala, in health
  and disease,'' \emph{Journal of Neuroscience Research}, vol.~94, no.~6, pp.
  590--605, 2016.

\bibitem{farahani2019application}
F.~V. Farahani, W.~Karwowski, and N.~R. Lighthall, ``Application of graph
  theory for identifying connectivity patterns in human brain networks: a
  systematic review,'' \emph{Frontiers in Neuroscience}, vol.~13, p. 585, 2019.

\bibitem{palombo2020sandi}
M.~Palombo, A.~Ianus, M.~Guerreri, D.~Nunes, D.~C. Alexander, N.~Shemesh, and
  H.~Zhang, ``Sandi: a compartment-based model for non-invasive apparent soma
  and neurite imaging by diffusion mri,'' \emph{Neuroimage}, vol. 215, p.
  116835, 2020.

\bibitem{szalkai2019high}
B.~Szalkai, C.~Kerepesi, B.~Varga, and V.~Grolmusz, ``High-resolution directed
  human connectomes and the consensus connectome dynamics,'' \emph{Plos One},
  vol.~14, no.~4, p. e0215473, 2019.

\bibitem{tadic2019functional}
B.~Tadi{\'c}, M.~Andjelkovi{\'c}, and R.~Melnik, ``Functional geometry of human
  connectomes,'' \emph{Scientific Reports}, vol.~9, no.~1, pp. 1--12, 2019.

\bibitem{deco2015rethinking}
G.~Deco, G.~Tononi, M.~Boly, and M.~L. Kringelbach, ``Rethinking segregation
  and integration: contributions of whole-brain modelling,'' \emph{Nature
  Reviews Neuroscience}, vol.~16, no.~7, pp. 430--439, 2015.

\bibitem{fornito2015connectomics}
A.~Fornito, A.~Zalesky, and M.~Breakspear, ``The connectomics of brain
  disorders,'' \emph{Nature Reviews Neuroscience}, vol.~16, no.~3, pp.
  159--172, 2015.

\bibitem{lacy2022cortical}
T.~C. Lacy, P.~A. Robinson, K.~M. Aquino, and J.~C. Pang, ``Cortical
  depth-dependent modeling of visual hemodynamic responses,'' \emph{Journal of
  Theoretical Biology}, vol. 535, p. 110978, 2022.

\bibitem{szalkai2017parameterizable}
B.~Szalkai, C.~Kerepesi, B.~Varga, and V.~Grolmusz, ``Parameterizable consensus
  connectomes from the human connectome project: the budapest reference
  connectome server v3. 0,'' \emph{Cognitive Neurodynamics}, vol.~11, no.~1,
  pp. 113--116, 2017.

\bibitem{mcnab2013human}
J.~A. McNab, B.~L. Edlow, T.~Witzel, S.~Y. Huang, H.~Bhat, K.~Heberlein,
  T.~Feiweier, K.~Liu, B.~Keil, J.~Cohen-Adad \emph{et~al.}, ``\mbox{The Human
  Connectome Project} and beyond: initial applications of 300 mt/m gradients,''
  \emph{Neuroimage}, vol.~80, pp. 234--245, 2013.

\bibitem{kerepesi2016direct}
C.~Kerepesi, B.~Szalkai, B.~Varga, and V.~Grolmusz, ``How to direct the edges
  of the connectomes: Dynamics of the consensus connectomes and the development
  of the connections in the human brain,'' \emph{Plos One}, vol.~11, no.~6, p.
  e0158680, 2016.

\bibitem{Seeley2010Neurodegenerative}
B.~Seeley, ``Neurodegenerative diseases target large-scale human brain
  networks,'' \emph{Alzheimer's \& Dementia}, vol.~4, no.~6, p. S167, 2010.

\bibitem{Erkkinen2018Clinical}
M.~G. Erkkinen, M.-O. Kim, and M.~D. Geschwind, ``Clinical neurology and
  epidemiology of the major neurodegenerative diseases,'' \emph{Cold Spring
  Harbor Perspectives in Biology}, vol.~10, no.~4, p. a033118, 2018.

\bibitem{Buckner2005Molecular}
R.~L. Buckner, A.~Z. Snyder, B.~J. Shannon, G.~LaRossa, R.~Sachs, A.~F.
  Fotenos, Y.~I. Sheline, W.~E. Klunk, C.~A. Mathis, J.~C. Morris
  \emph{et~al.}, ``Molecular, structural, and functional characterization of
  alzheimer's disease: evidence for a relationship between default activity,
  amyloid, and memory,'' \emph{Journal of Neuroscience}, vol.~25, no.~34, pp.
  7709--7717, 2005.

\bibitem{Palop2006Network}
J.~J. Palop, J.~Chin, and L.~Mucke, ``A network dysfunction perspective on
  neurodegenerative diseases,'' \emph{Nature}, vol. 443, no. 7113, pp.
  768--773, 2006.

\bibitem{Meyer1992Organizational}
J.~W. Meyer and W.~R. Scott, \emph{Organizational environments: Ritual and
  rationality}.\hskip 1em plus 0.5em minus 0.4em\relax SAGE Publications,
  Incorporated, 1992.

\bibitem{McGahan2020Mathematical}
K.~McGahan and J.~Keener, ``A mathematical model analyzing temperature
  threshold dependence in cold sensitive neurons,'' \emph{Plos One}, vol.~15,
  no.~8, p. e0237347, 2020.

\bibitem{Seeley2009Neurodegenerative}
W.~W. Seeley, R.~K. Crawford, J.~Zhou, B.~L. Miller, and M.~D. Greicius,
  ``Neurodegenerative diseases target large-scale human brain networks,''
  \emph{Neuron}, vol.~62, no.~1, pp. 42--52, 2009.

\bibitem{Supekar2008Network}
K.~Supekar, V.~Menon, D.~Rubin, M.~Musen, and M.~D. Greicius, ``Network
  analysis of intrinsic functional brain connectivity in \mbox {Alzheimer's}
  disease,'' \emph{PLoS Comput Biol}, vol.~4, no.~6, p. e1000100, 2008.

\bibitem{Dadario2022Should}
N.~B. Dadario and M.~E. Sughrue, ``Should neurosurgeons try to preserve
  non-traditional brain networks? a systematic review of the neuroscientific
  evidence,'' \emph{Journal of Personalized Medicine}, vol.~12, no.~4, p. 587,
  2022.

\bibitem{frank2003clinical}
R.~Frank and R.~Hargreaves, ``Clinical biomarkers in drug discovery and
  development,'' \emph{Nature Reviews Drug Discovery}, vol.~2, no.~7, pp.
  566--580, 2003.

\bibitem{Giampietri2022Fluid}
L.~Giampietri, E.~Belli, M.~F. Beatino, S.~Giannoni, G.~Palermo, N.~Campese,
  G.~Tognoni, G.~Siciliano, R.~Ceravolo, C.~De~Luca \emph{et~al.}, ``Fluid
  biomarkers in \mbox{Alzheimer's} disease and other neurodegenerative
  disorders: Toward integrative diagnostic frameworks and tailored
  treatments,'' \emph{Diagnostics}, vol.~12, no.~4, p. 796, 2022.

\bibitem{Young2020Imaging}
P.~N. Young, M.~Estarellas, E.~Coomans, M.~Srikrishna, H.~Beaumont, A.~Maass,
  A.~V. Venkataraman, R.~Lissaman, D.~Jim{\'e}nez, M.~J. Betts \emph{et~al.},
  ``Imaging biomarkers in neurodegeneration: current and future practices,''
  \emph{Alzheimer's Research \& Therapy}, vol.~12, no.~1, pp. 1--17, 2020.

\bibitem{jama1}
S.~Palmqvist, P.~Tideman, N.~Mattsson-Carlgren, and et~al, ``Blood biomarkers
  to detect alzheimer disease in primary care and secondary care,'' \emph{JAMA,
  doi:10.1001/jama.2024.13855}, 2024.

\bibitem{Ranson2023Harnessing}
J.~M. Ranson, M.~Bucholc, D.~Lyall, D.~Newby, L.~Winchester, N.~P. Oxtoby,
  M.~Veldsman, T.~Rittman, S.~Marzi, N.~Skene \emph{et~al.}, ``Harnessing the
  potential of machine learning and artificial intelligence for dementia
  research,'' \emph{Brain Informatics}, vol.~10, no.~1, p.~6, 2023.

\bibitem{Ewen2021Biomarkers}
J.~B. Ewen, W.~Z. Potter, and J.~A. Sweeney, ``Biomarkers and neurobehavioral
  diagnosis,'' \emph{Biomarkers in Neuropsychiatry}, vol.~4, p. 100029, 2021.

\bibitem{Peng2021Multiscale}
G.~C. Peng, M.~Alber, A.~Buganza~Tepole, W.~R. Cannon, S.~De, S.~Dura-Bernal,
  K.~Garikipati, G.~Karniadakis, W.~W. Lytton, P.~Perdikaris \emph{et~al.},
  ``Multiscale modeling meets machine learning: What can we learn?''
  \emph{Archives of Computational Methods in Engineering}, vol.~28, pp.
  1017--1037, 2021.

\bibitem{Babich2022Phytotherapeutic}
O.~Babich, V.~Larina, S.~Ivanova, A.~Tarasov, M.~Povydysh, A.~Orlova,
  J.~Strugar, and S.~Sukhikh, ``Phytotherapeutic approaches to the prevention
  of age-related changes and the extension of active longevity,''
  \emph{Molecules}, vol.~27, no.~7, p. 2276, 2022.

\bibitem{shaheen2021deep}
H.~Shaheen and R.~Melnik, ``Deep brain stimulation with a computational model
  for the cortex-thalamus-basal-ganglia system and network dynamics of
  neurological disorders,'' \emph{Computational and Mathematical Methods}, vol.
  2022, 2021.

\bibitem{Mengi2021Artificial}
M.~Mengi and D.~Malhotra, ``Artificial intelligence based techniques for the
  detection of socio-behavioral disorders: a systematic review,''
  \emph{Archives of Computational Methods in Engineering}, pp. 1--45, 2021.

\bibitem{vendel2019need}
E.~Vendel, V.~Rottsch{\"a}fer, and E.~C. de~Lange, ``The need for mathematical
  modelling of spatial drug distribution within the brain,'' \emph{Fluids and
  Barriers of the CNS}, vol.~16, no.~1, p.~12, 2019.

\bibitem{Guidoboni2020Neurodegenerative}
G.~Guidoboni, R.~Sacco, M.~Szopos, L.~Sala, A.~C. Verticchio~Vercellin,
  B.~Siesky, and A.~Harris, ``Neurodegenerative disorders of the eye and of the
  brain: a perspective on their fluid-dynamical connections and the potential
  of mechanism-driven modeling,'' \emph{Frontiers in Neuroscience}, vol.~14, p.
  566428, 2020.

\bibitem{Sacco2019Comprehensive}
R.~Sacco, G.~Guidoboni, and A.~G. Mauri, \emph{A Comprehensive Physically Based
  Approach to Modeling in Bioengineering and Life Sciences}.\hskip 1em plus
  0.5em minus 0.4em\relax Academic Press, 2019.

\bibitem{Zhang2020Data}
X.~Zhang, U.~Braun, H.~Tost, and D.~S. Bassett, ``Data-driven approaches to
  neuroimaging analysis to enhance psychiatric diagnosis and therapy,''
  \emph{Biological Psychiatry: Cognitive Neuroscience and Neuroimaging},
  vol.~5, no.~8, pp. 780--790, 2020.

\bibitem{vosoughi2020mathematical}
A.~Vosoughi, S.~Sadigh-Eteghad, M.~Ghorbani, S.~Shahmorad, M.~Farhoudi, M.~A.
  Rafi, and Y.~Omidi, ``Mathematical models to shed light on amyloid-beta and
  tau protein dependent pathologies in \mbox{Alzheimer's} disease,''
  \emph{Neuroscience}, vol. 424, pp. 45--57, 2020.

\bibitem{Marzetti2019Brain}
L.~Marzetti, A.~Basti, F.~Chella, A.~D'Andrea, J.~Syrj{\"a}l{\"a}, and
  V.~Pizzella, ``Brain functional connectivity through phase coupling of
  neuronal oscillations: a perspective from magnetoencephalography,''
  \emph{Frontiers in Neuroscience}, vol.~13, p. 964, 2019.

\bibitem{Battaglia2020Functional}
D.~Battaglia and A.~Brovelli, ``Functional connectivity and neuronal dynamics:
  insights from computational methods,'' \emph{The Cognitive Neurosciences,
  Sixth Edition}, 2020.

\bibitem{Bakshi2019Mathematical}
S.~Bakshi, V.~Chelliah, C.~Chen, and P.~H. van~der Graaf, ``Mathematical
  biology models of parkinson's disease,'' \emph{CPT: Pharmacometrics \&
  Systems Pharmacology}, vol.~8, no.~2, pp. 77--86, 2019.

\bibitem{Tandon2022Pathway}
G.~Tandon, S.~Yadav, and S.~Kaur, ``Pathway modeling and simulation analysis,''
  in \emph{Bioinformatics}.\hskip 1em plus 0.5em minus 0.4em\relax Elsevier,
  2022, pp. 409--423.

\bibitem{Surguchov2022Biomarkers}
A.~Surguchov, ``Biomarkers in \mbox{Parkinson’s} disease,''
  \emph{Neurodegenerative diseases biomarkers: Towards translating research to
  clinical practice}, pp. 155--180, 2022.

\bibitem{Rubinov2010Complex}
M.~Rubinov and O.~Sporns, ``Complex network measures of brain connectivity:
  uses and interpretations,'' \emph{Neuroimage}, vol.~52, no.~3, pp.
  1059--1069, 2010.

\bibitem{Suarez2020Linking}
L.~E. Su{\'a}rez, R.~D. Markello, R.~F. Betzel, and B.~Misic, ``Linking
  structure and function in macroscale brain networks,'' \emph{Trends in
  Cognitive Sciences}, vol.~24, no.~4, pp. 302--315, 2020.

\bibitem{Van2013Network}
M.~P. Van~den Heuvel and O.~Sporns, ``Network hubs in the human brain,''
  \emph{Trends in Cognitive Sciences}, vol.~17, no.~12, pp. 683--696, 2013.

\bibitem{pal2022nonlocal}
S.~Pal and R.~Melnik, ``Nonlocal models in the analysis of brain
  neurodegenerative protein dynamics with application to alzheimer’s
  disease,'' \emph{Scientific Reports}, vol.~12, no.~1, p. 7328, 2022.

\bibitem{Schilling2022Prevalence}
K.~G. Schilling, C.~M. Tax, F.~Rheault, B.~A. Landman, A.~W. Anderson,
  M.~Descoteaux, and L.~Petit, ``Prevalence of white matter pathways coming
  into a single white matter voxel orientation: The bottleneck issue in
  tractography,'' \emph{Human Brain Mapping}, vol.~43, no.~4, pp. 1196--1213,
  2022.

\bibitem{Del2017Mapping}
J.~Del~Gaizo, J.~Fridriksson, G.~Yourganov, A.~E. Hillis, G.~Hickok, B.~Misic,
  C.~Rorden, and L.~Bonilha, ``Mapping language networks using the structural
  and dynamic brain connectomes,'' \emph{Eneuro}, vol.~4, no.~5, 2017.

\bibitem{Keller2022Hierarchical}
A.~S. Keller, V.~J. Sydnor, A.~Pines, D.~A. Fair, D.~S. Bassett, and T.~D.
  Satterthwaite, ``Hierarchical functional system development supports
  executive function,'' \emph{Trends in Cognitive Sciences}, 2022.

\bibitem{Litwinczuk2022Combination}
M.~C. Litwi{\'n}czuk, N.~Muhlert, L.~Cloutman, N.~Trujillo-Barreto, and
  A.~Woollams, ``Combination of structural and functional connectivity explains
  unique variation in specific domains of cognitive function,''
  \emph{NeuroImage}, vol. 262, p. 119531, 2022.

\bibitem{Andica2020Mr}
C.~Andica, K.~Kamagata, T.~Hatano, Y.~Saito, K.~Ogaki, N.~Hattori, and S.~Aoki,
  ``Mr biomarkers of degenerative brain disorders derived from diffusion
  imaging,'' \emph{Journal of Magnetic Resonance Imaging}, vol.~52, no.~6, pp.
  1620--1636, 2020.

\bibitem{Cummings2019Role}
J.~Cummings, ``The role of biomarkers in \mbox{Alzheimer's} disease drug
  development,'' \emph{Reviews on Biomarker Studies in Psychiatric and
  Neurodegenerative Disorders}, pp. 29--61, 2019.

\bibitem{Oliveira2021Alpha}
L.~M. Oliveira, T.~Gasser, R.~Edwards, M.~Zweckstetter, R.~Melki, L.~Stefanis,
  H.~A. Lashuel, D.~Sulzer, K.~Vekrellis, G.~M. Halliday \emph{et~al.},
  ``Alpha-synuclein research: Defining strategic moves in the battle against
  \mbox{Parkinson’s} disease,'' \emph{NPJ \mbox {Parkinson's} Disease},
  vol.~7, no.~1, p.~65, 2021.

\bibitem{fjell2016brain}
A.~M. Fjell, M.~H. Sneve, A.~B. Storsve, H.~Grydeland, A.~Yendiki, and K.~B.
  Walhovd, ``Brain events underlying episodic memory changes in aging: a
  longitudinal investigation of structural and functional connectivity,''
  \emph{Cerebral Cortex}, vol.~26, no.~3, pp. 1272--1286, 2016.

\bibitem{Seidler2010Motor}
R.~D. Seidler, J.~A. Bernard, T.~B. Burutolu, B.~W. Fling, M.~T. Gordon, J.~T.
  Gwin, Y.~Kwak, and D.~B. Lipps, ``Motor control and aging: links to
  age-related brain structural, functional, and biochemical effects,''
  \emph{Neuroscience \& Biobehavioral Reviews}, vol.~34, no.~5, pp. 721--733,
  2010.

\bibitem{Zimmermann2016Structural}
J.~Zimmermann, P.~Ritter, K.~Shen, S.~Rothmeier, M.~Schirner, and A.~R.
  McIntosh, ``Structural architecture supports functional organization in the
  human aging brain at a regionwise and network level,'' \emph{Human Brain
  Mapping}, vol.~37, no.~7, pp. 2645--2661, 2016.

\bibitem{Schultz2017Phases}
A.~P. Schultz, J.~P. Chhatwal, T.~Hedden, E.~C. Mormino, B.~J. Hanseeuw,
  J.~Sepulcre, W.~Huijbers, M.~LaPoint, R.~F. Buckley, K.~A. Johnson
  \emph{et~al.}, ``Phases of hyperconnectivity and hypoconnectivity in the
  default mode and salience networks track with amyloid and tau in clinically
  normal individuals,'' \emph{Journal of Neuroscience}, vol.~37, no.~16, pp.
  4323--4331, 2017.

\bibitem{Fjell2017Relationship}
A.~M. Fjell, M.~H. Sneve, H.~Grydeland, A.~B. Storsve, I.~K. Amlien,
  A.~Yendiki, and K.~B. Walhovd, ``Relationship between structural and
  functional connectivity change across the adult lifespan: a longitudinal
  investigation,'' \emph{Human brain mapping}, vol.~38, no.~1, pp. 561--573,
  2017.

\bibitem{centofante2023specific}
E.~Centofante, L.~Fralleoni, C.~A. Lupascu, M.~Migliore, A.~Rinaldi, and
  A.~Mele, ``Specific patterns of neural activity in the hippocampus after
  massed or distributed spatial training,'' \emph{Scientific Reports}, vol.~13,
  no.~1, p. 13357, 2023.

\bibitem{lozano2019deep}
A.~M. Lozano, N.~Lipsman, H.~Bergman, P.~Brown, S.~Chabardes, J.~W. Chang,
  K.~Matthews, C.~C. McIntyre, T.~E. Schlaepfer, M.~Schulder \emph{et~al.},
  ``Deep brain stimulation: current challenges and future directions,''
  \emph{Nature Reviews Neurology}, vol.~15, no.~3, pp. 148--160, 2019.

\bibitem{Mestre2020Brain}
H.~Mestre, Y.~Mori, and M.~Nedergaard, ``The brain’s glymphatic system:
  current controversies,'' \emph{Trends in Neurosciences}, vol.~43, no.~7, pp.
  458--466, 2020.

\bibitem{Lu2019Application}
M.~Lu, X.~Wei, Y.~Che, J.~Wang, and K.~A. Loparo, ``Application of
  reinforcement learning to deep brain stimulation in a computational model of
  \mbox{Parkinson’s} disease,'' \emph{IEEE Transactions on Neural Systems and
  Rehabilitation Engineering}, vol.~28, no.~1, pp. 339--349, 2019.

\bibitem{Kaltenbacher2020Inverse}
B.~Kaltenbacher and W.~Rundell, ``The inverse problem of reconstructing
  reaction--diffusion systems,'' \emph{Inverse Problems}, vol.~36, no.~6, p.
  065011, 2020.

\bibitem{Isakov2006Inverse}
V.~Isakov, \emph{Inverse problems for partial differential equations}.\hskip
  1em plus 0.5em minus 0.4em\relax Springer, 2006, vol. 127.

\bibitem{adler2017solving}
J.~Adler and O.~{\"O}ktem, ``Solving ill-posed inverse problems using iterative
  deep neural networks,'' \emph{Inverse Problems}, vol.~33, no.~12, p. 124007,
  2017.

\bibitem{Uhlirova2016Roadmap}
H.~Uhlirova, P.~Tian, S.~Sakad{\v{z}}i{\'c}, L.~Gagnon, M.~Thunemann,
  M.~Desjardins, P.~A. Saisan, K.~Nizar, M.~A. Yaseen, D.~J. Hagler~Jr
  \emph{et~al.}, ``The roadmap for estimation of cell-type-specific neuronal
  activity from non-invasive measurements,'' \emph{Philosophical Transactions
  of the Royal Society B: Biological Sciences}, vol. 371, no. 1705, p.
  20150356, 2016.

\bibitem{Liu2017Detecting}
Q.~Liu, S.~Farahibozorg, C.~Porcaro, N.~Wenderoth, and D.~Mantini, ``Detecting
  large-scale networks in the human brain using high-density
  electroencephalography,'' \emph{Human Brain Mapping}, vol.~38, no.~9, pp.
  4631--4643, 2017.

\bibitem{Ronzitti2017Recent}
E.~Ronzitti, C.~Ventalon, M.~Canepari, B.~C. Forget, E.~Papagiakoumou, and
  V.~Emiliani, ``Recent advances in patterned photostimulation for
  optogenetics,'' \emph{Journal of Optics}, vol.~19, no.~11, p. 113001, 2017.

\bibitem{Raissi2019Physics}
M.~Raissi, P.~Perdikaris, and G.~E. Karniadakis, ``Physics-informed neural
  networks: A deep learning framework for solving forward and inverse problems
  involving nonlinear partial differential equations,'' \emph{Journal of
  Computational physics}, vol. 378, pp. 686--707, 2019.

\bibitem{Zuo2022Deep}
C.~Zuo, J.~Qian, S.~Feng, W.~Yin, Y.~Li, P.~Fan, J.~Han, K.~Qian, and Q.~Chen,
  ``Deep learning in optical metrology: a review,'' \emph{Light: Science \&
  Applications}, vol.~11, no.~1, p.~39, 2022.

\bibitem{Bergen2019Machine}
K.~J. Bergen, P.~A. Johnson, M.~V. de~Hoop, and G.~C. Beroza, ``Machine
  learning for data-driven discovery in solid earth geoscience,''
  \emph{Science}, vol. 363, no. 6433, p. eaau0323, 2019.

\bibitem{Zampieri2019Machine}
G.~Zampieri, S.~Vijayakumar, E.~Yaneske, and C.~Angione, ``Machine and deep
  learning meet genome-scale metabolic modeling,'' \emph{PLoS Computational
  Biology}, vol.~15, no.~7, p. e1007084, 2019.

\bibitem{Peng2020Multiscale}
G.~C. Peng, M.~Alber, A.~B. Tepole, W.~R. Cannon, S.~De, S.~Dura-Bernal,
  K.~Garikipati, G.~Karniadakis, W.~W. Lytton, P.~Perdikaris \emph{et~al.},
  ``Multiscale modeling meets machine learning: What can we learn?''
  \emph{Archives of Computational Methods in Engineering}, pp. 1--21, 2020.

\bibitem{agnati2018brain}
L.~F. Agnati, M.~Marcoli, G.~Maura, A.~Woods, and D.~Guidolin, ``The brain as a
  “hyper-network”: the key role of neural networks as main producers of the
  integrated brain actions especially via the “broadcasted”
  neuroconnectomics,'' \emph{Journal of Neural Transmission}, vol. 125, no.~6,
  pp. 883--897, 2018.

\bibitem{andjelkovic2020topology}
M.~Andjelkovi{\'c}, B.~Tadi{\'c}, and R.~Melnik, ``The topology of higher-order
  complexes associated with brain hubs in human connectomes,'' \emph{Scientific
  Reports}, vol.~10, no. 17320, pp. 1--10, 2020.

\bibitem{arendt2001alzheimer}
T.~Arendt, ``Alzheimer's disease as a disorder of mechanisms underlying
  structural brain self-organization,'' \emph{Neuroscience}, vol. 102, no.~4,
  pp. 723--765, 2001.

\bibitem{leistritz2015time}
L.~Leistritz, K.~Schiecke, L.~Astolfi, and H.~Witte, ``Time-variant modeling of
  brain processes,'' \emph{Proceedings of the IEEE}, vol. 104, no.~2, pp.
  262--281, 2015.

\bibitem{coronel2023whole}
C.~Coronel-Oliveros, C.~Gie{\ss}ing, V.~Medel, R.~Cofr{\'e}, and P.~Orio,
  ``Whole-brain modeling explains the context-dependent effects of cholinergic
  neuromodulation,'' \emph{NeuroImage}, vol. 265, p. 119782, 2023.

\bibitem{dai2023eight}
Y.-R. Dai, Y.-K. Wu, X.~Chen, Y.-W. Zeng, K.~Li, J.-T. Li, Y.-A. Su, L.-L. Zhu,
  C.-G. Yan, and T.-M. Si, ``Eight-week antidepressant treatment changes
  intrinsic functional brain topology in first-episode drug-na{\"\i}ve patients
  with major depressive disorder,'' \emph{Journal of Affective Disorders}, vol.
  329, pp. 225--234, 2023.

\bibitem{guo2023functional}
S.~Guo, L.~Feng, R.~Ding, S.~Long, H.~Yang, X.~Gong, J.~Lu, and D.~Yao,
  ``Functional gradients in prefrontal regions and somatomotor networks reflect
  the effect of music training experience on cognitive aging,'' \emph{Cerebral
  Cortex}, vol.~33, no.~12, pp. 7506--7517, 2023.

\bibitem{cohen2016segregation}
J.~R. Cohen and M.~D'Esposito, ``The segregation and integration of distinct
  brain networks and their relationship to cognition,'' \emph{Journal of
  Neuroscience}, vol.~36, no.~48, pp. 12\,083--12\,094, 2016.

\bibitem{shinn2023functional}
M.~Shinn, A.~Hu, L.~Turner, S.~Noble, K.~H. Preller, J.~L. Ji, F.~Moujaes,
  S.~Achard, D.~Scheinost, R.~T. Constable \emph{et~al.}, ``Functional brain
  networks reflect spatial and temporal autocorrelation,'' \emph{Nature
  Neuroscience}, pp. 1--12, 2023.

\bibitem{jin2022photoacoustic}
T.~Jin, W.~Qi, X.~Liang, H.~Guo, Q.~Liu, and L.~Xi, ``Photoacoustic imaging of
  brain functions: Wide filed-of-view functional imaging with high
  spatiotemporal resolution,'' \emph{Laser \& Photonics Reviews}, vol.~16,
  no.~2, p. 2100304, 2022.

\bibitem{shaheen2021mathematical}
H.~Shaheen, S.~Singh, and R.~Melnik, ``Mathematical modeling of
  calcium-mediated exosomal dynamics in neural cells,'' in \emph{Advances in
  Nonlinear Dynamics: Proceedings of the Second International Nonlinear
  Dynamics Conference (NODYCON 2021), Volume 3}.\hskip 1em plus 0.5em minus
  0.4em\relax Springer, 2021, pp. 83--92.

\bibitem{horstemeyer2021multiscale}
M.~F. Horstemeyer and R.~K. Prabhu, \emph{Multiscale Biomechanical Modeling of
  the Brain}.\hskip 1em plus 0.5em minus 0.4em\relax Academic Press, 2021.

\bibitem{hansen2017spatio}
S.~T. Hansen and L.~K. Hansen, ``Spatio-temporal reconstruction of brain
  dynamics from eeg with a markov prior,'' \emph{NeuroImage}, vol. 148, pp.
  274--283, 2017.

\bibitem{kass2023identification}
R.~E. Kass, H.~Bong, M.~Olarinre, Q.~Xin, and K.~N. Urban, ``Identification of
  interacting neural populations: methods and statistical considerations,''
  \emph{Journal of Neurophysiology}, vol. 130, no.~3, pp. 475--496, 2023.

\bibitem{murray2017working}
J.~D. Murray, J.~Jaramillo, and X.-J. Wang, ``Working memory and
  decision-making in a frontoparietal circuit model,'' \emph{Journal of
  Neuroscience}, vol.~37, no.~50, pp. 12\,167--12\,186, 2017.

\bibitem{jimenez2023macromolecular}
J.~S. Jim{\'e}nez, ``Macromolecular structures and proteins interacting with
  the microtubule associated tau protein,'' \emph{Neuroscience}, vol. 518, pp.
  70--82, 2023.

\bibitem{vecchio2017small}
F.~Vecchio, F.~Miraglia, F.~Piludu, G.~Granata, R.~Romanello, M.~Caulo,
  V.~Onofrj, P.~Bramanti, C.~Colosimo, and P.~M. Rossini, ``“small world”
  architecture in brain connectivity and hippocampal volume in
  \mbox{Alzheimer's} disease: a study via graph theory from eeg data,''
  \emph{Brain Imaging and Behavior}, vol.~11, no.~2, pp. 473--485, 2017.

\bibitem{yu2018fused}
D.~Yu, S.~H. Lee, J.~Lim, G.~Xiao, R.~C. Craddock, and B.~B. Biswal, ``Fused
  lasso regression for identifying differential correlations in brain
  connectome graphs,'' \emph{Statistical Analysis and Data Mining: The ASA Data
  Science Journal}, vol.~11, no.~5, pp. 203--226, 2018.

\bibitem{perdikaris2016visualizing}
P.~Perdikaris, J.~A. Insley, L.~Grinberg, Y.~Yu, M.~E. Papka, and G.~E.
  Karniadakis, ``Visualizing multiphysics, fluid-structure interaction
  phenomena in intracranial aneurysms,'' \emph{Parallel computing}, vol.~55,
  pp. 9--16, 2016.

\bibitem{han2018solving}
J.~Han, A.~Jentzen, and E.~Weinan, ``Solving high-dimensional partial
  differential equations using deep learning,'' \emph{Proceedings of the
  National Academy of Sciences}, vol. 115, no.~34, pp. 8505--8510, 2018.

\bibitem{weinan2017deep}
E.~Weinan, J.~Han, and A.~Jentzen, ``Deep learning-based numerical methods for
  high-dimensional parabolic partial differential equations and backward
  stochastic differential equations,'' \emph{Communications in Mathematics and
  Statistics}, vol.~5, no.~4, pp. 349--380, 2017.

\bibitem{zitnik2019machine}
M.~Zitnik, F.~Nguyen, B.~Wang, J.~Leskovec, A.~Goldenberg, and M.~M. Hoffman,
  ``Machine learning for integrating data in biology and medicine: Principles,
  practice, and opportunities,'' \emph{Information Fusion}, vol.~50, pp.
  71--91, 2019.

\bibitem{lu2018mathematical}
Y.~Lu, L.~Bi, J.~Lian, and H.~Li, ``Mathematical modeling of eeg signals-based
  brain-control behavior,'' \emph{IEEE Transactions on Neural Systems and
  Rehabilitation Engineering}, vol.~26, no.~8, pp. 1535--1543, 2018.

\bibitem{dumont2017stochastic}
G.~Dumont, A.~Payeur, and A.~Longtin, ``A stochastic-field description of
  finite-size spiking neural networks,'' \emph{PLoS Computational Biology},
  vol.~13, no.~8, p. e1005691, 2017.

\bibitem{kim2018effect}
S.-Y. Kim and W.~Lim, ``Effect of inhibitory spike-timing-dependent plasticity
  on fast sparsely synchronized rhythms in a small-world neuronal network,''
  \emph{Neural Networks}, vol. 106, pp. 50--66, 2018.

\bibitem{thieu2022coupled}
T.~K.~T. Thieu and R.~Melnik, ``Coupled effects of channels and synaptic
  dynamics in stochastic modelling of healthy and parkinson's-disease-affected
  brains.'' \emph{AIMS Bioengineering}, vol.~9, no.~2, pp. 213--238, 2022.

\bibitem{linninger2009mathematical}
A.~A. Linninger, M.~Xenos, B.~Sweetman, S.~Ponkshe, X.~Guo, and R.~Penn, ``A
  mathematical model of blood, cerebrospinal fluid and brain dynamics,''
  \emph{Journal of mathematical biology}, vol.~59, pp. 729--759, 2009.

\bibitem{alavijeh2005drug}
M.~S. Alavijeh, M.~Chishty, M.~Z. Qaiser, and A.~M. Palmer, ``Drug metabolism
  and pharmacokinetics, the blood-brain barrier, and central nervous system
  drug discovery,'' \emph{NeuroRx}, vol.~2, pp. 554--571, 2005.

\bibitem{pietras2022mesoscopic}
B.~Pietras, V.~Schmutz, and T.~Schwalger, ``Mesoscopic description of
  hippocampal replay and metastability in spiking neural networks with
  short-term plasticity,'' \emph{PLoS Computational Biology}, vol.~18, no.~12,
  p. e1010809, 2022.

\bibitem{jimenez2019metastable}
A.~Jimenez-Marin, I.~Diez, J.~M. Cortes, S.~Rodrigues \emph{et~al.},
  ``Metastable resting state brain dynamics,'' \emph{Frontiers in Computational
  Neuroscience}, vol.~13, p.~62, 2019.

\bibitem{colonna2017microglia}
M.~Colonna and O.~Butovsky, ``Microglia function in the central nervous system
  during health and neurodegeneration,'' \emph{Annual Review of Immunology},
  vol.~35, pp. 441--468, 2017.

\bibitem{csaszar2022microglia}
E.~Cs{\'a}sz{\'a}r, N.~L{\'e}n{\'a}rt, C.~Cser{\'e}p, Z.~K{\"o}rnyei,
  R.~Fekete, B.~P{\'o}sfai, D.~Bal{\'a}zsfi, B.~Hangya, A.~D. Schwarcz,
  E.~Szabadits \emph{et~al.}, ``Microglia modulate blood flow, neurovascular
  coupling, and hypoperfusion via purinergic actions,'' \emph{Journal of
  Experimental Medicine}, vol. 219, no.~3, 2022.

\bibitem{kozma2017cinematic}
R.~Kozma and W.~J. Freeman, ``Cinematic operation of the cerebral cortex
  interpreted via critical transitions in self-organized dynamic systems,''
  \emph{Frontiers in Systems Neuroscience}, p.~10, 2017.

\bibitem{tadic2021self}
B.~Tadi{\'c} and R.~Melnik, ``Self-organised critical dynamics as a key to
  fundamental features of complexity in physical, biological, and social
  networks,'' \emph{Dynamics}, vol.~1, no.~2, pp. 181--197, 2021.

\bibitem{tadic2024fundamental}
------, ``Fundamental interactions in self-organised critical dynamics on
  higher order networks,'' \emph{The European Physical Journal B}, vol.~97,
  no.~6, pp. 1--13, 2024.

\bibitem{bargmann2013connectome}
C.~I. Bargmann and E.~Marder, ``From the connectome to brain function,''
  \emph{Nature Methods}, vol.~10, no.~6, pp. 483--490, 2013.

\bibitem{benkarim2022riemannian}
O.~Benkarim, C.~Paquola, B.-y. Park, J.~Royer, R.~Rodr{\'\i}guez-Cruces, R.~V.
  de~Wael, B.~Misic, G.~Piella, and B.~C. Bernhardt, ``A riemannian approach to
  predicting brain function from the structural connectome,''
  \emph{NeuroImage}, vol. 257, p. 119299, 2022.

\bibitem{winding2023connectome}
M.~Winding, B.~D. Pedigo, C.~L. Barnes, H.~G. Patsolic, Y.~Park, T.~Kazimiers,
  A.~Fushiki, I.~V. Andrade, A.~Khandelwal, J.~Valdes-Aleman \emph{et~al.},
  ``The connectome of an insect brain,'' \emph{Science}, vol. 379, no. 6636, p.
  eadd9330, 2023.

\bibitem{lariviere2023brainstat}
S.~Larivi{\`e}re, {\c{S}}.~Bayrak, R.~V. de~Wael, O.~Benkarim, P.~Herholz,
  R.~Rodriguez-Cruces, C.~Paquola, S.-J. Hong, B.~Misic, A.~C. Evans
  \emph{et~al.}, ``Brainstat: A toolbox for brain-wide statistics and
  multimodal feature associations,'' \emph{NeuroImage}, vol. 266, p. 119807,
  2023.

\bibitem{zhang2020mathematical}
Y.~Zhang and W.~Wang, ``Mathematical analysis for stochastic model of
  \mbox{Alzheimer's} disease,'' \emph{Communications in Nonlinear Science and
  Numerical Simulation}, vol.~89, p. 105347, 2020.

\bibitem{fornari2019prion}
S.~Fornari, A.~Sch{\"a}fer, M.~Jucker, A.~Goriely, and E.~Kuhl, ``Prion-like
  spreading of \mbox{Alzheimer's} disease within the brain’s connectome,''
  \emph{Journal of the Royal Society Interface}, vol.~16, no. 159, p. 20190356,
  2019.

\bibitem{fornari2020spatially}
S.~Fornari, A.~Sch{\"a}fer, E.~Kuhl, and A.~Goriely, ``Spatially-extended
  nucleation-aggregation-fragmentation models for the dynamics of prion-like
  neurodegenerative protein-spreading in the brain and its connectome,''
  \emph{Journal of Theoretical Biology}, vol. 486, p. 110102, 2020.

\bibitem{thuraisingham2022kinetic}
R.~Thuraisingham, ``A kinetic scheme to examine the role of glial cells in the
  pathogenesis of \mbox{Alzheimer’s} disease,'' \emph{Metabolic Brain
  Disease}, vol.~37, no.~3, pp. 801--805, 2022.

\bibitem{bennett2021activated}
M.~L. Bennett and A.~N. Viaene, ``What are activated and reactive glia and what
  is their role in neurodegeneration?'' \emph{Neurobiology of Disease}, vol.
  148, p. 105172, 2021.

\bibitem{smethurst2022role}
P.~Smethurst, H.~Franklin, B.~E. Clarke, K.~Sidle, and R.~Patani, ``The role of
  astrocytes in prion-like mechanisms of neurodegeneration,'' \emph{Brain},
  vol. 145, no.~1, pp. 17--26, 2022.

\bibitem{prusiner1998prions}
S.~B. Prusiner, ``Prions,'' \emph{Proceedings of the National Academy of
  Sciences}, vol.~95, no.~23, pp. 13\,363--13\,383, 1998.

\bibitem{jucker2013self}
M.~Jucker and L.~C. Walker, ``Self-propagation of pathogenic protein aggregates
  in neurodegenerative diseases,'' \emph{Nature}, vol. 501, no. 7465, pp.
  45--51, 2013.

\bibitem{barros2018multi}
M.~T. Barros, W.~Silva, and C.~D.~M. Regis, ``The multi-scale impact of the
  \mbox{Alzheimer’s} disease on the topology diversity of astrocytes
  molecular communications nanonetworks,'' \emph{IEEE Access}, vol.~6, pp.
  78\,904--78\,917, 2018.

\bibitem{wang2018brain}
M.~Wang, Y.~He, T.~J. Sejnowski, and X.~Yu, ``Brain-state dependent astrocytic
  ca2+ signals are coupled to both positive and negative bold-fmri signals,''
  \emph{Proceedings of the National Academy of Sciences}, vol. 115, no.~7, pp.
  E1647--E1656, 2018.

\bibitem{selkoe2016amyloid}
D.~J. Selkoe and J.~Hardy, ``The amyloid hypothesis of alzheimer's disease at
  25 years,'' \emph{EMBO Molecular Medicine}, vol.~8, no.~6, pp. 595--608,
  2016.

\bibitem{buskila2019generating}
Y.~Buskila, A.~Bellot-Saez, and J.~W. Morley, ``Generating brain waves, the
  power of astrocytes,'' \emph{Frontiers in Neuroscience}, vol.~13, p. 1125,
  2019.

\bibitem{pallitto2001mathematical}
M.~M. Pallitto and R.~M. Murphy, ``A mathematical model of the kinetics of
  $\beta$-amyloid fibril growth from the denatured state,'' \emph{Biophysical
  Journal}, vol.~81, no.~3, pp. 1805--1822, 2001.

\bibitem{bertsch2017alzheimer}
M.~Bertsch, B.~Franchi, N.~Marcello, M.~C. Tesi, and A.~Tosin,
  ``\mbox{Alzheimer's} disease: a mathematical model for onset and
  progression,'' \emph{Mathematical Medicine and Biology: A Journal of the
  IMA}, vol.~34, no.~2, pp. 193--214, 2017.

\bibitem{pal2022influence}
S.~Pal, H.~Shaheen, and R.~Melnik, ``The influence of amyloid-beta on calcium
  dynamics in \mbox{Alzheimer’s} disease: A spatio-temporal study. in:
  Gervasi, o., murgante, b., misra, s., rocha, a.m.a.c., garau, c. (eds)
  computational science and its applications- \mbox{ICCSA 2022. Lecture Notes
  in Computer Science},'' in \emph{International Conference on Computational
  Science and Its Applications}, vol. 13377.\hskip 1em plus 0.5em minus
  0.4em\relax Springer, Cham, 2022, pp. 308--322.

\bibitem{hao2016mathematical}
W.~Hao and A.~Friedman, ``Mathematical model on \mbox{Alzheimer's} disease,''
  \emph{BMC Systems Biology}, vol.~10, no.~1, p. 108, 2016.

\bibitem{grasso2020molecular}
G.~Grasso and A.~Danani, ``Molecular simulations of amyloid beta assemblies,''
  \emph{Advances in Physics: X}, vol.~5, no.~1, p. 1770627, 2020.

\bibitem{kuznetsov2018formation}
I.~Kuznetsov and A.~Kuznetsov, ``How the formation of amyloid plaques and
  neurofibrillary tangles may be related: a mathematical modelling study,''
  \emph{Proceedings of the Royal Society A: Mathematical, Physical and
  Engineering Sciences}, vol. 474, no. 2210, p. 20170777, 2018.

\bibitem{dear2020catalytic}
A.~J. Dear, G.~Meisl, T.~C. Michaels, M.~R. Zimmermann, S.~Linse, and T.~P.
  Knowles, ``The catalytic nature of protein aggregation,'' \emph{The Journal
  of Chemical Physics}, vol. 152, no.~4, p. 045101, 2020.

\bibitem{vc2023comprehensive}
S.~S. VC, R.~Ponraj, K.~Kiran \emph{et~al.}, ``Comprehensive strategy for
  analyzing dementia brain images and generating textual reports through vit,
  faster r-cnn and gpt-2 integration,'' in \emph{2023 First International
  Conference on Advances in Electrical, Electronics and Computational
  Intelligence (ICAEECI)}.\hskip 1em plus 0.5em minus 0.4em\relax IEEE, 2023,
  pp. 1--10.

\bibitem{wang2018predictive}
T.~Wang, R.~G. Qiu, and M.~Yu, ``Predictive modeling of the progression of
  alzheimer’s disease with recurrent neural networks,'' \emph{Scientific
  reports}, vol.~8, no.~1, p. 9161, 2018.

\bibitem{bowles2018gan}
C.~Bowles, L.~Chen, R.~Guerrero, P.~Bentley, R.~Gunn, A.~Hammers, D.~A. Dickie,
  M.~V. Hern{\'a}ndez, J.~Wardlaw, and D.~Rueckert, ``Gan augmentation:
  Augmenting training data using generative adversarial networks,'' \emph{arXiv
  preprint arXiv:1810.10863}, 2018.

\bibitem{dragomir2018network}
A.~Dragomir, A.~G. Vrahatis, and A.~Bezerianos, ``A network-based perspective
  in alzheimer's disease: Current state and an integrative framework,''
  \emph{IEEE journal of biomedical and health informatics}, vol.~23, no.~1, pp.
  14--25, 2018.

\bibitem{de2020artificial}
S.~De~la Fuente~Garcia, C.~W. Ritchie, and S.~Luz, ``Artificial intelligence,
  speech, and language processing approaches to monitoring alzheimer’s
  disease: a systematic review,'' \emph{Journal of Alzheimer's Disease},
  vol.~78, no.~4, pp. 1547--1574, 2020.

\bibitem{monteverdi2023virtual}
A.~Monteverdi, F.~Palesi, M.~Schirner, F.~Argentino, M.~Merante, A.~Redolfi,
  F.~Conca, L.~Mazzocchi, S.~F. Cappa, M.~Cotta~Ramusino \emph{et~al.},
  ``Virtual brain simulations reveal network-specific parameters in
  neurodegenerative dementias,'' \emph{Frontiers in Aging Neuroscience},
  vol.~15, p. 1204134, 2023.

\bibitem{slanzi2020vitro}
A.~Slanzi, G.~Iannoto, B.~Rossi, E.~Zenaro, and G.~Constantin, ``In vitro
  models of neurodegenerative diseases,'' \emph{Frontiers in cell and
  developmental biology}, vol.~8, p. 328, 2020.

\bibitem{maestu2021neuronal}
F.~Maest{\'u}, W.~de~Haan, M.~A. Busche, and J.~DeFelipe, ``Neuronal
  excitation/inhibition imbalance: core element of a translational perspective
  on alzheimer pathophysiology,'' \emph{Ageing Research Reviews}, vol.~69, p.
  101372, 2021.

\bibitem{kashyap2019synapse}
G.~Kashyap, D.~Bapat, D.~Das, R.~Gowaikar, R.~Amritkar, G.~Rangarajan,
  V.~Ravindranath, and G.~Ambika, ``Synapse loss and progress of alzheimer’s
  disease-a network model,'' \emph{Scientific Reports}, vol.~9, no.~1, p. 6555,
  2019.

\bibitem{butcher2004systems}
E.~C. Butcher, E.~L. Berg, and E.~J. Kunkel, ``Systems biology in drug
  discovery,'' \emph{Nature biotechnology}, vol.~22, no.~10, pp. 1253--1259,
  2004.

\bibitem{yang2021bicoss}
S.~Yang, J.~Wang, X.~Hao, H.~Li, X.~Wei, B.~Deng, and K.~A. Loparo, ``Bicoss:
  toward large-scale cognition brain with multigranular neuromorphic
  architecture,'' \emph{IEEE Transactions on Neural Networks and Learning
  Systems}, vol.~33, no.~7, pp. 2801--2815, 2021.

\bibitem{argyle2023out}
L.~P. Argyle, E.~C. Busby, N.~Fulda, J.~R. Gubler, C.~Rytting, and D.~Wingate,
  ``Out of one, many: Using language models to simulate human samples,''
  \emph{Political Analysis}, vol.~31, no.~3, pp. 337--351, 2023.

\bibitem{shaheen2021analysis}
H.~Shaheen, R.~Melnik, and S.~Singh, ``Analysis of cortical spreading
  depression in brain with multiscale mathematical models,'' in \emph{Recent
  Developments in Mathematical, Statistical and Computational Sciences: The V
  AMMCS International Conference, Waterloo, Canada, August 18--23, 2019}.\hskip
  1em plus 0.5em minus 0.4em\relax Springer, 2021, pp. 197--207.

\bibitem{shaheen2023bayesian}
H.~Shaheen, R.~Melnik, and T.~A. D.~N. Initiative, ``Bayesian inference and
  role of astrocytes in amyloid-beta dynamics with modelling of alzheimer's
  disease using clinical data,'' \emph{ArXiv Preprint ArXiv:2306.12520}, 2023.

\bibitem{ajabani2023predicting}
D.~Ajabani, ``Predicting alzheimer's progression in mild cognitive impairment:
  Longitudinal mri with hmms and svm classifiers,'' \emph{IJACSA) International
  Journal of Advanced Computer Science and Applications}, vol.~14, no.~12,
  2023.

\bibitem{agha2021vitro}
H.~Agha and C.~R. McCurdy, ``In vitro and in vivo sigma 1 receptor imaging
  studies in different disease states,'' \emph{RSC Medicinal Chemistry},
  vol.~12, no.~2, pp. 154--177, 2021.

\bibitem{albahri2023systematic}
A.~Albahri, A.~M. Duhaim, M.~A. Fadhel, A.~Alnoor, N.~S. Baqer, L.~Alzubaidi,
  O.~Albahri, A.~Alamoodi, J.~Bai, A.~Salhi \emph{et~al.}, ``A systematic
  review of trustworthy and explainable artificial intelligence in healthcare:
  assessment of quality, bias risk, and data fusion,'' \emph{Information
  Fusion}, 2023.

\bibitem{valera2021multimodal}
J.~M. Valera~Bermejo, ``A multimodal approach to the study of self and
  others’ awareness in prodromal to mild alzheimer’s disease,'' Ph.D.
  dissertation, University of Sheffield, 2021.

\bibitem{wei2021self}
P.-H. Wei, H.~Chen, Q.~Ye, H.~Zhao, Y.~Xu, F.~Bai, and A.~D.~N. Initiative,
  ``Self-reference network-related interactions during the process of cognitive
  impairment in the early stages of alzheimer’s disease,'' \emph{Frontiers in
  Aging Neuroscience}, vol.~13, p. 666437, 2021.

\bibitem{golriz2020challenges}
S.~Golriz~Khatami, C.~Robinson, C.~Birkenbihl, D.~Domingo-Fern{\'a}ndez, C.~T.
  Hoyt, and M.~Hofmann-Apitius, ``Challenges of integrative disease modeling in
  alzheimer's disease,'' \emph{Frontiers in molecular biosciences}, vol.~6, p.
  158, 2020.

\bibitem{veitch2019understanding}
D.~P. Veitch, M.~W. Weiner, P.~S. Aisen, L.~A. Beckett, N.~J. Cairns, R.~C.
  Green, D.~Harvey, C.~R. Jack~Jr, W.~Jagust, J.~C. Morris \emph{et~al.},
  ``Understanding disease progression and improving alzheimer's disease
  clinical trials: Recent highlights from the alzheimer's disease neuroimaging
  initiative,'' \emph{Alzheimer's \& Dementia}, vol.~15, no.~1, pp. 106--152,
  2019.

\bibitem{bullmore2009complex}
E.~Bullmore and O.~Sporns, ``Complex brain networks: graph theoretical analysis
  of structural and functional systems,'' \emph{Nature Reviews Neuroscience},
  vol.~10, no.~3, pp. 186--198, 2009.

\bibitem{sporns2021dynamic}
O.~Sporns, J.~Faskowitz, A.~S. Teixeira, S.~A. Cutts, and R.~F. Betzel,
  ``Dynamic expression of brain functional systems disclosed by fine-scale
  analysis of edge time series,'' \emph{Network Neuroscience}, vol.~5, no.~2,
  pp. 405--433, 2021.

\bibitem{shine2019human}
J.~M. Shine, M.~Breakspear, P.~T. Bell, K.~A. Ehgoetz~Martens, R.~Shine,
  O.~Koyejo, O.~Sporns, and R.~A. Poldrack, ``Human cognition involves the
  dynamic integration of neural activity and neuromodulatory systems,''
  \emph{Nature neuroscience}, vol.~22, no.~2, pp. 289--296, 2019.

\bibitem{fox2018mapping}
M.~D. Fox, ``Mapping symptoms to brain networks with the human connectome,''
  \emph{New England Journal of Medicine}, vol. 379, no.~23, pp. 2237--2245,
  2018.

\bibitem{korhonen2021principles}
O.~Korhonen, M.~Zanin, and D.~Papo, ``Principles and open questions in
  functional brain network reconstruction,'' \emph{Human Brain Mapping},
  vol.~42, no.~11, pp. 3680--3711, 2021.

\bibitem{shine2021computational}
J.~M. Shine, E.~J. M{\"u}ller, B.~Munn, J.~Cabral, R.~J. Moran, and
  M.~Breakspear, ``Computational models link cellular mechanisms of
  neuromodulation to large-scale neural dynamics,'' \emph{Nature Neuroscience},
  vol.~24, no.~6, pp. 765--776, 2021.

\bibitem{jones2022computational}
D.~Jones, V.~Lowe, J.~Graff-Radford, H.~Botha, L.~Barnard, D.~Wiepert,
  M.~Murphy, M.~Murray, M.~Senjem, J.~Gunter \emph{et~al.}, ``A computational
  model of neurodegeneration in alzheimer’s disease,'' \emph{Nature
  communications}, vol.~13, no.~1, p. 1643, 2022.

\bibitem{sintini2021tau}
I.~Sintini, J.~Graff-Radford, D.~T. Jones, H.~Botha, P.~R. Martin, M.~M.
  Machulda, C.~G. Schwarz, M.~L. Senjem, J.~L. Gunter, C.~R. Jack~Jr
  \emph{et~al.}, ``Tau and amyloid relationships with resting-state functional
  connectivity in atypical alzheimer’s disease,'' \emph{Cerebral cortex},
  vol.~31, no.~3, pp. 1693--1706, 2021.

\bibitem{hijaz2020initiation}
B.~A. Hijaz and L.~A. Volpicelli-Daley, ``Initiation and propagation of
  $\alpha$-synuclein aggregation in the nervous system,'' \emph{Molecular
  Neurodegeneration}, vol.~15, pp. 1--12, 2020.

\bibitem{vaz2022extracellular}
M.~Vaz, T.~Soares~Martins, and A.~G. Henriques, ``Extracellular vesicles in the
  study of alzheimer's and parkinson's diseases: Methodologies applied from
  cells to biofluids,'' \emph{Journal of Neurochemistry}, vol. 163, no.~4, pp.
  266--309, 2022.

\bibitem{goloborshcheva2020reduced}
V.~V. Goloborshcheva, K.~D. Chaprov, E.~V. Teterina, R.~Ovchinnikov, and V.~L.
  Buchman, ``Reduced complement of dopaminergic neurons in the substantia nigra
  pars compacta of mice with a constitutive “low footprint” genetic
  knockout of alpha-synuclein,'' \emph{Molecular Brain}, vol.~13, no.~1, pp.
  1--4, 2020.

\bibitem{spillantini1998alpha}
M.~G. Spillantini, R.~A. Crowther, R.~Jakes, M.~Hasegawa, and M.~Goedert,
  ``$\alpha$-synuclein in filamentous inclusions of lewy bodies from
  \mbox{Parkinson’s} disease and dementia with lewy bodies,''
  \emph{Proceedings of the National Academy of Sciences}, vol.~95, no.~11, pp.
  6469--6473, 1998.

\bibitem{hirsch2009neuroinflammation}
E.~C. Hirsch and S.~Hunot, ``Neuroinflammation in parkinson's disease: a target
  for neuroprotection?'' \emph{The Lancet Neurology}, vol.~8, no.~4, pp.
  382--397, 2009.

\bibitem{bose2016mitochondrial}
A.~Bose and M.~F. Beal, ``Mitochondrial dysfunction in parkinson's disease,''
  \emph{Journal of Neurochemistry}, vol. 139, pp. 216--231, 2016.

\bibitem{goedert2013100}
M.~Goedert, M.~G. Spillantini, K.~Del~Tredici, and H.~Braak, ``100 years of
  lewy pathology,'' \emph{Nature Reviews Neurology}, vol.~9, no.~1, pp. 13--24,
  2013.

\bibitem{mckinnon2014ubiquitin}
C.~McKinnon and S.~J. Tabrizi, ``The ubiquitin-proteasome system in
  neurodegeneration,'' \emph{Antioxidants \& Redox Signaling}, vol.~21, no.~17,
  pp. 2302--2321, 2014.

\bibitem{testa2011germline}
J.~R. Testa, M.~Cheung, J.~Pei, J.~E. Below, Y.~Tan, E.~Sementino, N.~J. Cox,
  A.~U. Dogan, H.~I. Pass, S.~Trusa \emph{et~al.}, ``Germline bap1 mutations
  predispose to malignant mesothelioma,'' \emph{Nature Genetics}, vol.~43,
  no.~10, pp. 1022--1025, 2011.

\bibitem{yu2020potential}
H.~Yu, T.~Sun, J.~An, L.~Wen, F.~Liu, Z.~Bu, Y.~Cui, and J.~Feng, ``Potential
  roles of exosomes in \mbox{Parkinson’s} disease: from pathogenesis,
  diagnosis, and treatment to prognosis,'' \emph{Frontiers in Cell and
  Developmental Biology}, vol.~8, p.~86, 2020.

\bibitem{jain2019neuroprotection}
K.~K. Jain, ``Neuroprotection in alzheimer disease,'' in \emph{The Handbook of
  Neuroprotection}.\hskip 1em plus 0.5em minus 0.4em\relax Springer, 2019, pp.
  465--585.

\bibitem{liu2020stochastic}
R.-N. Liu and Y.-M. Kang, ``Stochastic master equation for early protein
  aggregation in the transthyretin amyloid disease,'' \emph{Scientific
  Reports}, vol.~10, no.~1, pp. 1--9, 2020.

\bibitem{vu2012progression}
T.~C. Vu, J.~G. Nutt, and N.~H. Holford, ``Progression of motor and nonmotor
  features of parkinson's disease and their response to treatment,''
  \emph{British journal of clinical pharmacology}, vol.~74, no.~2, pp.
  267--283, 2012.

\bibitem{khanna2018using}
S.~Khanna, D.~Domingo-Fern{\'a}ndez, A.~Iyappan, M.~A. Emon,
  M.~Hofmann-Apitius, and H.~Fr{\"o}hlich, ``Using multi-scale genetic,
  neuroimaging and clinical data for predicting alzheimer’s disease and
  reconstruction of relevant biological mechanisms,'' \emph{Scientific
  reports}, vol.~8, no.~1, p. 11173, 2018.

\bibitem{alexander2004biology}
G.~E. Alexander, ``Biology of parkinson's disease: pathogenesis and
  pathophysiology of a multisystem neurodegenerative disorder,''
  \emph{Dialogues in clinical neuroscience}, vol.~6, no.~3, pp. 259--280, 2004.

\bibitem{amartumur2024neuropathogenesis}
S.~Amartumur, H.~Nguyen, T.~Huynh, T.~S. Kim, R.-S. Woo, E.~Oh, K.~K. Kim,
  L.~P. Lee, and C.~Heo, ``Neuropathogenesis-on-chips for neurodegenerative
  diseases,'' \emph{Nature Communications}, vol.~15, no.~1, p. 2219, 2024.

\bibitem{ahmed2019machine}
A.~N. Ahmed, F.~B. Othman, H.~A. Afan, R.~K. Ibrahim, C.~M. Fai, M.~S. Hossain,
  M.~Ehteram, and A.~Elshafie, ``Machine learning methods for better water
  quality prediction,'' \emph{Journal of Hydrology}, vol. 578, p. 124084, 2019.

\bibitem{johnson2021precision}
K.~B. Johnson, W.-Q. Wei, D.~Weeraratne, M.~E. Frisse, K.~Misulis, K.~Rhee,
  J.~Zhao, and J.~L. Snowdon, ``Precision medicine, ai, and the future of
  personalized health care,'' \emph{Clinical and translational science},
  vol.~14, no.~1, pp. 86--93, 2021.

\bibitem{gupta2023perspective}
N.~S. Gupta and P.~Kumar, ``Perspective of artificial intelligence in
  healthcare data management: A journey towards precision medicine,''
  \emph{Computers in Biology and Medicine}, p. 107051, 2023.

\bibitem{dams2011modelling}
J.~Dams, B.~Bornschein, J.~P. Reese, A.~Conrads-Frank, W.~H. Oertel,
  U.~Siebert, and R.~Dodel, ``Modelling the cost effectiveness of treatments
  for parkinson’s disease: a methodological review,''
  \emph{Pharmacoeconomics}, vol.~29, pp. 1025--1049, 2011.

\bibitem{siebert2012state}
U.~Siebert, O.~Alagoz, A.~M. Bayoumi, B.~Jahn, D.~K. Owens, D.~J. Cohen, and
  K.~M. Kuntz, ``State-transition modeling: a report of the ispor-smdm modeling
  good research practices task force--3,'' \emph{Medical Decision Making},
  vol.~32, no.~5, pp. 690--700, 2012.

\bibitem{li2024cost}
N.~Li, J.~van~den Bergh, A.~Boonen, C.~Wyers, S.~Bours, and M.~Hiligsmann,
  ``Cost-effectiveness analysis of fracture liaison services: a markov model
  using dutch real-world data,'' \emph{Osteoporosis International}, vol.~35,
  no.~2, pp. 293--307, 2024.

\bibitem{kopec2010validation}
J.~A. Kopec, P.~Fin{\`e}s, D.~G. Manuel, D.~L. Buckeridge, W.~M. Flanagan,
  J.~Oderkirk, M.~Abrahamowicz, S.~Harper, B.~Sharif, A.~Okhmatovskaia
  \emph{et~al.}, ``Validation of population-based disease simulation models: a
  review of concepts and methods,'' \emph{BMC public health}, vol.~10, pp.
  1--13, 2010.

\bibitem{chiavenna2019estimating}
C.~Chiavenna, A.~M. Presanis, A.~Charlett, S.~de~Lusignan, S.~Ladhani, R.~G.
  Pebody, and D.~De~Angelis, ``Estimating age-stratified influenza-associated
  invasive pneumococcal disease in england: a time-series model based on
  population surveillance data,'' \emph{PLoS Medicine}, vol.~16, no.~6, p.
  e1002829, 2019.

\bibitem{chowell2016mathematical}
G.~Chowell, L.~Sattenspiel, S.~Bansal, and C.~Viboud, ``Mathematical models to
  characterize early epidemic growth: A review,'' \emph{Physics of life
  reviews}, vol.~18, pp. 66--97, 2016.

\bibitem{chen2014modeling}
D.~Chen, ``Modeling the spread of infectious diseases: A review,''
  \emph{Analyzing and modeling spatial and temporal dynamics of infectious
  diseases}, pp. 19--42, 2014.

\bibitem{dauer2003parkinson}
W.~Dauer and S.~Przedborski, ``Parkinson's disease: mechanisms and models,''
  \emph{Neuron}, vol.~39, no.~6, pp. 889--909, 2003.

\bibitem{deffains2019parkinsonism}
M.~Deffains and H.~Bergman, ``Parkinsonism-related $\beta$ oscillations in the
  primate basal ganglia networks--recent advances and clinical implications,''
  \emph{Parkinsonism \& Related Disorders}, vol.~59, pp. 2--8, 2019.

\bibitem{mamelak2018parkinson}
M.~Mamelak, ``\mbox{Parkinson’s} disease, the dopaminergic neuron and
  gammahydroxybutyrate,'' \emph{Neurology and Therapy}, vol.~7, no.~1, pp.
  5--11, 2018.

\bibitem{picca2020mitochondrial}
A.~Picca, R.~Calvani, H.~J. Coelho-Junior, F.~Landi, R.~Bernabei, and
  E.~Marzetti, ``Mitochondrial dysfunction, oxidative stress, and
  neuroinflammation: Intertwined roads to neurodegeneration,''
  \emph{Antioxidants}, vol.~9, no.~8, p. 647, 2020.

\bibitem{tan2009protein}
J.~M. Tan, E.~S. Wong, and K.-L. Lim, ``Protein misfolding and aggregation in
  parkinson's disease,'' \emph{Antioxidants \& redox signaling}, vol.~11,
  no.~9, pp. 2119--2134, 2009.

\bibitem{vidovic2022alpha}
M.~Vidovi{\'c} and M.~G. Rikalovic, ``Alpha-synuclein aggregation pathway in
  parkinson’s disease: current status and novel therapeutic approaches,''
  \emph{Cells}, vol.~11, no.~11, p. 1732, 2022.

\bibitem{muddapu2019computational}
V.~R. Muddapu, A.~Mandali, V.~S. Chakravarthy, and S.~Ramaswamy, ``A
  computational model of loss of dopaminergic cells in parkinson's disease due
  to glutamate-induced excitotoxicity,'' \emph{Frontiers in neural circuits},
  vol.~13, p.~11, 2019.

\bibitem{shaheen2022dbs}
H.~Shaheen and R.~Melnik, ``Deep brain stimulation with a computational model
  for the cortex-thalamus-basal-ganglia system and network dynamics of
  neurological disorders,'' \emph{Computational and Mathematical Methods}, vol.
  2022, pp. Article ID 8\,998\,150, 17, 2022.

\bibitem{barber2017neuroimaging}
T.~R. Barber, J.~C. Klein, C.~E. Mackay, and M.~T. Hu, ``Neuroimaging in
  pre-motor parkinson's disease,'' \emph{NeuroImage: Clinical}, vol.~15, pp.
  215--227, 2017.

\bibitem{leao1944spreading}
A.~A. Leao, ``Spreading depression of activity in the cerebral cortex,''
  \emph{Journal of Neurophysiology}, vol.~7, no.~6, pp. 359--390, 1944.

\bibitem{gerardo2017computational}
L.~Gerardo-Giorda and J.~M. Kroos, ``A computational multiscale model of
  cortical spreading depression propagation,'' \emph{Computers \& Mathematics
  with Applications}, vol.~74, no.~5, pp. 1076--1090, 2017.

\bibitem{cozzolino2018understanding}
O.~Cozzolino, M.~Marchese, F.~Trovato, E.~Pracucci, G.~M. Ratto, M.~G. Buzzi,
  F.~Sicca, and F.~M. Santorelli, ``Understanding spreading depression from
  headache to sudden unexpected death,'' \emph{Frontiers in Neurology}, vol.~9,
  p.~19, 2018.

\bibitem{chamanzar2018algorithm}
A.~Chamanzar, S.~George, P.~Venkatesh, M.~Chamanzar, L.~Shutter, J.~Elmer, and
  P.~Grover, ``An algorithm for automated, noninvasive detection of cortical
  spreading depolarizations based on eeg simulations,'' \emph{IEEE Transactions
  on Biomedical Engineering}, vol.~66, no.~4, pp. 1115--1126, 2018.

\bibitem{tuckwell2013stochastic}
H.~C. Tuckwell, ``Stochastic modeling of spreading cortical depression,'' in
  \emph{Stochastic Biomathematical Models}.\hskip 1em plus 0.5em minus
  0.4em\relax Springer, 2013, pp. 187--200.

\bibitem{tuckwell1981ion}
H.~C. Tuckwell and C.~L. Hermansen, ``Ion and transmitter movements during
  spreading cortical depression,'' \emph{International Journal of
  Neuroscience}, vol.~12, no.~2, pp. 109--135, 1981.

\bibitem{chang2013mathematical}
J.~C. Chang, K.~C. Brennan, D.~He, H.~Huang, R.~M. Miura, P.~L. Wilson, and
  J.~J. Wylie, ``A mathematical model of the metabolic and perfusion effects on
  cortical spreading depression,'' \emph{Plos One}, vol.~8, no.~8, p. e70469,
  2013.

\bibitem{duman2012synaptic}
R.~S. Duman and G.~K. Aghajanian, ``Synaptic dysfunction in depression:
  potential therapeutic targets,'' \emph{Science}, vol. 338, no. 6103, pp.
  68--72, 2012.

\bibitem{ellingsrud2022validating}
A.~J. Ellingsrud, D.~B. Dukefoss, R.~Enger, G.~Halnes, K.~Pettersen, and M.~E.
  Rognes, ``Validating a computational framework for ionic electrodiffusion
  with cortical spreading depression as a case study,'' \emph{Eneuro}, vol.~9,
  no.~2, 2022.

\bibitem{huang2011simplified}
H.~Huang, R.~M. Miura, and W.~Yao, ``A simplified neuronal model for the
  instigation and propagation of cortical spreading depression,''
  \emph{Advances in Applied Mathematics and Mechanics}, vol.~3, no.~6, pp.
  759--773, 2011.

\bibitem{kager2002conditions}
H.~Kager, W.~Wadman, and G.~Somjen, ``Conditions for the triggering of
  spreading depression studied with computer simulations,'' \emph{Journal of
  Neurophysiology}, vol.~88, no.~5, pp. 2700--2712, 2002.

\bibitem{shapiro2001osmotic}
B.~E. Shapiro, ``Osmotic forces and gap junctions in spreading depression: a
  computational model,'' \emph{Journal of Computational Neuroscience}, vol.~10,
  no.~1, pp. 99--120, 2001.

\bibitem{shukla2022molecular}
R.~Shukla, D.~F. Newton, A.~Sumitomo, H.~Zare, R.~Mccullumsmith, D.~A. Lewis,
  T.~Tomoda, and E.~Sibille, ``Molecular characterization of depression trait
  and state,'' \emph{Molecular psychiatry}, vol.~27, no.~2, pp. 1083--1094,
  2022.

\bibitem{kroos2017patient}
J.~M. Kroos, I.~Marinelli, I.~Diez, J.~M. Cortes, S.~Stramaglia, and
  L.~Gerardo-Giorda, ``Patient-specific computational modeling of cortical
  spreading depression via diffusion tensor imaging,'' \emph{International
  journal for numerical methods in biomedical engineering}, vol.~33, no.~11, p.
  e2874, 2017.

\bibitem{rovegno2018role}
M.~Rovegno and J.~C. Saez, ``Role of astrocyte connexin hemichannels in
  cortical spreading depression,'' \emph{Biochimica et Biophysica Acta
  (BBA)-Biomembranes}, vol. 1860, no.~1, pp. 216--223, 2018.

\bibitem{saetra2021electrodiffusive}
M.~J. S{\ae}tra, G.~T. Einevoll, and G.~Halnes, ``An electrodiffusive
  neuron-extracellular-glia model for exploring the genesis of slow potentials
  in the brain,'' \emph{PLoS Computational Biology}, vol.~17, no.~7, p.
  e1008143, 2021.

\bibitem{xu2020mathematical}
S.~Xu, J.~C. Chang, C.~C. Chow, K.~C. Brennan, and H.~Huang, ``A mathematical
  model for persistent post-csd vasoconstriction,'' \emph{PLoS computational
  biology}, vol.~16, no.~7, p. e1007996, 2020.

\bibitem{reshodko1975computer}
L.~Reshodko and J.~Bure{\v{s}}, ``Computer simulation of reverberating
  spreading depression in a network of cell automata,'' \emph{Biological
  cybernetics}, vol.~18, no.~3, pp. 181--189, 1975.

\bibitem{pettersen2012extracellular}
K.~H. Pettersen, H.~Lind{\'e}n, A.~M. Dale, and G.~T. Einevoll, ``Extracellular
  spikes and csd,'' \emph{Handbook of neural activity measurement}, vol.~1, pp.
  92--135, 2012.

\bibitem{desroches2019modeling}
M.~Desroches, O.~Faugeras, M.~Krupa, and M.~Mantegazza, ``Modeling cortical
  spreading depression induced by the hyperactivity of interneurons,''
  \emph{Journal of Computational Neuroscience}, vol.~47, no.~2, pp. 125--140,
  2019.

\bibitem{de2020neurons}
C.~De~Luca, A.~M. Colangelo, A.~Virtuoso, L.~Alberghina, and M.~Papa,
  ``Neurons, glia, extracellular matrix and neurovascular unit: a systems
  biology approach to the complexity of synaptic plasticity in health and
  disease,'' \emph{International journal of molecular sciences}, vol.~21,
  no.~4, p. 1539, 2020.

\bibitem{plenz2021self}
D.~Plenz, T.~L. Ribeiro, S.~R. Miller, P.~A. Kells, A.~Vakili, and E.~L. Capek,
  ``Self-organized criticality in the brain,'' \emph{Frontiers in Physics},
  vol.~9, p. 639389, 2021.

\bibitem{sun2021future}
K.~Sun, J.~Chen, and X.~Yan, ``The future of memristors: Materials engineering
  and neural networks,'' \emph{Advanced Functional Materials}, vol.~31, no.~8,
  p. 2006773, 2021.

\bibitem{chaisangmongkon2017computing}
W.~Chaisangmongkon, S.~K. Swaminathan, D.~J. Freedman, and X.-J. Wang,
  ``Computing by robust transience: how the fronto-parietal network performs
  sequential, category-based decisions,'' \emph{Neuron}, vol.~93, no.~6, pp.
  1504--1517, 2017.

\bibitem{einevoll2013modelling}
G.~T. Einevoll, C.~Kayser, N.~K. Logothetis, and S.~Panzeri, ``Modelling and
  analysis of local field potentials for studying the function of cortical
  circuits,'' \emph{Nature Reviews Neuroscience}, vol.~14, no.~11, pp.
  770--785, 2013.

\bibitem{oyetunde2018leveraging}
T.~Oyetunde, F.~S. Bao, J.-W. Chen, H.~G. Martin, and Y.~J. Tang, ``Leveraging
  knowledge engineering and machine learning for microbial bio-manufacturing,''
  \emph{Biotechnology advances}, vol.~36, no.~4, pp. 1308--1315, 2018.

\bibitem{michelson2018multi}
N.~J. Michelson, A.~L. Vazquez, J.~R. Eles, J.~W. Salatino, E.~K. Purcell,
  J.~J. Williams, X.~T. Cui, and T.~D. Kozai, ``Multi-scale, multi-modal
  analysis uncovers complex relationship at the brain tissue-implant neural
  interface: new emphasis on the biological interface,'' \emph{Journal of
  neural engineering}, vol.~15, no.~3, p. 033001, 2018.

\bibitem{nwagbo2024review}
C.~L. Nwagbo and E.~C. Genevra, ``Review of neuronal multiplexers with back
  propagation algorithm,'' \emph{IEEE-SEM}, vol.~2, 2024.

\bibitem{dhollander2021fixel}
T.~Dhollander, A.~Clemente, M.~Singh, F.~Boonstra, O.~Civier, J.~D. Duque,
  N.~Egorova, P.~Enticott, I.~Fuelscher, S.~Gajamange \emph{et~al.},
  ``Fixel-based analysis of diffusion mri: methods, applications, challenges
  and opportunities,'' \emph{Neuroimage}, vol. 241, p. 118417, 2021.

\bibitem{al2020giving}
L.~Al-Hassany, J.~Haas, M.~Piccininni, T.~Kurth, A.~Maassen Van Den~Brink, and
  J.~L. Rohmann, ``Giving researchers a headache--sex and gender differences in
  migraine,'' \emph{Frontiers in neurology}, vol.~11, p. 549038, 2020.

\bibitem{hampel2018revolution}
H.~Hampel, N.~Toschi, C.~Babiloni, F.~Baldacci, K.~L. Black, A.~L. Bokde, R.~S.
  Bun, F.~Cacciola, E.~Cavedo, P.~A. Chiesa \emph{et~al.}, ``Revolution of
  alzheimer precision neurology. passageway of systems biology and
  neurophysiology,'' \emph{Journal of Alzheimer's Disease}, vol.~64, no.~s1,
  pp. S47--S105, 2018.

\bibitem{sanacora2022stressed}
G.~Sanacora, Z.~Yan, and M.~Popoli, ``The stressed synapse 2.0:
  pathophysiological mechanisms in stress-related neuropsychiatric disorders,''
  \emph{Nature Reviews Neuroscience}, vol.~23, no.~2, pp. 86--103, 2022.

\bibitem{torres2020mir}
A.~Torres-Berr{\'\i}o, D.~Nouel, S.~Cuesta, E.~M. Parise, J.~M.
  Restrepo-Lozano, P.~Larochelle, E.~J. Nestler, and C.~Flores, ``Mir-218: a
  molecular switch and potential biomarker of susceptibility to stress,''
  \emph{Molecular psychiatry}, vol.~25, no.~5, pp. 951--964, 2020.

\bibitem{devor2012neuronal}
A.~Devor, D.~A. Boas, G.~T. Einevoll, R.~B. Buxton, and A.~M. Dale, ``Neuronal
  basis of non-invasive functional imaging: from microscopic neurovascular
  dynamics to bold fmri,'' \emph{Neural metabolism in vivo}, pp. 433--500,
  2012.

\bibitem{poort2016texture}
J.~Poort, M.~W. Self, B.~Van~Vugt, H.~Malkki, and P.~R. Roelfsema, ``Texture
  segregation causes early figure enhancement and later ground suppression in
  areas v1 and v4 of visual cortex,'' \emph{Cerebral cortex}, vol.~26, no.~10,
  pp. 3964--3976, 2016.

\bibitem{rajula2020comparison}
H.~S.~R. Rajula, G.~Verlato, M.~Manchia, N.~Antonucci, and V.~Fanos,
  ``Comparison of conventional statistical methods with machine learning in
  medicine: diagnosis, drug development, and treatment,'' \emph{Medicina},
  vol.~56, no.~9, p. 455, 2020.

\bibitem{alstadhaug2007periodicity}
K.~Alstadhaug, ``Periodicity of migraine,'' \emph{Migraine Disorders Research
  Trends. New York, NY: Nova Science}, pp. 107--44, 2007.

\bibitem{mubayi2019studying}
A.~Mubayi, C.~Kribs, V.~Arunachalam, and C.~Castillo-Chavez, ``Studying
  complexity and risk through stochastic population dynamics: Persistence,
  resonance, and extinction in ecosystems,'' in \emph{Handbook of
  Statistics}.\hskip 1em plus 0.5em minus 0.4em\relax Elsevier, 2019, vol.~40,
  pp. 157--193.

\bibitem{ball2000stochastic}
S.~Ball, ``Stochastic models of ion channels,'' Ph.D. dissertation, University
  of Nottingham, 2000.

\bibitem{de2014joint}
A.~de~Cheveign{\'e} and L.~C. Parra, ``Joint decorrelation, a versatile tool
  for multichannel data analysis,'' \emph{Neuroimage}, vol.~98, pp. 487--505,
  2014.

\bibitem{cipresso2023affects}
P.~Cipresso, F.~Borghesi, and A.~Chirico, ``Affects affect affects: A markov
  chain,'' \emph{Frontiers in Psychology}, vol.~14, p. 1162655, 2023.

\bibitem{Sevigny2016Antibody}
J.~Sevigny, P.~Chiao, T.~Bussi{\`e}re, P.~H. Weinreb, L.~Williams, M.~Maier,
  R.~Dunstan, S.~Salloway, T.~Chen, Y.~Ling \emph{et~al.}, ``The antibody
  aducanumab reduces a$\beta$ plaques in \mbox{Alzheimer's} disease,''
  \emph{Nature}, vol. 537, no. 7618, pp. 50--56, 2016.

\bibitem{Merchant2019Proposed}
K.~M. Merchant, J.~M. Cedarbaum, P.~Brundin, K.~D. Dave, J.~Eberling, A.~J.
  Espay, S.~J. Hutten, M.~Javidnia, J.~Luthman, W.~Maetzler \emph{et~al.}, ``A
  proposed roadmap for \mbox{Parkinson’s} disease proof of concept clinical
  trials investigating compounds targeting alpha-synuclein,'' \emph{Journal of
  Parkinson's Disease}, vol.~9, no.~1, pp. 31--61, 2019.

\bibitem{corti2024structure}
M.~Corti, F.~Bonizzoni, and P.~F. Antonietti, ``Structure preserving polytopal
  discontinuous galerkin methods for the numerical modeling of
  neurodegenerative diseases,'' \emph{Journal of Scientific Computing}, vol.
  100, no.~2, p.~39, 2024.

\bibitem{hadjichrysanthou2018development}
C.~Hadjichrysanthou, A.~K. Ower, F.~de~Wolf, R.~M. Anderson, and A.~D.~N.
  Initiative, ``The development of a stochastic mathematical model of
  \mbox{Alzheimer's} disease to help improve the design of clinical trials of
  potential treatments,'' \emph{Plos One}, vol.~13, no.~1, p. e0190615, 2018.

\bibitem{frisoni2022probabilistic}
G.~B. Frisoni, D.~Altomare, D.~R. Thal, F.~Ribaldi, R.~van~der Kant,
  R.~Ossenkoppele, K.~Blennow, J.~Cummings, C.~van Duijn, P.~M. Nilsson
  \emph{et~al.}, ``The probabilistic model of alzheimer disease: the amyloid
  hypothesis revised,'' \emph{Nature Reviews Neuroscience}, vol.~23, no.~1, pp.
  53--66, 2022.

\bibitem{Yamins2016Using}
D.~L. Yamins and J.~J. DiCarlo, ``Using goal-driven deep learning models to
  understand sensory cortex,'' \emph{Nature Neuroscience}, vol.~19, no.~3, pp.
  356--365, 2016.

\bibitem{Kriegeskorte2015Deep}
N.~Kriegeskorte, ``Deep neural networks: a new framework for modeling
  biological vision and brain information processing,'' \emph{Annual Review of
  Vision Science}, vol.~1, pp. 417--446, 2015.

\bibitem{Sussillo2009Generating}
D.~Sussillo and L.~F. Abbott, ``Generating coherent patterns of activity from
  chaotic neural networks,'' \emph{Neuron}, vol.~63, no.~4, pp. 544--557, 2009.

\bibitem{Barak2013Fixed}
O.~Barak, D.~Sussillo, R.~Romo, M.~Tsodyks, and L.~Abbott, ``From fixed points
  to chaos: three models of delayed discrimination,'' \emph{Progress in
  Neurobiology}, vol. 103, pp. 214--222, 2013.

\bibitem{Yang2020Artificial}
G.~R. Yang and X.-J. Wang, ``Artificial neural networks for neuroscientists: A
  primer,'' \emph{Neuron}, vol. 107, no.~6, pp. 1048--1070, 2020.

\bibitem{Lima2022Comprehensive}
A.~A. Lima, M.~F. Mridha, S.~C. Das, M.~M. Kabir, M.~R. Islam, and Y.~Watanobe,
  ``A comprehensive survey on the detection, classification, and challenges of
  neurological disorders,'' \emph{Biology}, vol.~11, no.~3, p. 469, 2022.

\bibitem{Sarishma2022Review}
D.~Sarishma, S.~Sangwan, R.~Tomar, and R.~Srivastava, ``A review on cognitive
  computational neuroscience: overview, models, and applications,''
  \emph{Innovative Trends in Computational Intelligence}, pp. 217--234, 2022.

\bibitem{Battaglia2022Functional}
S.~Battaglia and J.~F. Thayer, ``Functional interplay between central and
  autonomic nervous systems in human fear conditioning,'' \emph{Trends in
  Neurosciences}, 2022.

\bibitem{Wang2022Hierarchical}
S.~Wang, T.~H. Cheng, and M.~H. Lim, ``A hierarchical taxonomic survey of
  spiking neural networks,'' \emph{Memetic Computing}, vol.~14, no.~3, pp.
  335--354, 2022.

\bibitem{Bihl2023Artificial}
T.~Bihl, W.~A. Young~II, A.~Moyer, and S.~Frimel, ``Artificial neural networks
  and data science,'' in \emph{Encyclopedia of Data Science and Machine
  Learning}.\hskip 1em plus 0.5em minus 0.4em\relax IGI Global, 2023, pp.
  899--921.

\bibitem{Thukroo2022Review}
I.~A. Thukroo, R.~Bashir, and K.~J. Giri, ``A review into deep learning
  techniques for spoken language identification,'' \emph{Multimedia Tools and
  Applications}, vol.~81, no.~22, pp. 32\,593--32\,624, 2022.

\bibitem{ghebrehiwet2024revolutionizing}
I.~Ghebrehiwet, N.~Zaki, R.~Damseh, and M.~S. Mohamad, ``Revolutionizing
  personalized medicine with generative ai: a systematic review,''
  \emph{Artificial Intelligence Review}, vol.~57, no.~5, pp. 1--41, 2024.

\bibitem{storm2024integrative}
J.~F. Storm, P.~C. Klink, J.~Aru, W.~Senn, R.~Goebel, A.~Pigorini, P.~Avanzini,
  W.~Vanduffel, P.~R. Roelfsema, M.~Massimini \emph{et~al.}, ``An integrative,
  multiscale view on neural theories of consciousness,'' \emph{Neuron}, vol.
  112, no.~10, pp. 1531--1552, 2024.

\bibitem{liang2021explaining}
Y.~Liang, S.~Li, C.~Yan, M.~Li, and C.~Jiang, ``Explaining the black-box model:
  A survey of local interpretation methods for deep neural networks,''
  \emph{Neurocomputing}, vol. 419, pp. 168--182, 2021.

\bibitem{basheer2000artificial}
I.~A. Basheer and M.~Hajmeer, ``Artificial neural networks: fundamentals,
  computing, design, and application,'' \emph{Journal of microbiological
  methods}, vol.~43, no.~1, pp. 3--31, 2000.

\bibitem{irimata2020fundamental}
K.~E. Irimata, B.~N. Dugger, and J.~R. Wilson, \emph{Fundamental Statistical
  Methods for Analysis of \mbox{Alzheimer's} and Other Neurodegenerative
  Diseases}.\hskip 1em plus 0.5em minus 0.4em\relax Johns Hopkins University
  Press, 2020.

\bibitem{raket2020statistical}
L.~L. Raket, ``Statistical disease progression modeling in \mbox{Alzheimer's}
  disease,'' \emph{Frontiers in Big Data}, vol.~3, p.~24, 2020.

\bibitem{davenport2023neurodegenerative}
F.~Davenport, J.~Gallacher, Z.~Kourtzi, I.~Koychev, P.~M. Matthews, N.~P.
  Oxtoby, L.~M. Parkes, V.~Priesemann, J.~B. Rowe, S.~W. Smye \emph{et~al.},
  ``Neurodegenerative disease of the brain: a survey of interdisciplinary
  approaches,'' \emph{Journal of the Royal Society Interface}, vol.~20, no.
  198, p. 20220406, 2023.

\bibitem{adhikari2022exploiting}
S.~Adhikari, S.~Thapa, U.~Naseem, P.~Singh, H.~Huo, G.~Bharathy, and M.~Prasad,
  ``Exploiting linguistic information from nepali transcripts for early
  detection of alzheimer's disease using natural language processing and
  machine learning techniques,'' \emph{International Journal of Human-Computer
  Studies}, vol. 160, p. 102761, 2022.

\bibitem{Van2009Efficiency}
M.~P. Van Den~Heuvel, C.~J. Stam, R.~S. Kahn, and H.~E.~H. Pol, ``Efficiency of
  functional brain networks and intellectual performance,'' \emph{Journal of
  Neuroscience}, vol.~29, no.~23, pp. 7619--7624, 2009.

\bibitem{Farzan2016Enhancing}
F.~Farzan, A.~Pascual-Leone, J.~D. Schmahmann, and M.~Halko, ``Enhancing the
  temporal complexity of distributed brain networks with patterned cerebellar
  stimulation,'' \emph{Scientific Reports}, vol.~6, no.~1, pp. 1--9, 2016.

\bibitem{Ito2020Discovering}
T.~Ito, L.~Hearne, R.~Mill, C.~Cocuzza, and M.~W. Cole, ``Discovering the
  computational relevance of brain network organization,'' \emph{Trends in
  Cognitive Sciences}, vol.~24, no.~1, pp. 25--38, 2020.

\bibitem{Dubois2018Distributed}
J.~Dubois, P.~Galdi, L.~K. Paul, and R.~Adolphs, ``A distributed brain network
  predicts general intelligence from resting-state human neuroimaging data,''
  \emph{Philosophical Transactions of the Royal Society B: Biological
  Sciences}, vol. 373, no. 1756, p. 20170284, 2018.

\bibitem{Li2018Brain}
B.-J. Li, K.~Friston, M.~Mody, H.-N. Wang, H.-B. Lu, and D.-W. Hu, ``A brain
  network model for depression: From symptom understanding to disease
  intervention,'' \emph{CNS Neuroscience \& Therapeutics}, vol.~24, no.~11, pp.
  1004--1019, 2018.

\bibitem{Feng2021Distributed}
G.~Feng, Z.~Gan, F.~Llanos, D.~Meng, S.~Wang, P.~C. Wong, and
  B.~Chandrasekaran, ``A distributed dynamic brain network mediates linguistic
  tone representation and categorization,'' \emph{NeuroImage}, vol. 224, p.
  117410, 2021.

\bibitem{Xu2022Physics}
Y.~Xu, S.~Kohtz, J.~Boakye, P.~Gardoni, and P.~Wang, ``Physics-informed machine
  learning for reliability and systems safety applications: State of the art
  and challenges,'' \emph{Reliability Engineering \& System Safety}, p. 108900,
  2022.

\bibitem{Nabati2023Real}
M.~Nabati and S.~A. Ghorashi, ``A real-time fingerprint-based indoor
  positioning using deep learning and preceding states,'' \emph{Expert Systems
  with Applications}, vol. 213, p. 118889, 2023.

\bibitem{Whittington2018Spatiotemporal}
A.~Whittington, D.~J. Sharp, and R.~N. Gunn, ``Spatiotemporal distribution of
  $\beta$-amyloid in \mbox{Alzheimer} disease is the result of heterogeneous
  regional carrying capacities,'' \emph{Journal of Nuclear Medicine}, vol.~59,
  no.~5, pp. 822--827, 2018.

\bibitem{Chen2022Challenges}
G.~T. Chen and D.~H. Geschwind, ``Challenges and opportunities for precision
  medicine in neurodevelopmental disorders,'' \emph{Advanced Drug Delivery
  Reviews}, p. 114564, 2022.

\bibitem{Bender2021Artificial}
A.~Bender and I.~Cort{\'e}s-Ciriano, ``Artificial intelligence in drug
  discovery: what is realistic, what are illusions? part 1: ways to make an
  impact, and why we are not there yet,'' \emph{Drug Discovery Today}, vol.~26,
  no.~2, pp. 511--524, 2021.

\bibitem{Ding2020Towards}
K.~Ding, A.~Dragomir, R.~Bose, L.~E. Osborn, M.~S. Seet, A.~Bezerianos, and
  N.~V. Thakor, ``Towards machine to brain interfaces: Sensory stimulation
  enhances sensorimotor dynamic functional connectivity in upper limb
  amputees,'' \emph{Journal of Neural Engineering,}, vol.~17, no.~3, p. 035002,
  2020.

\bibitem{Niarakis2022Addressing}
A.~Niarakis, D.~Waltemath, J.~Glazier, F.~Schreiber, S.~M. Keating,
  D.~Nickerson, C.~Chaouiya, A.~Siegel, V.~No{\"e}l, H.~Hermjakob
  \emph{et~al.}, ``Addressing barriers in comprehensiveness, accessibility,
  reusability, interoperability and reproducibility of computational models in
  systems biology,'' \emph{Briefings in Bioinformatics}, vol.~23, no.~4, p.
  bbac212, 2022.

\bibitem{Schafer2021Bayesian}
A.~Sch{\"a}fer, M.~Peirlinck, K.~Linka, E.~Kuhl, \mbox{Alzheimer's} Disease
  Neuroimaging Initiative~(ADNI \emph{et~al.}, ``Bayesian physics-based
  modeling of tau propagation in \mbox{Alzheimer's} disease,'' \emph{Frontiers
  in Physiology}, p. 1081, 2021.

\bibitem{bilgel2019predicting}
M.~Bilgel, B.~M. Jedynak, \mbox{Alzheimer's} Disease Neuroimaging~Initiative
  \emph{et~al.}, ``Predicting time to dementia using a quantitative template of
  disease progression,'' \emph{\mbox{Alzheimer's} \& \mbox{Dementia}:
  Diagnosis, Assessment \& Disease Monitoring}, vol.~11, pp. 205--215, 2019.

\bibitem{ghazi2021robust}
M.~M. Ghazi, M.~Nielsen, A.~Pai, M.~Modat, M.~J. Cardoso, S.~Ourselin, and
  L.~S{\o}rensen, ``Robust parametric modeling of \mbox{Alzheimer's} disease
  progression,'' \emph{NeuroImage}, vol. 225, p. 117460, 2021.

\bibitem{Beaumont2019Approximate}
M.~A. Beaumont, ``Approximate bayesian computation,'' \emph{Annual Review of
  Statistics and its Application}, vol.~6, pp. 379--403, 2019.

\bibitem{Christopher2018Parameter}
J.~D. Christopher, N.~T. Wimer, C.~Lapointe, T.~R. Hayden, I.~Grooms, G.~B.
  Rieker, and P.~E. Hamlington, ``Parameter estimation for complex
  thermal-fluid flows using approximate bayesian computation,'' \emph{Physical
  Review Fluids}, vol.~3, no.~10, p. 104602, 2018.

\end{thebibliography}
\bibliographystyle{IEEEtran}

\EOD

\end{document}